\documentclass[a4paper,fleqn,usenatbib]{mnras}
\usepackage{newtxtext,newtxmath}
\usepackage[T1]{fontenc}
\usepackage{ae,aecompl}

\usepackage{graphicx}	
\usepackage{amssymb}	
\usepackage{enumitem} 
\usepackage{color} 
\usepackage[export]{adjustbox}
\usepackage{subfig}
\usepackage{mathtools} 
\usepackage{bm} 
\usepackage{hyperref} 
\usepackage{amsmath} 
\usepackage{algorithm} 
\usepackage[noend]{algpseudocode} 
\usepackage{pdflscape}
\usepackage{listings}   
\usepackage{verbatim}   
\usepackage{pgfplotstable}  
\usepackage{tikz}
\usepackage{capt-of} 
\usepackage{rotating}  
\usepackage{array} 
\usepackage{multirow} 
\usepackage{diagbox} 
\usetikzlibrary{backgrounds,shapes,arrows,positioning,calc,snakes,fit, chains}
\usepgflibrary{decorations.markings}

\renewcommand{\d}[1]{\ensuremath{\operatorname{d}\!{#1}}}
\makeatletter 
\newcommand\norm[1]{\left\lVert#1\right\rVert}                            
\def\BState{\State\hskip-\ALG@thistlm}
\newcommand{\ALOOP}[1]{\ALC@it\algorithmicloop\ #1%
  \begin{ALC@loop}}
\newcommand{\ENDALOOP}{\end{ALC@loop}\ALC@it\algorithmicendloop}

\makeatother
\algnewcommand{\Step}[1]{%
  \State \textbf{Step:}{\Statex}
  \Statex \hspace*{\algorithmicindent}\parbox[t]{.8\linewidth}{\raggedright #1}
}
\algnewcommand{\Initialize}[1]{%
  \State \textbf{Initialize:}
  \Statex \hspace*{\algorithmicindent}\parbox[t]{.8\linewidth}{\raggedright #1}
}
\newcolumntype{L}[1]{>{\raggedright\let\newline\\\arraybackslash\hspace{0pt}}m{#1}}
\newcolumntype{C}[1]{>{\centering\let\newline\\\arraybackslash\hspace{0pt}}m{#1}}
\newcolumntype{R}[1]{>{\raggedleft\let\newline\\\arraybackslash\hspace{0pt}}m{#1}}

\title[Covariant polarized radiative transfer on cosmological scales]{Covariant polarized radiative transfer on cosmological scales for 
investigating large-scale magnetic field structures}

\author[J. Y. H. Chan, K. Wu, A. Y. L. On, D. J. Barnes,  J. D. McEwen, and T. D. Kitching]
{Jennifer Y. H. Chan$^{1,2}$\thanks{E-mail: y.chan.12@ucl.ac.uk (JYHC), kinwah.wu@ucl.ac.uk (KW), alvina.on.09@ucl.ac.uk (AYLO), djbarnes@mit.edu (DJB), jason.mcewen@ucl.ac.uk (JDM), t.kitching@ucl.ac.uk (TDK)}, 
Kinwah Wu$^{1}$, Alvina Y. L. On$^{1}$, David J. Barnes$^{3}$, 
 \newauthor
Jason D. McEwen$^{1}$, Thomas D. Kitching$^{1}$\\
$^{1}$Mullard Space Science Laboratory, University College London, Holmbury St Mary, Surrey, RH5 6NT, UK\\
$^{2}$Department of Physics and Astronomy, University College London, Gower Street, London, WC1E 6BT, UK \\
$^{3}$Kavli Institute for Astrophysics and Space Research, Massachusetts Institute of Technology, Cambridge, MA 02139, USA}

\date{Accepted 2018 December 18. Received 2018 November 25; in original form 2018 July 3}

\pubyear{2018}

\begin{document}
\onecolumn
\label{firstpage}
\pagerange{\pageref{firstpage}--\pageref{lastpage}}
\maketitle

\begin{abstract}
Polarization of radiation is a powerful tool to study cosmic magnetism and analysis of polarization can be used as a diagnostic tool for large-scale structures. In this paper, we present a solid theoretical foundation for using polarized light to investigate large-scale magnetic field structures: the cosmological polarized radiative transfer (CPRT) formulation. The CPRT formulation is fully covariant. It accounts for cosmological and relativistic effects in a self-consistent manner and explicitly treats Faraday rotation, as well as Faraday conversion, emission, and absorption processes. The formulation is derived from the first principles of conservation of phase--space volume and photon number.  Without loss of generality, we consider a flat Friedmann--Robertson--Walker (FRW) space--time metric and construct the corresponding polarized radiative transfer equations. We propose an all-sky CPRT calculation algorithm, based on a ray-tracing method, which allows cosmological simulation results to be incorporated and, thereby, model templates of polarization maps to be constructed. Such maps will be crucial in our interpretation of polarized data, such as those to be collected by the Square Kilometer Array (SKA). We describe several tests which are used for verifying the code and demonstrate applications in the study of the polarization signatures in different distributions of electron number density and magnetic fields. We present a pencil-beam CPRT calculation and an all-sky calculation, using a simulated galaxy cluster or a model magnetized universe obtained from GCMHD+ simulations as the respective input structures. The implications on large-scale magnetic field studies are discussed; remarks on the standard methods using rotation measure are highlighted.
\end{abstract} 

\begin{keywords}
polarization -- radiative transfer -- magnetic fields 
-- large-scale structure of Universe -- radiation mechanisms: thermal -- radiation mechanisms: non-thermal 
\end{keywords}

\section{Introduction}\label{sec:Intro} 

Signatures of magnetic fields are seen on all scales, 
from planets \citep[see e.g.\,][]{Stevenson03, schubert11} 
and stars \citep[see e.g.\,][]{Parker70, Brandenburg05, beck08ISM, schrijver08bk, Vallee98, Vallee11a}, 
to galaxies \citep{Ferriere09, Vallee2011b, Beck13galacticBfields, Planck15dust, Planck16dust} 
and galaxy clusters \citep[see e.g.\,][]{Govoni06, Guidetti08, Vacca10, Pratley13, kronberg2016cosmic}. Magnetic fields should also permeate large-scale structures such as superclusters \citep[see e.g.\,][]{Xu06}, filaments \citep[see e.g.\,][]{Ryu98, Bruggen05, ryu2008turbulence}, 
walls, and voids \citep[see e.g.\,][]{Beck12Void}, 
as the early-time magnetic seeds get amplified during the structure formation and evolution processes in the Universe \citep[see e.g.\,][for comprehensive reviews]{Widrow02, Durrer13, kronberg2016cosmic}. However, observational evidence of these weak large-scale magnetic fields is scarce. Magnetic fields must have played a pivotal role in: 
(i) star formation by transporting angular momentum out from accretion discs and so allowing materials to accrete onto proto-stars \citep[see e.g.\,][]{Ballbus11}, 
(ii) jet production by affecting the central accretion, as well as accelerating and collimating the materials that form jets \citep[see e.g.][]{pudritz2012review}, 
(iii) cosmic-ray production through the acceleration of charged particles \citep{Fermi49}, and 
(iv) cosmic-ray propagation through deflecting the ray or confining the charged particles \citep[see e.g.][]{Jokipii66, Jokipii67}. However, the origins of large-scale magnetic fields, their co-evolution with astrophysical structures, and their properties at present are as of yet to be determined.
How magnetic fields impact the formation of the first structures 
and their subsequent evolution remains a pressing problem in contemporary astrophysics and cosmology. 
The current understanding, however, will be revolutionized as high-quality all-sky polarization data set will become available from upcoming radio telescopes, such as the Square Kilometer Array (SKA)\footnote{\url{https://www.skatelescope.org}} \citep[see e.g.][]{BeckGaensler04, Feretti04, Gaensler04, Melanie15}.

Polarization surveys by emerging generation of radio telescopes, such as the GaLactic and Extragalactic All-sky MWA (GLEAM) Survey \citep{Wayth15} on the Murchison Widefield Array (MWA\footnote{\url{http://www.mwatelescope.org/}}) \citep{Tingay13}, 
the polarization Sky Survey of the Universe's Magnetism (POSSUM) \citep{Gaensler10Possum} on the Australian SKA Pathfinder (ASKAP\footnote{\url{http://www.atnf.csiro.au/projects/askap/index.html}}) \citep{Hotan14ASKAP}, as well as the Multifrequency Snapshot Sky Survey (MSSS) \citep{Heald15MWSS} on the Low-Frequency Array (LOFAR\footnote{\url{http://www.lofar.org/}}) \citep{Haarlem13LOFAR}, 
due to their improved sensitivities and resolutions, can already access a domain of weak magnetic field strengths that was unexplored before. 
They enable investigations of magnetism in a variety of astrophysical sources. These experiments pave the way for broad-band spectro-polarimetric surveys to be performed by the SKA, which will be a game-changer. The SKA is an interferometric radio telescope designated to have a total collecting area of about a square kilometer in its complete configuration. Its sensitivity, bandwidths, and field-of-view will provide a transformational polarization data set with which the detection of the very weak magnetic field of the cosmic web\footnote{Current upper limits on the intergalactic field strength are all model-dependent but generally fall within the range of $|\mathbfit{B}^{\rm IGM}| \leq 10^{-8}$ to $10^{-9}$~G \citep[see e.g.][]{Kronberg94, Blasi99, Brown17}.} may become possible \citep{Giovannini15, Vazza15} -- with the SKA the evolution of magnetism in galaxies and galaxy clusters may be traced \citep{Gaensler15}, and the detailed internal structure of the magnetized cosmic plasmas (both across the medium and along the line-of-sight) may be mapped or imaged \citep[][]{Han15, Heald15}. An all-sky polarization survey performed with the SKA will significantly increase the density of known polarized background sources on the sky \citep{BeckGaensler04, Feretti04}. These sources serve as distant radio backlights, illuminating the magnetized Universe via the effect of Faraday rotation, i.e. rotation of the polarization plane of radiation as it travels and interacts with the magnetic fields threaded in an ionized medium. Such a data set will be immensely rich, containing information of the polarized sources themselves, as well as the foreground sources lying along the line-of-sight. These sources can be the Milky Way, nearby or distant galaxies, galaxy clusters, and even the cosmic filaments connecting clusters of galaxies. 
With all the exciting opportunities opened up by observational advances, 
the pressing questions to be addressed now are: 
how do we uncover and characterize the polarization signals from data, 
and ultimately, use them to infer and quantify magnetic field properties? 
How do we confront our theoretical models of cosmic magnetism against observations? 
More specifically, how do we compare simulation results which encode physical model predictions to 
the results obtained by observational experiments?  

This paper aims to address the second and third questions by providing a solid theoretical foundation and a 
polarized radiative transfer tool to investigate cosmic magnetism on large scales (i.e. Mpc scales and beyond). We present a new formulation of cosmological polarized radiative transfer (CPRT), which is fully covariant and is valid for polarization transfer in flat space--time. Our derivation is based on a covariant general relativistic radiative formulation stemmed from the first principles of conservation of phase--space volume and photon number \citep[][]{Fuerst04, Younsi12}. The covariant CPRT equation allows the properties of the magnetic fields to be captured as they co-evolve with the structures in the expanding Universe. Furthermore, since our formulation accounts for the relativistic and cosmological effects in a self-consistent manner, polarization evolution in various cosmic media as a function of redshift can be investigated. The formulation preserves the basic structure of the conventional polarized radiative transfer \citep[see e.g.][]{SazonovTsytovich68, Sazonov69a, 
JonesOdell77_transfer, JonesOdell77_inhomo, Pacholczyk77, Deglinnocenti85_PRT_FormalSol}, making it easy to implement for practical calculations, 
as we demonstrate in example problems and applications. 
Moreover, the formulation is general: 
it can be reduced to the form from which 
the conventional rotation measure (RM) quantity \citep[see e.g.][]{Rybicki86} is derived, 
assuming the absence of emission and absorption, insignificant Faraday conversion, and negligible effects of non-thermal electrons in the medium \citep[see][for details and the generalization of the standard RM expression to account for an isotropic distribution of non-thermal relativistic electrons with a power-law energy spectrum]{On18}. 
At the same time, 
since the CPRT formulation explicitly accounts for absorption, emission, and Faraday processes, 
its application is not restricted to any special cases. 
To our knowledge, our formulation of CPRT, which is applicable to study large-scale cosmic magnetism, is the first of its kind\footnote{Formulations and codes capable of computing general relativistic polarized radiative transfer (GRPRT) in 
the (curved) Kerr space--time metric have been extensively studied and presented \cite[][]{Broderick03, Broderick04, Shcherbakov11, Gammie12, Dexter16, MoscibrodzkaGammie17_IPOLE}. Their applications primary concern polarized emissions from magnetized accretion flows and jets around a spinning black hole. 
}. 

The CPRT formulation serves as a solid platform whereby, given some input distributions of electron number densities and magnetic fields, 
one can trace the rays and compute their intensities and polarization over redshifts. These inputs can either be generated by simple modeling or cosmological simulations. We devise and construct a ray-tracing algorithm that solves the CPRT equation, thereby constructing model templates and making theoretical all-sky intensity and polarization maps. These data outputs, when combined with advanced statistical methods for data analysis and characterization, will help us achieve a reliable interpretation of observational data, crucial for scientific  extraction. 
Results obtained from such a forward approach also provide an experimental test-bed for assessing line-of-sight component separation methods and methods used for characterizing signals themselves and the underlying physical processes. 

This paper focuses on laying the foundation of the cosmological polarized radiative transfer approach to study the structure of large-scale magnetic fields. In Section 2, we present the CPRT formulation and its derivation. Our ray-tracing algorithms for solving the CPRT equation are given in Section 3. Verification tests for the code implementation and their results are described in Section 4. Demonstrations of applying the CPRT calculations to practical astrophysical applications are discussed in Section 5. We perform a set of single-ray CPRT calculations, showcasing the ability of the tool to study the cosmological evolution of polarization with or without bright radio sources along the line-of-sight. We also demonstrate how to compute polarization maps of an astrophysical object and an entire polarized sky, interfacing cosmological MHD simulation results with the CPRT calculations. We highlight the implications of these calculations on large-scale magnetic field studies. In Section 6, we summarize the whole paper.

Unless otherwise specified, c.g.s. units and a $[\,-, +, +, +\,]$ signature are used throughout this work. 

\section{Cosmological polarized radiative transfer}\label{sec:CPRTderivation} 

The CPRT formulation is derived based on a covariant general relativistic radiative transfer (GRRT) formulation \cite[][]{Fuerst04, Younsi12}, stemming from the first principles of conservation of phase--space volume and photon number. We start off by reviewing the polarized radiative transfer equation and the GRRT formulation to derive the covariant CPRT formulation. Then, we construct the corresponding CPRT equations assuming a flat Friedmann--Robertson--Walker (FRW) space--time metric, without loss of generality. 

\subsection[]{Conventional polarized radiative transfer}\label{sec:Conveneqn} 
We first set out the polarized radiative transfer (PRT) equation and show how the 
covariant formulation of radiative transfer can be directly generalized to that of the PRT. 

In the absence of scattering, the transfer equation of polarized radiation, in tensor representation, can be written as 
\begin{eqnarray}
\frac{\d I_{i, \nu}}{\d s}=-\kappa_{ij, \nu}I_{j, \nu}+\epsilon_{i, \nu}  \   , 
\label{eq:tensorRT}
\end{eqnarray} 
or in the matrix form, 
\begin{eqnarray}
 \frac{\rm d}{\d s}\left[ \begin{array}{c}
I_{\nu} \\
Q_{\nu}\\
U_{\nu}\\
V_{\nu} \end{array} \right]= - \left[ \begin{array}{cccc}
\kappa_{\nu} & q_{\nu} & u_{\nu} & v_{\nu} \\
q_{\nu} & \kappa_{\nu} & f_{\nu} & -g_{\nu}  \\
u_{\nu} & -f_{\nu} & \kappa_{\nu} & h_{\nu}  \\
v_{\nu} & g_{\nu} & -h_{\nu} &\kappa_{\nu} \end{array} \right]\left[ \begin{array}{c}
I_{\nu} \\
Q_{\nu}\\
U_{\nu}\\
V_{\nu}   \end{array} \right]
+\left[ \begin{array}{c}
\epsilon_{I, \nu} \\
\epsilon_{Q, \nu}  \\
\epsilon_{U, \nu} \\
\epsilon_{V, \nu} \end{array} \right]    
\label{eq:PRT}
\end{eqnarray} 
\citep{Sazonov69a, Pacholczyk70, Pacholczyk77, JonesOdell77_transfer, Deglinnocenti85_PRT_FormalSol, HuangShcherbakov11, Janett17_PRTI,Janett17_PRTII, Janett18_PRTIII}, 
where $s$ is the path length of the radiation; the tensor index $i$ or $j$ in equation\,(\ref{eq:tensorRT}) runs from 1 to 4, 
denoting the Stokes parameters $I_{\nu}$, $Q_{\nu}$, $U_{\nu}$ and $V_{\nu}$, respectively. The coefficient tensor $\kappa_{ij, \nu}$ accounts for the amount of absorption (through $\kappa_{\nu}$, $q_{\nu}$, $u_{\nu}$ and $v_{\nu}$), rotation (through $f_{\nu}$) and conversion (through $h_{\nu}$ and $g_{\nu}$) of the radiation along its direction of propagation, and $\epsilon_{i, \nu}$ accounts for the amount of emission. Essentially, absorption acts as a sink; emission serves as a source. Propagation effects of Faraday rotation and Faraday conversion are non-dispersive. Faraday rotation, due to circular birefringence (i.e. the slightly different speeds at which the left and right circularly waves travel in a magneto-ionic medium), results in the change of polarization angles as radiation propagates (i.e. $Q_{\nu} \leftrightarrow U_{\nu}$). Faraday conversion, due to linear birefringence, concerns with the interconversion between the linear and circular polarization modes of the radiation (i.e. $Q_{\nu} \leftrightarrow V_{\nu}$; $U_{\nu} \leftrightarrow V_{\nu}$). The equation for the transfer of polarized radiation presented at above, with variable transfer coefficients, is suitable for transport in a homogeneous or weakly anisotropic medium \citep{SazonovTsytovich68, Sazonov69a, Pacholczyk77, JonesOdell77_transfer}. Note that all quantities in equations\,(\ref{eq:tensorRT}) and (\ref{eq:PRT}) depend on the frequency of the radiation $\nu$. 

It is useful to note that the Stokes parameters are observables fully describing the properties of light but are {\it coordinate-system dependent} quantities. They can be combined in the complex forms, i.e. $(Q_{\nu} \pm iU_{\nu})$, 
and be linearly transformed to so-called $E$- and $B$- modes, 
which describe, respectively, parity-odd polarization and parity-even polarization, and so are invariant under transform of coordinate systems. 
Stokes parameters alone are not rotationally invariant.
Therefore, coordinate systems adopted, as well as the definitions and conventions of polarization, must be explicitly stated to remove any ambiguities in the interpretation of the Stokes results. We note that different handedness of coordinate systems (right-handed or left-handed), as well as the geometry of the problem, have been used in the literature that derived the polarized radiative transfer equations and the (thermal and non-thermal) transfer coefficients \citep[see e.g.][]{Sazonov69a, Pacholczyk70, Melrose91Book, HuangShcherbakov11}. The sign of Stokes $V_{\nu}$ that describes the sense of the circular polarization also varies from paper to paper. Furthermore, different conventions have been used in the literature regarding the definition of the polarization angle\footnote{
Investigations of the polarization of the cosmic microwave background adopt the opposite convention 
to the International Astronomical Union (IAU) standard, 
for which polarization angle increases clockwise (counterclockwise) when looking at the source for the former (latter). 
To rectify the discrepancy requires an opposite sign applied to Stokes $U_{\nu}$ 
(see \url{https://aas.org/posts/news/2015/12/iau-calls-consistency-use-polarization-angle}).}
, the definition of the handedness of circular polarization, and the definition of $V_{\nu}$ \citep[see][for a compilation of the conventions used in radio polarization work]{Robishaw08}.  
We thus define in Appendix~\ref{app:adoptedCoordsys} the coordinate systems and the geometry of the problem considered in this work, and discuss in 
Appendix~\ref{app:Convention} the intricacies of keeping a consistent polarization convention. 

Here, we note that the $U_{\nu}$ components $u_{\nu}$, $g_{\nu}$ and $\epsilon_{U, \nu}$ can vanish 
(so $V_{\nu}$ couples only to $U_{\nu}$, i.e. $U_{\nu} \leftrightarrow V_{\nu}$ but $Q_{\nu} \nleftrightarrow V_{\nu}$) 
by a choice of a local coordinate system \citep[see e.g.][]{Sazonov69a, Pacholczyk77}. With the geometry defined in Fig.~\ref{fig:Coord} in Appendix~\ref{app:adoptedCoordsys}, 
$u_{\nu}$, $g_{\nu}$ and $\epsilon_{U, \nu}$ become zero in the basis $(x, y)$ since the projection of the magnetic field onto the $(x, y)$-plane is parallel to $y$.

Another useful remark concerns the features of equations\,(\ref{eq:tensorRT}) and (\ref{eq:PRT}). 
They reduce to the usual scalar radiative transfer equation only when a specific intensity $I_{\nu}$ is considered, 
i.e. ${\rm d}I_{\nu}/{\d s}=-\kappa_{\nu}\,I_{\nu}+\epsilon_{\nu}$. 
Conversely, one can utilize the fact that all the Stokes parameters have the same physical units to easily include polarization in the covariant formulation of radiative transfer, 
as is outlined in the subsequent subsection. 

\subsection[]{Covariant general relativistic radiative transfer}\label{sec:CovariantGRFormu} 

From the first principles of conservation of photon number and phase space volume, 
it can be shown that 
the Lorentz-invariant intensity is given by $\mathcal{I}_{\nu} \equiv \ {I_{\nu}}/{\nu^{3}} $, and that 
the covariant formulation of the radiative transfer takes the form 
\begin{eqnarray} 
\frac{\d {\mathcal I}_{\nu} }{\d \tau_{\nu}}
=-{\mathcal I}_{\nu} +\frac{\xi_{\nu}}{\zeta_{\nu}}= -{\mathcal I}_{\nu} + {\mathcal S}_{\nu}    
\label{eq:covarRTmain}
\end{eqnarray} 
(see Appendix~\ref{App:Covariant}), 
where  $\tau_{\nu} = \int \kappa_{\nu}\,{\rm d}s$ is the optical depth, 
$\zeta_{\nu} = \nu\,\kappa_{\nu}$ and $\xi_{\nu} = {\epsilon_{\nu}}/\nu^{2}$ are 
the Lorentz-invariant coefficients of absorption and emission 
respectively, and 
the Lorentz-invariant source function is defined by ${\mathcal S_{\nu}} \equiv {\xi_{\nu}}/{\zeta_{\nu}} $. 

In relativistic settings, 
we want the covariant radiative transfer equation to be evaluated in space--time intervals instead of 
optical depth or path length. 
This can be achieved by introducing the mathematical affine parameter $\lambda_{\rm a}$. 
The problem is then translated into an evaluation of ${\d s}/{\d \lambda_{\rm a}}$ (i.e.\,the variation in the path length $s$ with respect to  $\lambda_{\rm a}$), 
and asking the question of what is the co-moving 4-velocity $v^{\beta}$  of a photon traveling in a fluid that has 4-velocity $u^{\beta}$. 

Assuming the photon has a 4-momentum $k^{\alpha}$, then the co-moving 4-velocity $v^{\beta}$ can be obtained by the projection of $k^{\alpha}$ on to the fluid frame, i.e. 
\begin{eqnarray}
v^{\beta} =  P^{\alpha \beta}k_{\alpha}  = k^{\beta}+(k_{\alpha}u^{\alpha})u^{\beta}  
\label{eq:v1}
\end{eqnarray}
\citep{Fuerst04}, where we have used the projection tensor $P^{\alpha \beta}=g^{\alpha \beta}+u^{\alpha}u^{\beta}$, with $g^{\alpha \beta}$ as the space--time metric tensor. 
The variation in $s$ with respect to  $\lambda_{\rm a}$ is therefore 
\begin{eqnarray}
\frac{\d s}{\d \lambda_{\rm a}} 
&=&-\norm{v^{\beta}}\Big|_{\lambda_{\rm a,obs}}           \nonumber \\
&=& - \sqrt{g_{\alpha \beta}(k^{\beta}+(k_{\alpha}u^{\alpha})u^{\beta})(k^{\alpha}+(k_{\beta}u^{\beta})u^{\alpha})}\Big|_{\lambda_{\rm a,obs}}                                     \nonumber \\
    &=& - k_{\alpha}u^{\alpha}\Big|_{\lambda_{\rm a,obs}}   
\label{eq:kaua}
\end{eqnarray} 
\citep[]{Younsi12}. 
Note that for a stationary observer positioned at infinity 
$k_{\beta}u^{\beta}= -E_{\rm obs}$. 
It follows that the ratio 
\begin{eqnarray}
\frac{k_{\alpha}u^{\alpha}\Big|_{\lambda_{\rm a, co}}}{k_{\beta}u^{\beta}\Big|_{\lambda_{\rm a, obs}}}= \frac{\nu_{\rm co}}{\nu_{\rm obs}}  \    , 
\label{eq:ratiorel}
\end{eqnarray} 
which corresponds to the relative energy shift of the photon between the observer's frame and the comoving frame. Using the Lorentz-invariant properties of $\mathcal{I}_{\nu}$, $\zeta_{\nu}$ and $\xi_{\nu}$ yields the 
covariant relativistic radiative transfer equation 
\begin{eqnarray} 
\frac{{\rm d}{\mathcal I_{\nu}}}{\d \lambda_{\rm a}} 
&= &  -k_{\alpha}u^{\alpha}\Big|_{\lambda_{\rm a, co}} 
\Big(-\kappa_{{\rm co}, \nu}\,{\mathcal I}_{\nu}+\frac{\epsilon_{{\rm co}, \nu}}{\nu_{\rm co}^3}\Big)   
\label{eq:covarRT}
\end{eqnarray}
\citep[]{Younsi12}, 
where all the quantities are frequency dependent and are evaluated along the path of a photon, i.e. comoving as denoted by the subscript ``co".

\subsection[]{Cosmological polarized radiative transfer formulation}\label{sec:CovariantFormu} 

The CPRT formulation is constructed by making two generalizations to the GRRT:
(i) by accounting for the polarization of the radiation and 
(ii) by incorporating a cosmological model to describe the space--time geometry of the Universe in which the radiation propagates. 
The former generalization is straightforward in the sense that 
the PRT equation takes the general form of radiative transfer\,(see Section \ref{sec:Conveneqn}) and that 
all the Stokes parameters have the same physical units. 
Therefore, similar to how one can obtain the Lorentz-invariant intensity by taking $\mathcal{I}_{\nu} \equiv {I_{\nu}}/{\nu^{3}}$, 
the invariant Stokes parameters are obtained by 
$\mathcal{I}_{\nu, i}= [{\mathcal I}_{\nu}, {\mathcal Q}_{\nu}, {\mathcal U}_{\nu}, {\mathcal V}_{\nu}]^{\rm T}  = [I_{\nu},Q_{\nu},U_{\nu},V_{\nu}]^{\rm T} /\nu^{3}$ 
where the tensor index $i$ runs from 1 to 4, and the superscript ``T" denotes the transpose (for notational simplicity we drop the subscript $\nu$ of 
the Stokes parameters and in the coefficients of absorption and emission hereafter). It follows that the covariant polarized radiative transfer equation, in tensor notation, takes the form 
\begin{eqnarray}
\frac{\d (\mathcal{I}_{i, \rm co})}{\d \lambda_{\rm a}}= 
\frac{\d (I_{i, \rm co}/\nu_{\rm co}^{3})}{\d \lambda_{\rm a}}= 
-k_{\alpha}u^{\alpha}\Big|_{\lambda_{\rm a, co}} 
\left\{-\kappa_{ij,{\rm co}}\, \left(\frac{I_{j}}{\nu_{\rm co}^{3}}\right)+
\frac{\epsilon_{i,{\rm co}}}{\nu_{\rm co}^{3}}  \right\}  \   .
\label{eq:covarPRT123}
\end{eqnarray}  

Next, to make the formulation appropriate in cosmological settings and, therefore, suitable for (but not limited to) the investigation of cosmological magnetic fields, 
the factor $k_{\alpha}u^{\alpha}$ is to be evaluated using the space--time metric of a chosen cosmological model such that equation\,(\ref{eq:covarPRT123}) is evaluated in terms of a cosmological variable, e.g.\, the redshift $z$, instead of the mathematical affine parameter $\lambda_{\rm a}$. 

Without loss of generality, we consider a flat FRW universe whose space--time metric 
has the diagonal elements \,($-1 , a^2 , a^2 , a^2$), where $a = 1/(1+z)$ is the cosmological scale factor describing the expansion of the universe. 
For simplicity, we consider a photon propagating radially in a cosmological medium with 4-velocity $u^{\beta}$, i.e. 
\begin{eqnarray}
k^{\alpha} =  \left[ \begin{array}{c}
E \\
p_{r}\\
p_{\theta}\\
p_{\phi} \end{array} \right] = \nu
\left[ \begin{array}{c}
1 \\
1\\
0\\
0 \end{array} \right]  \  ;  \hspace*{0.8cm}
u^{\beta} = \gamma 
\left[ \begin{array}{c}
1 \\
\beta_{r}\\
\beta_{\theta}\\
\beta_{\phi} \end{array} \right]  \   ,
\end{eqnarray}
where ${\mathbfit{p}} = (p_{r}, p_{\theta}, p_{\phi}) $ denotes the 3-velocity of the photon, {\boldmath${\beta}$} $ =(\beta_{r}, \beta_{\theta}, \beta_{\phi})$ denotes the 3-velocity of the medium, and $\gamma = 1/\sqrt{(1+ \beta^2)}$ is the corresponding Lorentz factor\,(here, we use $c=h=1$). Evaluating $k_{\alpha}u^{\alpha}$ then yields  
\begin{eqnarray}
{k_{\alpha}u^{\alpha}\Big|_{z}} &=& \gamma_{z} \nu_{z} (-1+a^{2}\beta_{r, z})   \   ,
\end{eqnarray}
and the ratio 
\begin{eqnarray}
\frac{{k_{\alpha}u^{\alpha}\Big|_{z}}}{k_{\beta}u^{\beta}\Big|_{z_{\rm obs}}} &=& \frac{\nu_{z}}{\nu_{z_{\rm obs}}}\,\left(\frac{\gamma_{z}}{\gamma_{z_{\rm obs}}} \frac{(a^{2}\,\beta_{r, z}-1)}{(a_{\rm obs}^{2}\,\beta_{r, {z_{\rm obs}}}-1)}\right)  \  .
\label{eq:ratioku}
\end{eqnarray}
If the motion of the medium can be neglected\,(i.e. $\beta$=0, $\gamma$=1), 
the ratio is simplified to 
\begin{eqnarray}
\frac{{k_{\alpha}u^{\alpha}\Big|_{z}}}{k_{\beta}u^{\beta}\Big|_{z_{\rm obs}}}
= \frac{\nu_{\rm z}}{\nu_{z_{\rm obs}}}  \  , 
\label{eq:ratioCPRT}
\end{eqnarray} 
which is the relative shift of energy (or frequency) of the photon, as one expects from equation\,(\ref{eq:ratiorel}). 
By defining $k^{\alpha} = (E, {\mathbfit{p}}) = {\d {x^{\alpha}}}/{\d \lambda_{\rm a}}$, one may also obtain 
\begin{eqnarray} 
 \frac{\rm d}{{\rm d} \lambda_{\rm a}} =  \frac{{\rm d}\,x^{0}}{{\rm d}\lambda_{\rm a}}  \frac{\rm d}{{\rm d}\,x^{0}} = E \frac{\rm d}{{\rm d}s}  = E\,\frac{{\rm d}z}{{\rm d}s} \frac{\rm d}{{\rm d}z}  
 \label{eq:chainrule1} \   , 
\end{eqnarray} 
and use this to also show that the photon's energy $E \propto {a}^{-1}$ and thus 
\begin{eqnarray} 
\frac{\nu_{z}}{\nu_{z_{\rm obs}}}= \frac{a_{\rm obs}}{a}= \frac{1+z}{1+z_{\rm obs}}\    ,
\end{eqnarray} 
in a flat FRW universe \citep[see e.g.\,][]{dodelson2003modern}.   
In other words, the ratio in equation\,(\ref{eq:ratioCPRT}) corresponds to the relative energy shift of the photon due to the cosmic expansion. 

Finally, by applying the chain rule given in equation\,(\ref{eq:chainrule1}) to equation\,(\ref{eq:covarPRT123}), we obtain the CPRT equation defined in redshift space: 
\begin{eqnarray}
\frac{\rm d}{\d z}\left[ \begin{array}{c}
{\mathcal I}\\
{\mathcal Q}\\
{\mathcal U}\\
{\mathcal V}\end{array} \right]  
= ({1+z})
 \left\{ - \left[ \begin{array}{cccc}
\kappa & q & u & v \\
q & \kappa & f & -g  \\
u & -f & \kappa & h  \\
v & g & -h &\kappa \end{array} \right] \left[ \begin{array}{c}
{\mathcal I}\\
{\mathcal Q}\\
{\mathcal U}\\
{\mathcal V}\end{array} \right]    \right . +  \left.   \left[ \begin{array}{c}
\epsilon_{I} \\
\epsilon_{Q}  \\
\epsilon_{U}  \\
\epsilon_{V} \end{array} \right]\frac{1}{\nu^3} \right\} \frac{\d s}{\d z}    \  ,    
\label{eq:covarPRTinz}
\end{eqnarray} 
where all the quantities are Lorentz invariant 
and ${\d s}/{\d z}$ for a flat FRW universe is given by 
\begin{eqnarray} 
\frac{\d s}{\d z}=\frac{c}{H_{0}}\,({1+z})^{-1}\left[\Omega_{{\rm r},0}(1+z)^{4}+\Omega_{{\rm m},0}(1+z)^{3}+\Omega_{\Lambda,0}\right]^{-\frac{1}{2}}  \    ,
\label{eq:dldz}
\end{eqnarray} 
\citep[see e.g.\,][]{peacock1999}, 
where $H_{0}$ is the standard Hubble parameter, $\Omega_{{\rm r},0}$, $\Omega_{{\rm m},0}$ and $\Omega_{\Lambda,0}$ are the 
dimensionless energy densities of relativistic matter and radiation, non-relativistic matter, 
and a cosmological constant (dark energy with an equation of state of $w \equiv -1$), respectively. 
The subscript ``0" denotes that the quantities are measured at the present epoch\,(i.e. $z=0$).  

Note that the CPRT formulation is general 
and can adopt different cosmological models with flat space--time geometry through the $k_{\alpha}u^{\alpha}$ factor. 
Ray-tracing calculation for equation\,(\ref{eq:covarPRTinz}) can then be performed for arbitrary photon geodesics. 
For clarity, we reiterate that a flat space--time is considered in our derivation such that straightforward parallel transport of the polarization Stokes vector 
$S_{\nu, i} = [I_{\nu}, Q_{\nu}, U_{\nu}, V_{\nu}]^{\rm T}$ of 
the radiation along the photon geodesics is enabled\footnote{
For completeness, we note that polarized radiative transfer in Kerr space--time 
has been extensively studied \citep[][]{Broderick03, Broderick04, Shcherbakov11, Gammie12, Dexter16, MoscibrodzkaGammie17_IPOLE}, 
for which the standard approach involves solving the photon geodesic, 
keeping track of the local coordinate system such that polarized emission is being added appropriately 
in the presence of a rotation of the coordinate system propagated along the ray, and finally, 
connecting these frames to the polarization frame at the point of observation. 
Difficulties stem from that the Stokes parameters are {\it not} rotationally invariant quantities. 
Working with rotationally invariant quantities, e.g. the spin-2 signals of $E$- and $B$-modes, might therefore be more favourable.
}. 
For radiation propagating in a curved space--time, 
the rotation of its polarization vector measured by the observer has a contribution caused not only by the Faraday rotation 
but also by the curvature of the embedded manifold, i.e. angle is not preserved transporting along the line-of-sight. 
Taking advantage of the flat geometry of the Universe \citep{Planck15}, 
we therefore limit our evaluation and discussion to a cosmological model describing a flat universe only. 
The flatness of space--time ensures that the angles measured in the local comoving frame would be the same everywhere along the geodesic. 

We highlight that the covariant nature of the CPRT formulation 
allows a straightforward transform of an observable between the comoving frame 
and the observer's frame. 
Computation from the invariant Stokes parameters to the observable Stokes parameters in the comoving frame 
requires only a scalar multiplication of the cube of the radiation frequency, 
i.e. $ [I_{\nu} (z), Q_{\nu}(z), U_{\nu}(z), V_{\nu}(z)]^{\rm T} = [{\mathcal I}_{\nu}(z), {\mathcal Q}_{\nu}(z), {\mathcal U}_{\nu}(z), {\mathcal V}_{\nu}(z)]^{\rm T}  \times \nu(z)^{3}$. 
The results at $z=0$ are then what would be measured in the observer's frame at the present time, 
provided that the transform of the local polarization frame to the instrument's polarization frame are properly handled 
(as is noted in Appendix \ref{app:adoptedCoordsys}), 
along with the corrections of instrumental effects and foregrounds, such as ionospheric effects. 

\subsection[]{Polarized transfer coefficients}\label{subsec:TransCoef}

Following the derivation of the CPRT equation, i.e. equation (\ref{eq:covarPRTinz}),   
in this subsection we discuss the corresponding transfer coefficients appropriate for the context of cosmic plasmas. 
The expressions of the coefficients considered in this paper are explicitly specified in Appendix~\ref{App:transferCoef}. 

An astrophysical plasma generally consists of 
a population of thermal electrons and a population of non-thermal electrons, 
which could be relativistic electrons that gyrate around magnetic field lines, 
electrons accelerated by shocks, or electrons injected by cosmic rays. 
Given that dielectric suppression\footnote{Dielectric suppression, 
or known as the Razin effect or Razin-Tsytovich effect \citep{Razin60, Ramaty68}, is a plasma effect on synchrotron emission. 
Synchrotron radiation is suppressed exponentially below the Razin frequency $\omega_{\rm R} = \omega_{\rm p}^{2}/\omega_{\rm B}$, 
where $\omega_{\rm p}$ is the plasma angular frequency and $\omega_{\rm B}$ is the electron angular gyrofrequency, since 
the electrons can no longer maintain the phase with the emitted radiation as the wave phase velocity would increase to above the speed of light \citep[see e.g.][]{Melrose80Plasma}.} \citep[see e.g.][]{Bekefi66, Rybicki86} 
is insignificant, which is generally true for cosmic plasmas \citep[see e.g.][]{Melrose91Book}, 
here the transfer coefficients are expressed as the sum of their respective thermal and non-thermal components, i.e. $\kappa_{ij}=(\kappa_{ij, {\rm th}}+\kappa_{ij, {\rm nt}})$ and $\epsilon_{i} = (\epsilon_{i, {\rm th}}+\epsilon_{i, {\rm nt}})$, where ``th" and ``nt" denote the thermal and non-thermal components of the absorption and emission coefficients respectively. 

In this work, we consider thermal bremsstrahlung and non-thermal synchrotron radiation process\footnote{
In addition to thermal bremsstrahlung and non-thermal synchrotron radiation process considered in this work, 
we note that transfer coefficients appropriate for different astrophysical environments have been extensively studied 
in the literature. Accurate expressions for the coefficients of Faraday rotation and Faraday conversion 
in uniformly magnetized relativistic plasmas, such as those in jets and hot accretion flows around black-holes, are reported in \citet{Huang11}. 
Expressions of the transfer coefficients in the case of ultra-relativistic plasma that is 
permeated by a static uniform magnetic fields, for frequencies of high harmonic number limits, 
and for a number of distribution functions (isotropic, thermal, or power law) are presented in \citet{Heyvaerts13}. 
Emission and absorption coefficients for cyclotron process, 
that is important in accretion discs of compact objects, 
have also been studied by \citet{Chanmugam89, Vaeth95}. 
Careful incorporation of the above would be a useful improvement to the current CPRT implementation, 
expanding the range of its applications and enabling a realistic modeling of the magnetized Universe.  
}. 
%
For thermal bremsstrahlung, expressions of the 
Faraday rotation coefficient $f_{\rm th}$ and Faraday conversion coefficient $h_{\rm th}$, 
as well as the expressions of the absorption coefficients 
$\kappa_{\rm th}$, $q_{\rm th}$ and $v_{\rm th}$ 
follow \citet{Pacholczyk77}\footnote{The same expressions of 
$\kappa_{\rm th}$, $q_{\rm th}$ and $v_{\rm th}$ are provided in \citet{Wickramasinghe85} 
but a typo of an extra factor of the square of angular frequency is found in the denominator of $v_{\rm th}$ via dimensional analysis. 
We also note that the sign of $q_{\rm th}$ in \citet{Wickramasinghe85} is also different to \citet{Pacholczyk77}, which might be due to different 
polarization sign conventions or a sign error.}. The emission coefficients are computed via Kirchoff's law accordingly. 
For non-thermal synchrotron emission, we consider relativistic electrons that have a power-law energy distribution. 
We use the expressions of the transfer coefficients that follow \citet{JonesOdell77_transfer} and consider an isotropic distribution of relativistic electrons' momentum direction. 

As detailed in Appendix~\ref{app:Convention}, the sign of Stokes $V$ depends on its definition, 
polarization conventions, handedness of the coordinate systems used, 
as well as the time dependence of the electromagnetic wave (i.e. whether the exponent has $+i\omega t$ or $-i\omega t$) 
and the definition of the relative phase between the $x$ and $y$-components of the electric field of the radiation. 
However, some of these information were not explicitly stated in \citet{JonesOdell77_transfer}, 
and inconsistent definitions of the time dependence of the electromagnetic wave were used in \citet{Pacholczyk77} 
in deriving the radiative transfer coefficients for bremsstrahlung and synchrotron radiation process (see their equations 3.33 and 3.93). 
We therefore expound on the strategy to eliminate ambiguity in Appendix~\ref{app:Convention} and 
present a consistent set of expressions of all the transfer coefficients in Appendix~\ref{App:transferCoef}, 
given the geometry explicitly defined in Appendix~\ref{app:adoptedCoordsys} 
and the polarization convention conforming to the IEEE/IAU standard. 

\section{Algorithms and Numerical Calculation}\label{sec:algo} 

The CPRT equations given in equation\,(\ref{eq:covarPRTinz}) can, in principle, be either solved by 
direct integration via numerical methods, or by diagonalizing and determining the inverse of the transfer matrix operator. 
We adopt the former approach and employ a ray-tracing method in this paper. 
In this section, we present algorithms to solve the CPRT equation numerically. 
We first present the algorithm for computing the CPRT for a single ray, followed by 
the algorithm for an all-sky setting wherein cosmological MHD simulation results may be incorporated 
to generate a set of theoretical all-sky intensity and polarization maps. 

\subsection{Ray-tracing}\label{subsec:algoray}
The CPRT algorithm consists of three basic components concerning 
(i) the interaction of radiation with the line-of-sight plasmas, 
(ii) the cosmological effects on radiation and the co-evolution of plasmas with the Universe's history, and 
(iii) numerical computation of the CPRT equation, which is a set of four coupled differential equations 
evaluated in the redshift $z$-space. In the following, we discuss each of these components, 
starting with the numerical method. We describe the implementation of the algorithm and 
highlight its specific designs to accommodate the inclusion of line-of-sight astrophysical sources and intervening plasmas of different properties. 

\subsubsection{Numerical method}\label{subsec:zsampling}
The radiation propagation is parameterized by redshift $z$ and is sampled discretely into 
$N_{\rm cell}$ number of cells. We adopt a sampling scheme such that each $z$-interval corresponds to 
an approximately equal light travel distance. 
That is, between the initial redshift $z_{\rm init}$ at which 
we start evaluating the CPRT equation and 
the final redshift $z=0$ at which observation is made, 
the total light travel distance $s_{\rm tot}$ is first computed by 
solving equation\,(\ref{eq:dldz}) followed by finding the corresponding lower and upper boundary values of $z$; 
where in each $z$-interval, light travels a distance as close to $s_{\rm eq} = s_{\rm tot}/N_{\rm cell}$ as possible. 
Note that the light travel distance acts as a scaling factor in the context of numerical evaluation. 
For efficient numerical computation, its multiplications with the transfer coefficients would ideally be close to unity. 

Each $z$ interval can be further refined to incorporate astrophysical structure(s) and their sub-structure(s). Our code implementation allows an option to switch on/off such a refinement scheme, as well as to incorporate multiple structures at different redshifts. 
Within the refinement zone, we employ a uniform sampling in the $\log_{10}{(1+z)}$ space 
which has the advantage of preserving the profile shape when multi-frequency calculations are to be carried out. 
In algorithmic terms, 
at the cell of index $ind^{\rm refine}$, 
the increment over each refined cell is given by 
$[\log_{10}{(1+z')}-\log_{10}{(1+z)}]/N^{\rm refine}_{\rm cell}$, 
where $z'$ and $z$ are, respectively, 
the upper and lower boundaries of the $z$-interval to be refined, 
and $N^{\rm refine}_{\rm cell}$ is the total number of refined cells. 

A fourth-order Runge?Kutta (RK) differential equation solver 
is used to integrate equation\,(\ref{eq:dldz}),  and to solve equation\,(\ref{eq:covarPRTinz}), 
which ultimately gives us the Stokes parameters $\{I, Q, U, V\}$ at $z=0$ in the observer's frame. 
Parameters to be set for the solver include the total number of (coupled) differential equations to be solved $N_{\rm eqn}$, 
the number of steps for the RK solver $N_{\rm step}$; 
and the error tolerance level {\tt eps}. 
The error estimation of the solver is carried out by comparing the solution obtained with a fourth-order RK formula to that obtained with a fifth-order RK formula. 
If the computed error is less than {\tt eps} then the calculation proceeds; 
otherwise the algorithm halts, reports errors of non-convergence, and returns without further computation. 

The upper and lower limits of the $z$-variables are updated along the ray. The outputs are passed into the next computation as the inputs (i.e. as updated initial conditions). Since the evaluation of the CPRT starts from a higher $z$ to a lower z value until the present $z_{0}=0$ is reached, 
a substitution of $z \rightarrow -z $ is made in equation\,(\ref{eq:covarPRTinz}) as we set that as the function to be evaluated by the RK solver. 

\subsubsection{Interaction of radiation with plasmas}
Radiation is parameterized by frequency $\nu(z)$, which has a redshift dependence of $\nu(z) = \nu_{\rm obs}(1+z)$, 
where $\nu_{\rm obs}$ is the observed frequency at the present epoch $z =0$. 
Its intensity and polarization properties change when passing through the magnetized intervening plasmas. 
The strength of the radiative processes, captured through transfer coefficients in the CPRT equation, depends on 
the physical properties of the plasmas, in addition to the frequency of the radiation. 
In general, both thermal and non-thermal electrons are present in astrophysical plasmas. Parameters describing them include: 
$n_{\rm e, tot}$, 
fraction of non-thermal electrons ${\mathcal F}_{\rm nt}$ (and thus $n_{\rm e, th}$ and $n_{\rm e, nt}$), 
temperatures $T_{\rm e}$ for thermal electrons, 
power-law index of the non-thermal electrons' energy spectrum $p$ and 
the electrons' low energy cutoff described by the Lorentz factor $\gamma_{i}$. 
Added to this list are parameters describing the strength and direction of magnetic fields, 
$\mathbfit{B}$, which can be decomposed into two components.  
One component is decomposed along the line-of-sight direction $\mathbfit{B}_\parallel= |\mathbfit{B}| \cos{\theta}$, 
and another component is decomposed in the plane normal to the line-of-sight $\mathbfit{B}_\perp = |\mathbfit{B}| \sin{\theta}$, 
where $\theta$ is the angle between the direction of the magnetic field and the line-of-sight. 

By specifying the observed frequency of radiation $\nu_{\rm obs}$ at $z=0$ and the 
radiation background at an initial redshift $z_{\rm init}$,  
and given some input distributions of electron number density 
$n_{\rm e}(z)$ and magnetic field strength $|\mathbfit{B}(z)|$ through which light travels, 
solving the CPRT equations 
yields the evolution of the intensity and polarization of the radiation as a function of $z$. 

\subsubsection{Cosmological effects}\label{subsubsec:CosmoEvo}

In this paper, we adopt the maximum likelihood cosmological parameters obtained by the \citet{Planck15} with 
the present Hubble constant $H_{0} = 100\,h_{0} = 67.74 ~\rm{kms^{-1}Mpc^{-1}}$, 
the matter density today $\Omega_{\rm m,0} = 0.3089$, 
and the cosmological constant or vacuum density today $\Omega_{\Lambda, 0}= 0.6911$ \citep{Planck15}. 
The radiation density today is given by $\Omega_{\rm r,0}=  4.1650\times 10^{-5}(h_0)^{-2}$ \citep{Wright06}. 
 
We have already noted the frequency shift of the radiation due to the expansion of the Universe, 
i.e. $\nu(z) = \nu_{\rm obs}(1+z)$. 
The cosmological (expansion) effects on the temperatures, electron number densities, as well as the strengths of magnetic fields are given by, respectively, 
$T_{\rm e}(z) =T_{\rm e,0} (1+z)^{2}$, 
$n_{\rm e}(z) =n_{\rm e,0}  (1+z)^{3}$, and 
$|\mathbfit{B}(z)| =|\mathbfit{B}_{0}| (1+z)^{2}$, assuming frozen-in flux condition. 
These properties, as well as the structures of magnetic fields, are also subjected to local structure formation, evolution and outflows, as well as to influences by external injections, such as cosmic rays. 
Consequently, the inter-stellar medium (ISM), 
intra-cluster medium (ICM), and intergalactic media (IGM) all exhibit 
different characteristic properties. 
The CPRT formulation is covariant and 
accounts for cosmological and relativistic effects self-consistently. 
Because of these advantages, 
theoretical predictions of the intensity and polarization of the radiation can be computed straightforwardly 
by incorporating simulation results describing the cosmic plasmas into the computation of the transfer coefficients, 
and then solving the CPRT equation. 

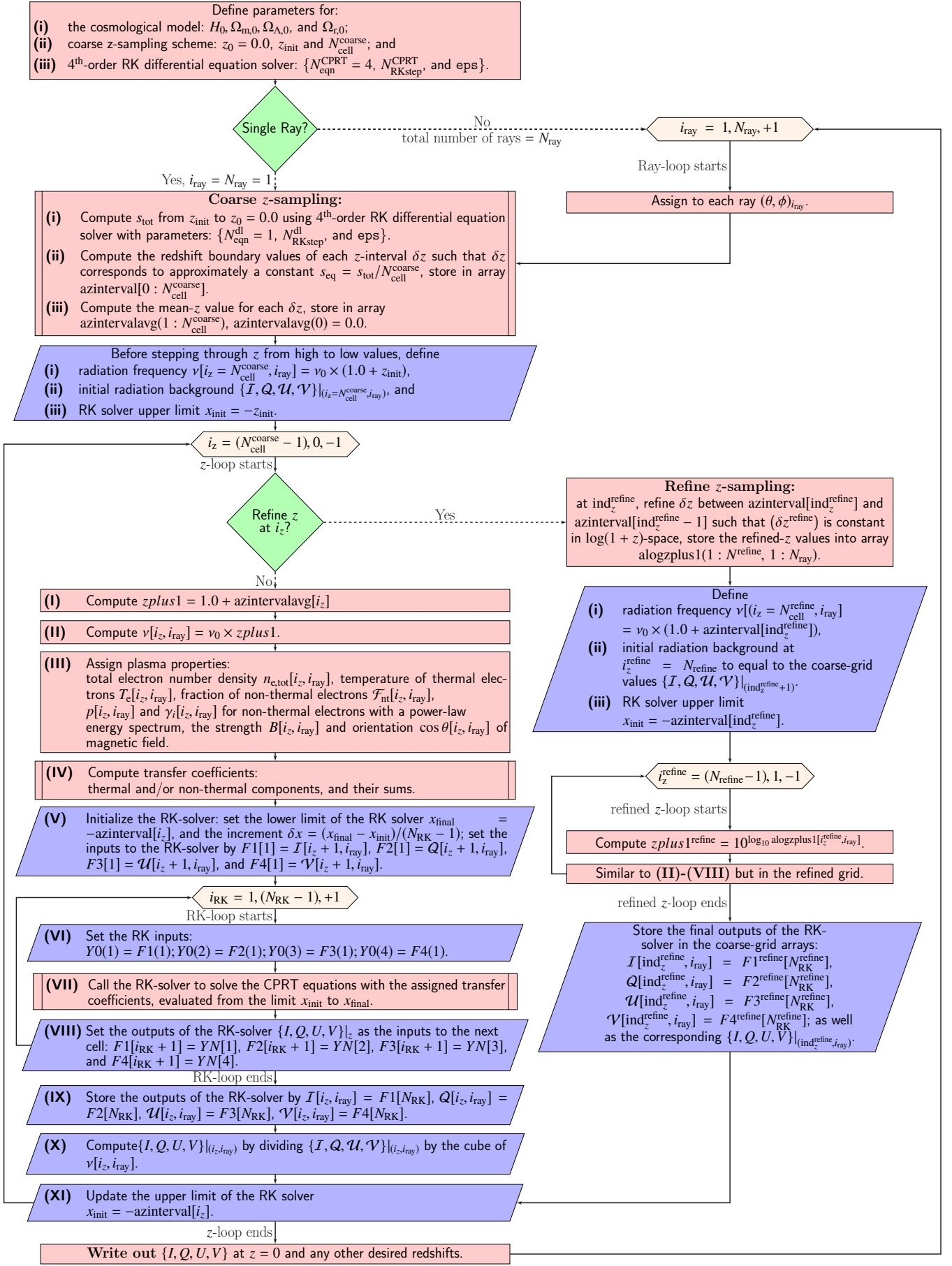
\begin{figure}
  \centering
\tikzstyle{decision} = [diamond, draw, fill=green!30,
    text width=10.0em, text badly centered, node distance=2.5cm, font=\sffamily\huge,  inner sep=0pt]
\tikzstyle{information} = [text width=7em, text badly centered, node distance=2.5cm, font=\sffamily\huge, inner sep=0pt]    
\tikzstyle{bigblock} = [rectangle, draw, fill=red!20, text width=68em, text centered,  font=\sffamily\huge, minimum height=2em, scale=1.0]  
\tikzstyle{medblockrefined} = [rectangle, draw, fill=red!20,  text badly centered, text width=45em,  font=\sffamily\huge, minimum height=2.7em]  
\tikzstyle{medblock} = [rectangle, draw, fill=red!20,  text badly centered, text width=65em,  font=\sffamily\huge, minimum height=2.7em]  
    \tikzstyle{smallblock} = [rectangle, draw, fill=red!20,   text badly centered, text width=20em, font=\sffamily\huge, minimum height=2em]        
\tikzstyle{block} = [rectangle, draw, fill=blue!20, text width=18.5em, text centered,  font=\sffamily\huge, minimum height=2em]
\tikzstyle{subblock} = [rectangle, draw, fill=orange!10,
    text width=25em, text centered, font=\sffamily\huge, minimum height=4em]
\tikzstyle{ssubblock} = [rectangle, draw, fill=orange!10,
    text width=10em, text centered, rounded corners, font=\sffamily\huge, minimum height=4em]    
\tikzstyle{line} = [draw, very thick, color=black!90, -latex']  
\tikzstyle{cloud} = [draw, ellipse,fill=red!20, node distance=2.5cm,  font=\sffamily\huge,
    minimum height=2em]
\tikzstyle{loop}= [draw, fill=orange!10, 
        chamfered rectangle, text centered, 
        chamfered rectangle xsep=2cm,
                 font=\sffamily\huge,
        text width=20em] 
\tikzstyle{loop2}= [draw, fill=orange!10, 
        chamfered rectangle, text centered, 
        chamfered rectangle xsep=2cm,
        font=\sffamily\huge,
        text width=20em]   
\tikzstyle{func}=[draw, fill=red!20,
         rectangle split,
         rectangle split horizontal,
         rectangle split parts = 3,
         align = center, 
         inner sep = 3 pt,
         font=\sffamily\huge, text badly centered, 
         minimum height = 7mm] 
\tikzstyle{ioblock} = [draw=black, fill=blue!30, trapezium, 
trapezium left angle=70, text centered, trapezium right angle=-70, 
minimum height=7mm, font=\sffamily\huge,
text width=65em]  
\tikzstyle{subioblock} = [draw=black, fill=blue!30, trapezium, 
trapezium left angle=70, trapezium right angle=-70, 
minimum height=7mm, text centered,
font=\sffamily\huge,
text width=65em]  %
\tikzstyle{subioblockrefined} = [draw=black, fill=blue!30, trapezium, 
trapezium left angle=70, trapezium right angle=-70, 
minimum height=7mm, text centered,
font=\sffamily\huge,
text width=40em]  %

\resizebox{1.0\textwidth}{1.0\textheight}{%
\begin{tikzpicture}[scale=0.25, node distance = 2.0cm, auto]
\linespread{0.5}
    \node [bigblock] (p1) {{Define parameters} for: \\ 
    \begin{enumerate}[leftmargin=*, labelindent= 1 pt, label = {\bfseries (\roman*)},align=left]
    \item the cosmological model:             
              $H_{0},  \Omega_{\rm m,0}, 
              \Omega_{\Lambda, 0}$, and $\Omega_{\rm r,0}$; 
     \item coarse z-sampling scheme: 
                                          $z_{0} =0.0$,  
                                          $z_{\rm init}$ and 
                                          $N_{\rm cell}^{\rm coarse}$; and
     \item  $4^{\rm th}$-order RK differential equation solver: \{$N_{\rm eqn}^{\rm CPRT} =4$,  $N_{\rm RKstep}^{\rm CPRT}$, and {\tt eps}\}.               
    \end{enumerate}
 };
    \node [decision, below of=p1, yshift=-1.13cm] (d1) {Single Ray?};
    \node [loop, right of=d1, xshift=17cm] (aY2) {$i_{\rm ray} = 1 , N_{\rm ray}, +1$};
    \node [medblockrefined, below of=aY2, yshift=-1.0cm] (subp1) {Assign to each ray $(\theta, \phi)_{i_{\rm ray}}$.};
    \node [func, below of=d1, yshift=-3.5cm] (aY1) {
    \nodepart[text width= 0.25em]{one} 
    \nodepart[text width=64.5em]{two}
    {\bf Coarse $z$-sampling:} \\
    \begin{enumerate}[leftmargin=*, labelindent= 1 pt, label = {\bfseries (\roman*)}, align=left]
	\item Compute $s_{\rm tot}$ from $z_{\rm init}$ to $z_{0} =0.0$ 
              using $4^{\rm th}$-order RK differential equation solver with parameters: 
               \{$N_{\rm eqn}^{\rm dl} = 1$, $N_{\rm RKstep}^{\rm dl}$, 
               and {\tt eps}\}.
    \item Compute the redshift boundary values of each $z$-interval 
             $\delta z$ such that $\delta z$ corresponds to approximately a constant $s_{\rm eq} = s_{\rm tot}/N_{\rm cell}^{\rm coarse}$, 
             store in array ${\rm azinterval}[0:N_{\rm cell}^{\rm coarse}]$.  
    \item  Compute the mean-$z$ value for each $\delta z$, 
    store in array \\ ${\rm azintervalavg}(1:N_{\rm cell}^{\rm coarse})$, 
    ${\rm azintervalavg}(0)=0.0$.    
    \end{enumerate}
    \nodepart[text width= 0.25em]{three} 
    };
    \node [ioblock, below of=aY1, yshift=-2.9cm] (p2) {Before stepping through $z$ from high to low values, {define} \\
   \begin{enumerate}[leftmargin=*, labelindent= 1 pt, label = {\bfseries (\roman*)}, align=left]
	\item radiation frequency  $\nu[{i_{\rm z}=N_{\rm cell}^{\rm coarse}, i_{\rm ray}]} = \nu_{0}\times(1.0 +z_{\rm init})$, 
    \item  initial radiation background $\{{\mathcal I}, {\mathcal Q}, {\mathcal U}, {\mathcal V} \}|_{(i_{\rm z}=N_{\rm cell}^{\rm coarse}, i_{\rm ray})}$, and 
    \item RK solver upper limit $x_{\rm init} = -z_{\rm init}$.         
    \end{enumerate} 
    };  
    \node [loop2, below of=p2, yshift= -0.5cm] (loopZ) {$i_{\rm z} = (N_{\rm cell}^{\rm coarse}-1), 0, -1$};
    \node [decision, below of=loopZ, yshift=-0.7cm] (d2) {Refine $z$ \\ at $i_{z}$?};    
    \node [medblockrefined, right of=d2, yshift=0.0cm, xshift=17cm] (bY2) {
    {\bf Refine $z$-sampling:} \\
    								              at ${\rm ind}_{z}^{\rm refine}$, refine $\delta z$ between ${\rm azinterval}[{\rm ind}_{z}^{\rm refine}]$ and 
								              ${\rm azinterval}[{\rm ind}_{z}^{\rm refine}-1]$ 
								              such that ($\delta z^{\rm refine}$) is 
								              constant in $\log({1+z})$-space, 
								              store the refined-$z$ values into array ${\rm alogzplus1}(1:N^{\rm refine}, \, 1: N_{\rm ray})$.};
								              
    \node [subioblockrefined, below of=bY2, yshift= -3.6cm] (initsub) {{Define} 
   \begin{enumerate}[leftmargin=*, labelindent= 1 pt, label = {\bfseries (\roman*)}, align=left]
   \item radiation frequency 
            $\nu[{(i_{\rm z}=N_{\rm cell}^{\rm refine}, i_{\rm ray}]}$ \\
            $ = \nu_{0}\times(1.0 +{\rm azinterval}[{\rm ind}_{z}^{\rm refine}])$,  
   \item initial radiation background at \\
            $i_{z}^{\rm refine} = N_{\rm refine}$
             to equal to the coarse-grid values $\{{\mathcal I}, {\mathcal Q}, {\mathcal U}, {\mathcal V} \}|_{({\rm ind}_{\rm z}^{\rm refine}+1)}$. 
   \item  RK solver upper limit \\ 
             $x_{\rm init} = -{\rm azinterval}[{\rm ind}_{z}^{\rm refine}]$. 
   \end{enumerate}
   
   };         
    \node [loop2, below of=initsub, yshift= -2.8cm] (loopRefine) {$i_{\rm z}^{\rm refine} = (N_{\rm refine}-1), 1, -1$};   
    \node [medblockrefined, below of=loopRefine, yshift= -0.7cm] (ptb2) {Compute $zplus1^{\rm refine}= 10^{{\log_{10}{{\rm alogzplus1}[i_{z}^{\rm refine}, i_{\rm ray}]}}}$.};
    \node [medblockrefined, below of=ptb2, yshift= 0.7 cm] (ptb3) {Similar to {\bf (II)-(VIII)} but in the refined grid.};
     \node [subioblockrefined, below of=ptb3, yshift= -2.64cm] (ptb4) {Store the final outputs of the RK-solver in the coarse-grid arrays: 
                                                                                 ${\mathcal I}[{\rm ind}_{z}^{\rm refine}, i_{\rm ray}]= F1^{\rm refine}[N^{\rm refine}_{\rm RK}] $, 
                                                                                 ${\mathcal Q}[{\rm ind}_{z}^{\rm refine}, i_{\rm ray}]= F2^{\rm refine}[N^{\rm refine}_{\rm RK}] $, 
                                                                                 ${\mathcal U}[{\rm ind}_{z}^{\rm refine}, i_{\rm ray}]= F3^{\rm refine}[N^{\rm refine}_{\rm RK}] $, 
                                                                                 ${\mathcal V}[{\rm ind}_{z}^{\rm refine}, i_{\rm ray}]= F4^{\rm refine}[N^{\rm refine}_{\rm RK}] $;
                                                                                 as well as the corresponding  $\{I, Q, U, V \}|_{({\rm ind}_{z}^{\rm refine}, i_{\rm ray})}$. 
                                                                                 };                       
   \node [medblock, below of=d2, yshift= -1.2cm] (bY1a){
    \begin{enumerate}[leftmargin=*, labelindent= 1 pt, label = {\bfseries (\Roman*)}, align=left]
    \setcounter{enumi}{0}
     \item  {Compute $zplus1= 1.0+{\rm azintervalavg}[{i_{z}}]$}
    \end{enumerate}   
    { }
       };
       \node [medblock, below of=bY1a, yshift= 0.7cm] (bY1b){
    \begin{enumerate}[leftmargin=*, labelindent= 1 pt, label = {\bfseries (\Roman*)}, align=left]
    \setcounter{enumi}{1}
     \item  {Compute $\nu[{i_{z}}, i_{\rm ray}] = \nu_{0} \times zplus1$}. 
    \end{enumerate}   
    { }
   };
    \node [medblock, below of=bY1b, yshift= -0.87cm] (bY1c) {
    \begin{enumerate}[leftmargin=*, labelindent= 1 pt, label = {\bfseries (\Roman*)}, align=left]
    \setcounter{enumi}{2}
    \item Assign plasma properties:\\ 
             total electron number density 
             $n_{\rm e, tot}[{i_{z}}, i_{\rm ray}]$, temperature of thermal electrons $T_{\rm e}[{i_{z}}, i_{\rm ray}]$, 
            fraction of non-thermal electrons ${\mathcal F}_{\rm nt}[{i_{z}}, i_{\rm ray}]$, \\
            $p[{i_{z}}, i_{\rm ray}]$ and $\gamma_{i}[{i_{z}}, i_{\rm ray}]$ for non-thermal electrons with a power-law \\
            energy spectrum, the strength ${B}[{i_{z}}, i_{\rm ray}]$ and orientation $\cos{\theta}[{i_{z}}, i_{\rm ray}]$ of magnetic field. 
    \end{enumerate}   
    { }
   };
    \node [func, below of=bY1c, yshift=-4.0em] (p3) {
    \nodepart[text width= 0.25em]{one} 
    \nodepart[text width=64.5em]{two}{
    \begin{enumerate}[leftmargin=*, labelindent= 1 pt, label = {\bfseries (\Roman*)}, align=left]
    \setcounter{enumi}{3}
    \item {Compute transfer coefficients: \\ thermal and/or non-thermal components, and their sums.}
    \end{enumerate}}   
    \nodepart[text width= 0.25em]{three} 
    };

    \node [ioblock, below of=p3, yshift= -1.7em] (p4) {
    \begin{enumerate}[leftmargin=*, labelindent= 1 pt, label = {\bfseries (\Roman*)}, align=left]
    \setcounter{enumi}{4}
     \item  {Initialize the RK-solver: 
                                                                                 set the lower limit of the RK solver $x_{\rm final}= -{\rm azinterval}[i_{z}]$, 
                                                                                 and the increment $\delta x= (x_{\rm final} - x_{\rm init})/({N_{\rm RK}}-1)$; 
                                                                                 set the inputs to the RK-solver by 
                                                                                 $F1[1] = {\mathcal I}[i_{z}+1, i_{\rm ray}]$, 
                                                                                 $F2[1] = {\mathcal Q}[i_{z}+1, i_{\rm ray}]$, 
                                                                                 $F3[1] = {\mathcal U}[i_{z}+1, i_{\rm ray}]$, and 
                                                                                 $F4[1] = {\mathcal V}[i_{z}+1, i_{\rm ray}]$. 
}
    \end{enumerate}   
    { }
    };
    \node [loop2, below of=p4, yshift=-0.3cm] (loopRK) {$i_{\rm RK} = 1, (N_{\rm RK}-1), +1$};   
    \node [ioblock, below of=loopRK, yshift= 0.15cm] (p5a){
    \begin{enumerate}[leftmargin=*, labelindent= 1 pt, label = {\bfseries (\Roman*)}, align=left]
    \setcounter{enumi}{5}
     \item{Set the RK inputs: \\
     $Y0(1)=F1(1); Y0(2)=F2(1); Y0(3)=F3(1); Y0(4)=F4(1)$.}
   \end{enumerate}};     
    \node [func, below of=p5a, yshift= 0.0cm] (p5b) {
   \nodepart[text width= 0.5em]{one} 
    \nodepart[text width= 64.5em]{two} {
    \begin{enumerate}[leftmargin=*, labelindent= 1 pt, label = {\bfseries (\Roman*)}, align=left]
    \setcounter{enumi}{6}
   \item {Call the RK-solver 
    to solve the CPRT equations with the assigned transfer coefficients, 
    evaluated from the limit $x_{\rm init}$  to $x_{\rm final}$}.    \end{enumerate}}
    \nodepart[text width= 0.5em]{three} }; 
    \node [ioblock, below of=p5b, yshift= -0.2cm] (p6a){      
    \begin{enumerate}[leftmargin=*, labelindent= 1 pt, label = {\bfseries (\Roman*)}, align=left]
    \setcounter{enumi}{7}
     \item{Set the outputs of the RK-solver $\{I, Q, U, V\}|_{z}$ as the inputs to the next cell: 
                                                                                 $F1[i_{\rm RK} +1] = YN[1]$, 
                                                                                 $F2[i_{\rm RK} +1] = YN[2]$, 
                                                                                 $F3[i_{\rm RK} +1] = YN[3]$, and 
                                                                                 $F4[i_{\rm RK} +1] = YN[4]$.}
    \end{enumerate}   
    { }
    };       
    \node [ioblock, below of=p6a, yshift= -0.45cm] (p6b){
     \begin{enumerate}[leftmargin=*, labelindent= 1 pt, label = {\bfseries (\Roman*)}, align=left]
    \setcounter{enumi}{8}
     \item{Store the outputs of the RK-solver by ${\mathcal I}[i_{z}, i_{\rm ray}]= F1[N_{\rm RK}] $, 
                                                                                 ${\mathcal Q}[i_{z}, i_{\rm ray}]= F2[N_{\rm RK}] $, 
                                                                                 ${\mathcal U}[i_{z}, i_{\rm ray}]= F3[N_{\rm RK}] $, 
                                                                                 ${\mathcal V}[i_{z}, i_{\rm ray}]= F4[N_{\rm RK}] $.}
     \end{enumerate}   
    { }
    };            
    \node [ioblock, below of=p6b, yshift= 0.0cm] (p6c){ 
    \begin{enumerate}[leftmargin=*, labelindent= 1 pt, label = {\bfseries (\Roman*)}, align=left]
    \setcounter{enumi}{9}
     \item {Compute$\{I, Q, U, V \}|_{(i_{z}, i_{\rm ray})}$ by dividing 
                                                                                          $\{{\mathcal I}, {\mathcal Q}, {\mathcal U}, {\mathcal V} \}|_{(i_{z}, i_{\rm ray})}$ 
                                                                                          by the cube of $\nu[{i_{z}}, i_{\rm ray}]$.}
   \end{enumerate}   
    { }
    };                             .  
    \node [ioblock, below of=p6c, yshift= 0.0cm] (p6d) {
    \begin{enumerate}[leftmargin=*, labelindent= 1 pt, label = {\bfseries (\Roman*)}, align=left]
    \setcounter{enumi}{10}
     \item  {Update the upper limit of the RK solver \\ $x_{\rm init}= -{\rm azinterval}[i_{z}]$.}
    \end{enumerate}   
    { }
    };
    \node [medblock, below of=p6d, yshift=-0.1cm] (p7) {{\bf Write out} $\{I, Q, U, V \}$ at $z=0$ and any other desired redshifts.};         
                                                                                                  
    \path [line] (p1) -- (d1);
    \path [line, dashed] (d1) -- node[anchor=east, font=\huge]{Yes, $i_{\rm ray}=N_{\rm ray}=1$}(aY1);
    \path [line, dashed] (d1) -- node[anchor=north, font=\huge]{total number of rays $=N_{\rm ray}$}(aY2);
    \path [line, dashed] (d1) -- node[anchor=south, font=\huge]{No}(aY2);
    \path [line] (aY1) -- (p2);    
    \path [line] (p2) -- node[anchor=east, font=\huge]{}(loopZ);    
    \path [line] (loopZ) -- node[anchor=east, font=\huge]{$z$-loop starts}(d2);    
      \path [line] (aY2) -- node[anchor=east, font=\huge]{Ray-loop starts}(subp1);
      \path [line] (subp1) |- node[anchor=east]{}(aY1);
    \path [line, dashed] (d2) -- node[anchor=east, font=\huge]{No}(bY1a);
    \path [line, dashed] (d2) -- node[anchor=south, font=\huge]{Yes}(bY2);
    \path [line] (bY1a) -- node[anchor=east]{}(bY1b);
    \path [line] (bY1b) -- node[anchor=east]{}(bY1c);
     \path [line] (bY1c) -- node[anchor=south]{}(p3);  
     \path [line] (p3) -- node[anchor=east]{}(p4);  
     \path [line] (p4) -- node[anchor=east, font=\huge]{}(loopRK);  
     \path [line] (loopRK) -- node[anchor=east, font=\huge]{RK-loop starts}(p5a);  
     \path [line] (p5a) -- node[anchor=east]{}(p5b);  
     \path [line] (p5b) -- node[anchor=east]{}(p6a);  
     \path [line] (p6a) -- node[anchor=east, font=\huge]{RK-loop ends}(p6b);           
     \path [line] (p6b) -- node[anchor=east]{}(p6c);      
     \path [line] (p6c) -- node[anchor=east]{}(p6d);           
     \path [line] (p6d) -- node[anchor=east, font=\huge]{$z$-loop ends}(p7);      
     \path [line] (bY2) -- node[anchor=east]{}(initsub);  
     \path [line] (initsub) -- node[anchor=east, font=\huge]{}(loopRefine);  
     \path [line] (loopRefine) -- node[anchor=east, font=\huge]{refined $z$-loop starts}(ptb2);  
     \path [line] (ptb2) -- node[anchor=south]{}(ptb3);  
     \path [line] (ptb3) -- node[anchor=east, font=\huge]{refined $z$-loop ends}(ptb4);  
      \path [line] (ptb4) |-  node[anchor=east]{}(p6d);  
      \path [line] (p6d.west)  --+(-5cm,0) |- (loopZ.west);   
     \path [line] (p6a.west)  --+(-2.5cm,0) |- (loopRK.west);       
     \path [line] (ptb3.west)  --+(-2.5cm,0) |- (loopRefine.west);       
      \path [line] (p7.east)  --+(+72.0cm,0) |- node[anchor=east]{}(aY2.east);       
\end{tikzpicture}
} 
  \caption{The CPRT algorithm.} 
  \label{fig:algoflow}
\end{figure}


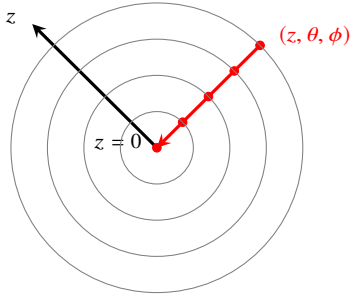
\begin{figure}
  \centering
  \begin{tikzpicture}[scale=0.8, node distance = 0.25, auto]
    \coordinate (1rayNode) at (135:2.9cm);
    \coordinate (2rayNode) at (45:2.4cm);
    \coordinate (3rayNode) at (225:2.4cm);
    \coordinate (4rayNode) at (315:2.4cm);
    \coordinate (originNode) at (0:0cm);

 \draw[<-, >=stealth, very thick] (1rayNode.south) -- (0,0) node[left,pos=-.05]{$z$} 
 node[left,pos=0.2]{}  
 node[left,pos=0.4]{} 
 node[left,pos=0.6]{} 
 node[left,pos=0.8]{} 
 node[left,pos=0.95]{$z=0$};
 \draw[-> , >=stealth, red, very thick] (2rayNode.south) -- (0,0) 
  node[ red, right, pos=-0.1]{{$(z, \theta, \phi)$}}
 node[ red, left,pos=0]{} 
 node[left,pos=0.2]{} 
 node[ red, left,pos=0.4]{} 
 node[ red, left,pos=0.6]{} node[left,pos=0.8]{};

    \draw[fill,color=red] (barycentric cs:2rayNode=1.0,originNode=0) circle (2pt);
    \draw[fill,color=red] (barycentric cs:2rayNode=0.75,originNode=0.25) circle (2pt);
    \draw[fill,color=red] (barycentric cs:2rayNode=0.5,originNode=0.5) circle (2pt);
    \draw[fill,color=red] (barycentric cs:2rayNode=0.25,originNode=0.75) circle (2pt);
    \draw[fill,color=red] (barycentric cs:2rayNode=0,originNode=1.0) circle (2pt);

    \draw[black!50] (0,0) circle (2.4cm);
    \draw[black!50] (0,0) circle (1.8cm);
    \draw[black!50] (0,0) circle (1.2cm);
    \draw[black!50] (0,0) circle (0.6cm);

  \end{tikzpicture}
  \caption{Illustration of the concept of the all-sky algorithm based on a ray-tracing technique: the CPRT equation is solved for each light ray (indicated in red) that is parameterized by $(z, \theta, \phi)$. The radial direction coincides with the direction of redshift $z$ while $(\theta, \phi)$ maps to the coordinates of the celestial sky. The observer is positioned at the center of the circles, i.e. at $z=0$. Note that the co-moving Hubble radius is represented inside-out. That is, the co-moving Hubble sphere expands as we approach the center\,($z=0$) due to the expansion of the Universe. This set-up is applicable for a universe that has a simple topology like ours, as is suggested by measurements of the cosmic microwave background \citep{Planck13Topology}.} 
  \label{fig:algo}
\end{figure}

\subsection{All-sky polarization calculation}\label{subsec:algoallsky}

We construct an all-sky CPRT algorithm that can 
interface with cosmological simulation results, 
numerically solve the CPRT equation, and thereby generate 
theoretical all-sky polarization maps that serve as model templates.  A schematic of the algorithm is shown in Fig.~\ref{fig:algoflow}. 

Fig.~\ref{fig:algo} illustrates the concept of the all-sky algorithm, 
in which the CPRT equation is solved in a spherical polar coordinate system $(r, \theta, \phi)$, 
where $(\theta, \phi)$ corresponds to the celestial sky coordinates and 
the radial axis $r$ corresponds to the redshift axis $z$. 
Note that outputs of the cosmological evolutions of plasma properties, 
e.g. $n_{\rm e}(z)$ and $|\mathbfit{B}(z)|$, obtained from a cosmological MHD simulation can be inputted to the CPRT calculations through the transfer coefficients. 
Spatial fluctuations of the plasma properties in a finite simulation volume, usually in Cartesian coordinate system $(i,j,k)$, can also be mapped to 
the spherical polar coordinate system $(r, \theta, \phi)$ at each sampled redshift $z$. A more rigorous treatment that guarantees the magnetic field is divergence-free is also possible within our all-sky framework. We will present these details in a forthcoming paper. 

We add a remark here on the sampling scheme over a sphere for efficient follow-on data analysis. 
There is an option to compute rays that are randomly positioned over the entire celestial sphere. Alternatively, one may utilize the advantages of efficient spherical sampling schemes, 
such as the HEALPix sampling \citep[][]{gorski:2005} and the sampling scheme devised by \citet{mcewen:fssht} which affords exact numerical quadrature. In such a case, ray-tracing CPRT calculation is performed at each grid point on the sphere. 
Map data constructed this way allows 
efficient power spectrum analyses and 
spherical wavelet analyses \citep[e.g.][]{mcewen:2006:cswt2, sanz:2006, starck:2006, geller:2008, marinucci:2008, 
wiaux:2007:sdw, leistedt:2013:s2let_axisym, mcewen:2013:waveletsxv,
mcewen:2015:s2let_spin, 
mcewen:2016:s2let_localisation, 
chan17:curvelets} 
to characterize the spatial fluctuations of polarization, 
crucial for searching polarization signatures imprinted by large-scale magnetic fields in observational data. 

\section{Code verification}\label{sec:codeveri} 

In this section, we present the single-ray and multiple-ray experiments performed for code verification\footnote{Consistency test is 
also performed by comparing the results of light-travel time obtained by integrating equation\,(\ref{eq:dldz}) 
using our code (then dividing by the speed of light) to those that are 
obtained using the publicly available cosmological calculator 
by \citet{Wright06}, \url{http://www.astro.ucla.edu/~wright/CosmoCalc.html}. 
The results agree with each other, up to the maximum digits displayed in \citet{Wright06}, i.e. three decimal places.}. 
The $z$-sampling scheme follows the recipe described in Section \ref{subsec:zsampling} (or see the related red boxes in Fig.~\ref{fig:algoflow}). We consider polarized radiative transfer at frequencies $\nu_{\rm obs} = 1.42$~GHz and $5.00$~GHz for illustrative purposes\footnote{$\nu_{\rm obs} = 1.4$~GHz is chosen since it lies within the operating range of many current and upcoming radio telescopes, such as the Arecibo radio telescope (\url{http://www.naic.edu/}), the Five hundred meter Aperture Spherical Telescope (FAST, \url{http://fast.bao.ac.cn}), the Australia Telescope Compact Array (ATCA, \url{https://www.narrabri.atnf.csiro.au/}), LOFAR, MWA, ASKAP, SKA, etc.}.
Properties of the intervening plasma considered are listed in Table~\ref{tab:plasmaprop}, which can be IGM-like (model A) or ICM-like (model B) with magnetic field directions along the line-of-sight set at a fixed angle (models A-I and B-I) or set as randomly oriented (models A-II and B-II). 
Thermal bremsstrahlung and non-thermal synchrotron radiation process are accounted for.

\begin{table}
\centering
 \begin{tabular}{|p{3.1cm}|*4{C{1.5cm}|}}  
  \hline
   & \multicolumn{2}{c|}{IGM-like plasma}& \multicolumn{2}{c|}{ICM-like plasma}\\ 
  \hline
   \diagbox{Properties}{Model} &A-I& A-II & B-I &B-II \\
   \hline
  $n_{\rm e, tot}$~(${\rm cm}^{-3}$) & \multicolumn{2}{c|}{$2.1918 \times 10^{-7}$} &  \multicolumn{2}{c|}{$1.00 \times 10^{-3}$}\\     \hline
  ${\mathcal F}_{\rm nt}$~($\%$)  & \multicolumn{2}{c|}{$1.00$} &  \multicolumn{2}{c|}{$1.00$}\\     \hline
  $T_{\rm e, th}$~(K)  & \multicolumn{2}{c|}{$1.875  \times 10^{3} $} &  \multicolumn{2}{c|}{$5.00 \times 10^{5}$}\\ \hline
  $p$ & \multicolumn{2}{c|}{$4.00$} &  \multicolumn{2}{c|}{$2.50$}\\   
   corresponds to  $\alpha$ & \multicolumn{2}{c|}{$1.50$} &  \multicolumn{2}{c|}{$0.75$}\\      \hline
  $\gamma_{\rm i}$  & \multicolumn{2}{c|}{$10.0$} &  \multicolumn{2}{c|}{$30.0$}\\     \hline
  $|\mathbfit{B}|$~(G)  & \multicolumn{2}{c|}{$1.00 \times 10^{-9}$} &  \multicolumn{2}{c|}{$1.00 \times 10^{-6}$}\\      \hline
  $\cos{\theta}$ & 0.5 & $[-1.0,1.0]$ & 0.5 & $[-1.0,1.0]$ \\
  \hline
 \end{tabular}
 \caption{Properties of different intervening plasma models used in this paper. 
 To test the ability of our CPRT equation solver to handle the extreme limits, 
 the total electron number density $n_{\rm e, tot}$ for models A is set equal to the mean electron number density of the Universe (see Appendix~\ref{app:neIGM} for details); 
 temperature of the thermal electrons in the IGM-like and ICM-like plasma models are assumed to take the lower-end values typical to IGM and ICM. 
 ${\mathcal F}_{\rm nt}$ denotes the non-thermal relativistic electron fraction, $p$ denotes the power-law index of the energy spectrum of the non-thermal relativistic electrons, which relates to the spectral index of the synchrotron radiation $\alpha= (p-1)/2$. 
 $\gamma_{i}$ is the electrons' low-energy cutoff Lorentz factor. 
 $|\mathbfit{B}| $ denotes the magnetic field strength. 
 The magnetic field direction along the line-of-sight is described by $\cos{\theta} \in [-1.0, 1.0] $ which is set random for Models A-II and B-II. 
 }
 \label{tab:plasmaprop}
\end{table}

\subsection{Single-ray tests} \label{subsec:SRTest}

To test the accuracy and precision of our CPRT integrator in handling polarized radiative transfer in scenarios investigated in this paper, we repeat the two integration tests presented in Section 3.2 in \citet{Dexter16}. We solve the standard PRT equation (equation\,(\ref{eq:PRT})) that is reduced from the general CPRT equation ((\ref{eq:covarPRTinz}), and compare the numerical solutions that we obtained with the analytic solutions, which are explicitly given in Appendix C of \citet{Dexter16} for the idealized situations with constant transfer coefficients along a ray.

In the first test, we consider pure emission and absorption in Stokes $I$ and $Q$. The light ray travels through the Faraday-thin IGM-like plasma or the Faraday-thick ICM-like plasma (models A-I or B-I) over a cosmological distance from $z=6$ to $z=0$. Detailed values of both the (thermal and non-thermal) emission and absorption transfer coefficients, as well as the optical depths used in the calculations are given in Table \ref{tab:emi_absOnly} in Appendix \ref{app:extrainfor}. 
As is seen in Fig.~\ref{fig:SRTest_IQ}, the numerical solution obtained by our CPRT integrator agrees with the analytical solution up to the machine floating-point precision throughout the entire light path.

In the second test, we consider radiation of  observed frequencies $\nu_{\rm obs} = 1.4$~GHz and $\nu_{\rm obs} = 5.0$~GHz. The radiation travels through the Faraday-thick ICM-like plasma (B-I) of a few Mpc in length scale. Only pure Faraday rotation and Faraday conversion and polarized emission in $Q$ and $V$ are considered (note that $\epsilon_{U}$ is set to zero due to the choice of coordinate systems (see Appendix \ref{sec:Conveneqn})). To ease checking the oscillatory behavior of the resulting $V$, we boost the Faraday conversion effect artificially by setting its transfer coefficient to the same order magnitude as the Faraday rotation coefficient. The results of the second test is presented in Fig.~\ref{fig:SRTest_QUV}. An excellent agreement between the numerical and analytic solutions is obtained in both cases of different radiation frequencies. Machine floating-point precision is maintained over the ray despite that the residuals in $Q$, $U$ and $V$ increase with each oscillation. Similar trend is also found in Fig.~4 in \citet{Dexter16} and that in \citet{MoscibrodzkaGammie17_IPOLE}.

\begin{landscape}
\begin{figure}
\noindent\begin{minipage}{\linewidth}
\centering
    \vspace{-15pt}
    \hspace{-10pt}
    \mbox{\includegraphics[trim={0.2cm 0.0cm 0.0cm 0.0cm},clip, width=.49\linewidth]{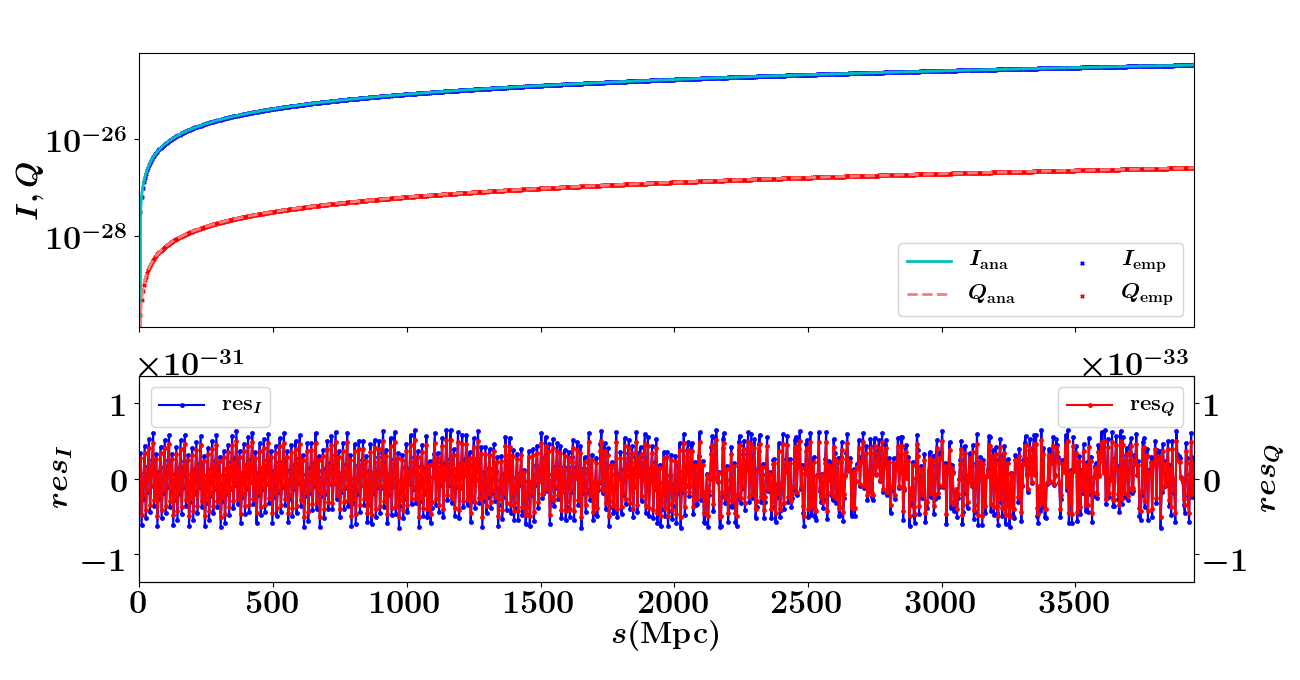}}
    \hspace{5pt}
    \mbox{\includegraphics[width=.49\linewidth]{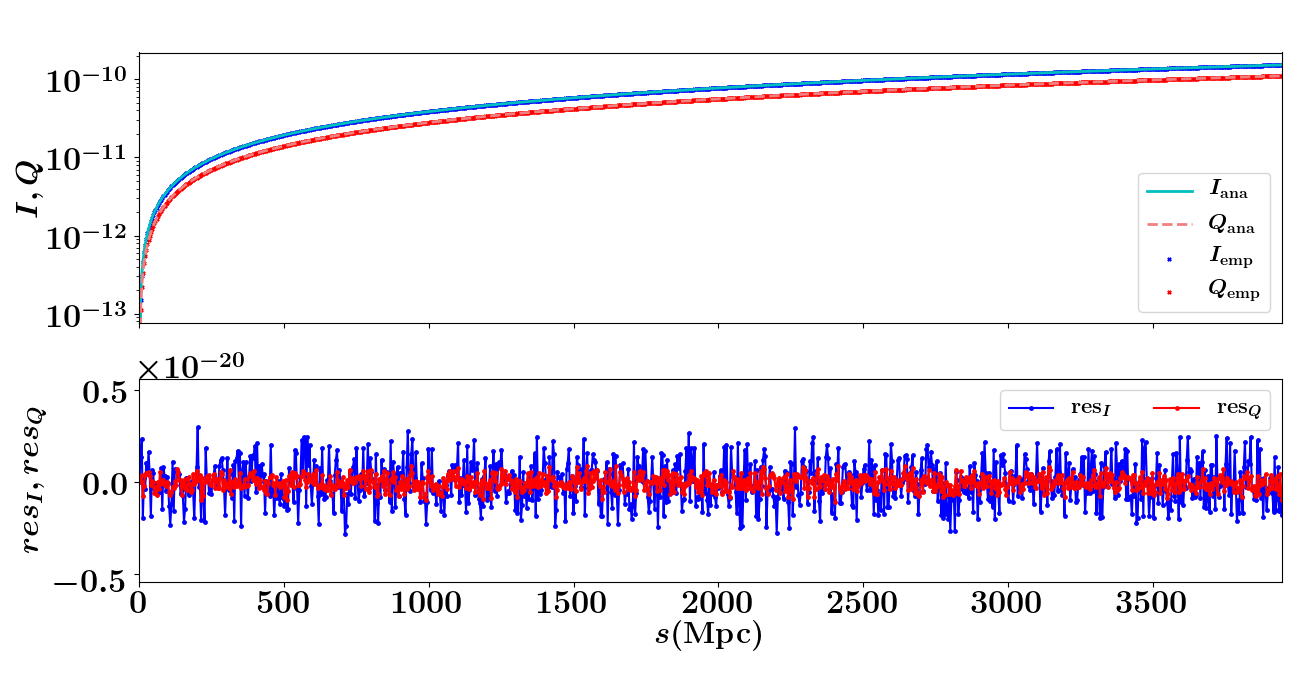}}   
    \vspace{-12pt}
    \captionof{figure}
     {\small{
     Plots of the analytic solutions (computed using equation C2, and C3 in \citet{Dexter16}; denoted by line) and numerical solutions (obtained from our CPRT code in Fortran; denoted in star) to the test problem with pure emission and absorption in $I$ and $Q$. Transfer coefficients are constant over the entire ray. 
The left-hand panels show the results using the IGM-like plasma model (A-I), where we use $(\epsilon_{I, {\rm tot}}, \epsilon_{Q, {\rm tot}})= (2.62 \times 10^{-53}, 2.06 \times 10^{-55})$\,${\rm erg}\,{\rm s}^{-1}\,{\rm cm}^{-3}\,{\rm Hz}^{-1}\,{\rm str}^{-1}$, 
$(\kappa_{\rm tot}, q_{\rm tot}) = (2.23 \times 10^{-38}, 7.07 \times 10^{-52})$\,${\rm cm}^{-1}$. 
The right-hand panels show the results using the ICM-like plasma model (B-I), where we use $(\epsilon_{I, {\rm tot}}, \epsilon_{Q, {\rm tot}}) = (1.25 \times 10^{-38}, 9.05 \times 10^{-39})$\,${\rm erg}\,{\rm s}^{-1}\,{\rm cm}^{-3}\,{\rm Hz}^{-1}\,{\rm str}^{-1}$, 
$\kappa_{\rm tot}, q_{\rm tot} = (9.34 \times 10^{-34}, 5.49 \times 10^{-34})$\,${\rm cm}^{-1}$. 
All the other transfer coefficients are set to zero. 
Note that the resulting $I$ and $Q$ have a very small order of magnitude, and thus their residual ${res}_{x} = x_{\rm emp} - x_{\rm ana}$ too, with $x = \{ I, Q\}$; dividing ${res}_{x}$ by the order of magnitude of quantity $x$ gives  machine floating-point precision. Note that such a precision is attained over the entire light path in both models.
     } \label{fig:SRTest_IQ}} 
\end{minipage}
\end{figure} 

\vspace{-45pt}
    
\begin{figure}
\noindent\begin{minipage}{\linewidth}
\centering
    \vspace{-15pt}
     \hspace{-10pt}
    \mbox{\includegraphics[trim={0.2cm 0.0cm 0.0cm 0.0cm},clip, width=.49\linewidth]{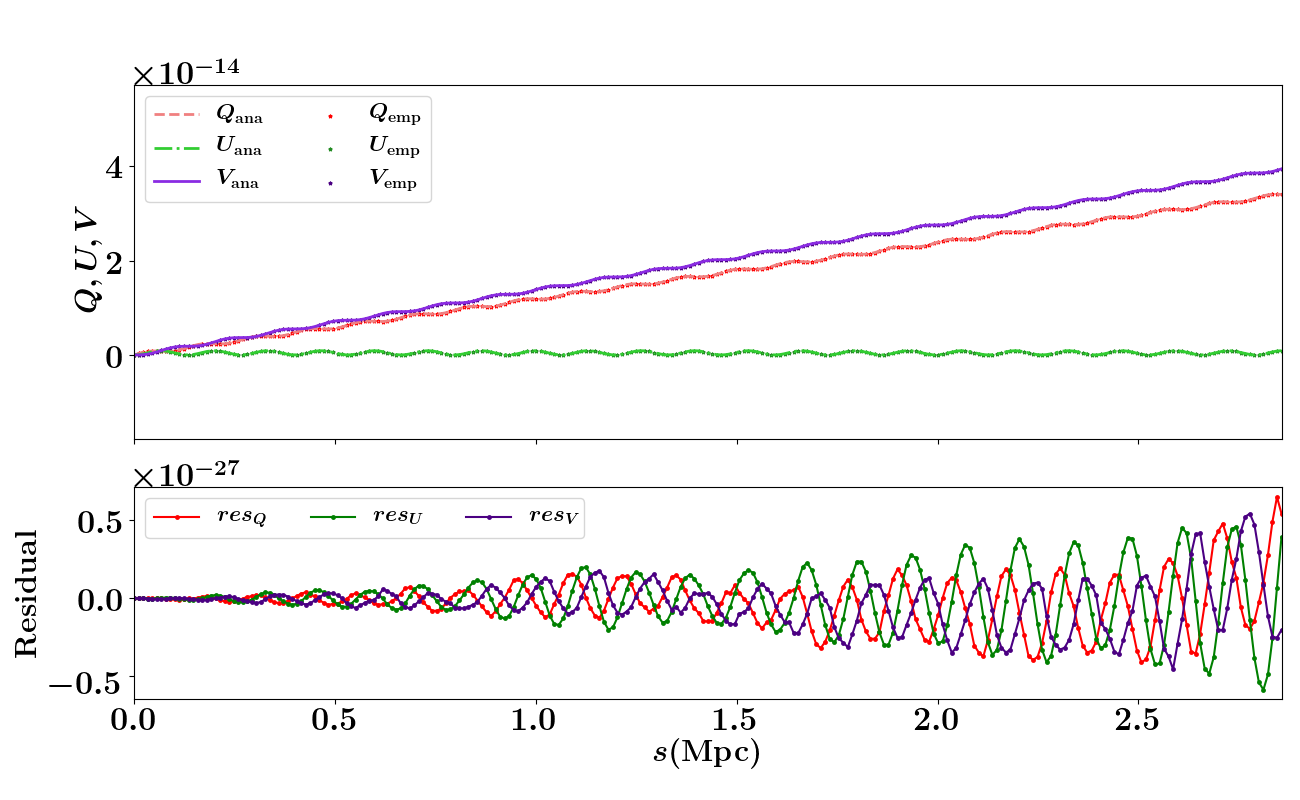}}
    \hspace{5pt}
    \mbox{\includegraphics[width=.49\linewidth]{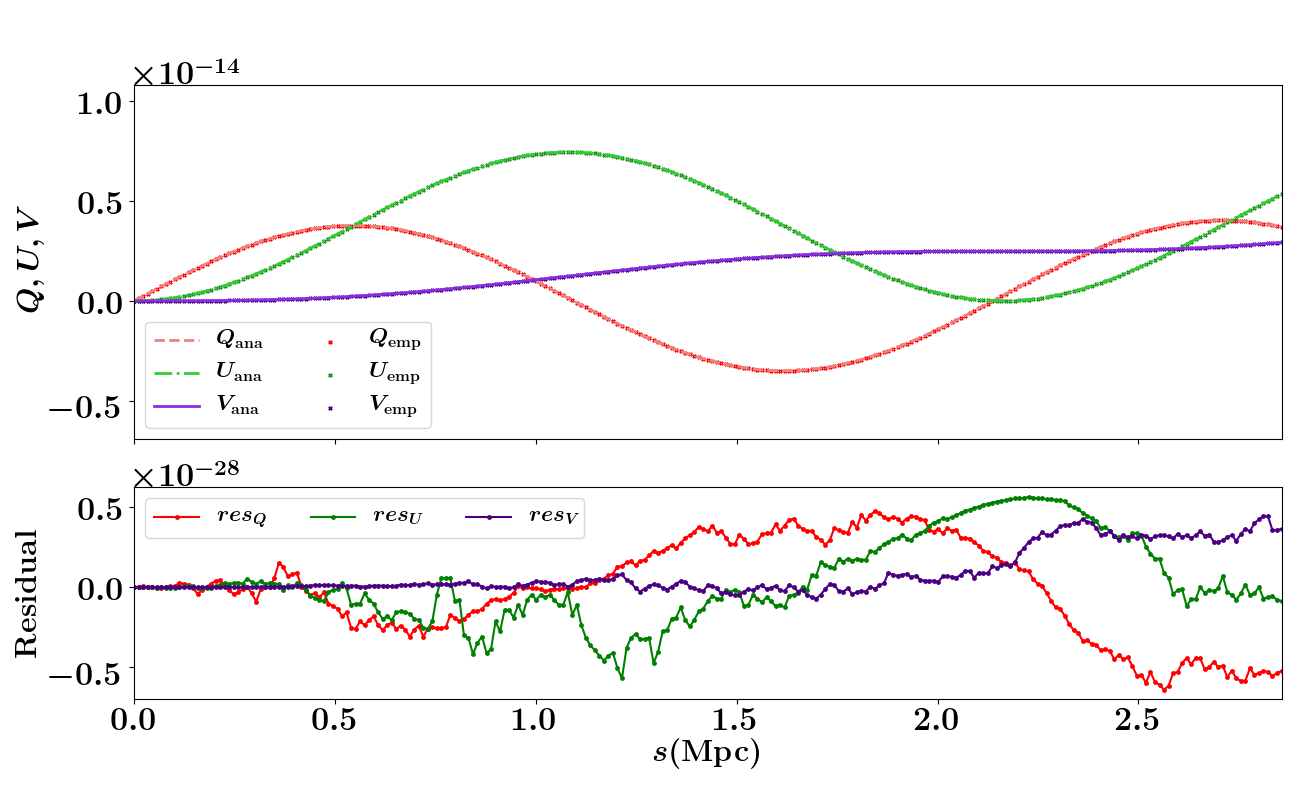}}   
    \vspace{-10pt}
    \captionof{figure}
     {\small{
     Plots of the analytic solutions (computed using equation C6, C7, C8 in \citet{Dexter16}; denoted by line) and numerical solutions (obtained from our CPRT code in Fortran; denoted in star) to the test problem with pure constant Faraday rotation, Faraday conversion and emission in $Q$ and $V$. ICM-like plasma parameters (model B-I) are used to compute the coefficients $f$, $\epsilon_{Q}$, and $\epsilon_{V}$ while $h$ is set to be of the same order of magnitude of $f$ to make the oscillatory behavior in $V$ apparent. 
     The left- and right-hand panels show the results using $\nu_{\rm obs} = 1.4$~GHz and $\nu_{\rm obs} = 5.0$~GHz, respectively. 
     At $\nu_{\rm obs} = 1.4$~GHz, the non-zero transfer coefficients are $(f_{\rm tot}, h_{\rm tot}) = (1.16 \times 10^{-23}, 1.00 \times 10^{-23}$)\,${\rm cm}^{-1}$, 
     $(\epsilon_{Q, {\rm tot}}, \epsilon_{V, {\rm tot}}) = (9.05 \times 10^{-39}, 5.51 \times 10^{-43})$\,${\rm erg}\,{\rm s}^{-1}\,{\rm cm}^{-3}\,{\rm Hz}^{-1}\,{\rm str}^{-1}$. 
     At $\nu_{\rm obs} = 5.0$~GHz, the non-zero transfer coefficients are $(f_{\rm tot}, h_{\rm tot}) = (9.37 \times 10^{-25}, 1.00 \times 10^{-25})$\,${\rm cm}^{-1}$, 
     $(\epsilon_{Q, {\rm tot}}, \epsilon_{V, {\rm tot}}) = (3.52 \times 10^{-39}, 1.14 \times 10^{-43})$\,${\rm erg}\,{\rm s}^{-1}\,{\rm cm}^{-3}\,{\rm Hz}^{-1}\,{\rm str}^{-1}$. 
     Residuals grow with each oscillation, yet, machine floating-point precision is attained (with residual divided by the order of magnitude of the corresponding Stokes parameter) over the entire light path in both cases.
     } \label{fig:SRTest_QUV}} 
\end{minipage}
\end{figure} 
\end{landscape}

\subsection{Multiple-ray test} \label{subsec:Spectrumz0}

To verify the redshift-refinement scheme and the whole code, we performed multiray calculations evaluating the cosmological radiative transfer of two Gaussian profiles that centered at two different redshifts. The two input profiles, originating at $z_{\rm ori} = 5.94$ and $1.00$, have frequency samples assigned through the redshift-refinement scheme at that specific $z_{\rm ori}$. The central frequency of the profiles is then given by $\nu^{\rm in}_{\rm central}=\nu_{\rm obs}(1+z^{\rm refine}_{\rm central})$, where $z^{\rm refine}_{\rm central}$ is the redshift value of the $N^{\rm refine}/2$ cell, for $N^{\rm refine} = 500$. The ray then freely propagates in vacuum afterwards, i.e. all the transfer coefficients are set to zero when computing the CPRT equation. As such, frequency shift of the radiation is the only cosmological effect which modifies the radiation properties in its transport. 
We compare the values of four quantities obtained from the CPRT calculations against the theoretical expected values. These quantities are 
(i) the frequency at which the resulting profile peaks, $\nu_{\rm peak, 0}$, 
(ii) the standard deviation of the resulting profile, $\sigma_{0}$, 
(iii) the empirical ratio of the output to the input peak intensity 
$r^{\rm emp}_{I}= I^{\rm in}_{\rm peak}/I_{\rm peak, 0}$, for each Gaussian profile, and 
(iv) the power-law index of the ratio of the output peak intensities of the two profiles, $m^{\rm emp}$. 
Analytically, 
the resulting profile obtained from the CPRT of each case (i.e. emission at $z_{\rm ori} = 5.94$ or at $z_{\rm ori}=1.00$) should remain Gaussian 
and peak at the frequency of $\nu_{\rm obs} \times (1+z^{\rm refine}_{\rm central})/(1+z_{\rm ori})$ with $\nu_{\rm obs}= 1.42$~GHz.
The standard deviation of the normalized input and the output Gaussian profiles should remain the same. 
The ratio of the peak intensity of the output emission profile to that of the input profile follows $r^{\rm ana}_{I}= 1/(1+z_{\rm ori})^{3}$. 
Furthermore, comparing the outputs of the two cases, the ratio of the peak intensity at zero redshift should follow a power law of $[(1+z_{\rm ori}'')/(1+z_{\rm ori}')]^3$, where  $z''$ denotes the higher redshift. That is, the power-law index $m^{\rm ana} = 3.0$. 

We summarize the results in Table~\ref{tab:Veri_table}, from which one can see that the empirical results are consistent with the theoretical expectation up to machine floating-point precision. Furthermore, consistent results are obtained using the parallelized code (i.e.\, with multiple threading using OpenMP) as those obtained by the serial execution. 

\begin{table}
\centering
 \begin{tabular}{|p{1.8cm}|C{3.7cm}|*2{C{5.5cm}|}}  
  \hline
  \multicolumn{2}{|c|}{}& Profile I & Profile II \\ 
  \hline   
  \multicolumn{2}{|c|}{$z_{\rm ori}$} & \multicolumn{1}{c|}{$5.936 234 097 751 42$}  &   \multicolumn{1}{c|}{$1.00082825012323$}  \\ \hline 
  \multicolumn{2}{|c|}{$z^{\rm refine}_{\rm central}$} & \multicolumn{1}{c|}{$5.904 994 497 304 97$}  &   \multicolumn{1}{c|}{$0.9992434724235666$}  \\ \hline 
  \multicolumn{1}{|C{2.0cm}|}{{Peak frequency}} & {Input $\nu^{\rm in}_{\rm central}$~(GHz)}& $1.414 008 488 810 21$ &  ${1.419 280 703 231 15}$\\ \cline{2-4}
  \multicolumn{1}{|C{2.0cm}|}{}   & Output {$\nu_{\rm peak, 0}$~(GHz)} &  $1.414 008 488 810 21 $ &   $ {1.419 280 703 231 15}$  \\   \cline{2-4}  
  \multicolumn{1}{|C{2.0cm}|}{{}}  & Fractional difference $(\nu_{\rm peak, 0} - \nu^{\rm in}_{\rm central})/ \nu^{\rm in}_{\rm central}$& $-8.99280649946616 \times 10^{-15}$  & $ -1.110223037256493 \times 10^{-16}$ \\     
    \hline
  \multicolumn{1}{|C{2.0cm}|}{Dispersion} & Input $\sigma^{\rm in}$  & ${0.000866648853601}$ &  $-0.000123250853536$ \\ \cline{2-4}
  \multicolumn{1}{|C{2.0cm}|}{} & Output $\sigma_{0}$ &  {$0.000866648853601$} &   $ -0.000123250853536 $  \\   \cline{2-4}
  \multicolumn{1}{|C{2.0cm}|}{{}}  & Fractional difference $(\sigma_{0} - \sigma^{\rm in})/\sigma^{\rm in}$ & $1.59377719355 \times 10^{-17}$  & $ 5.66495635124 \times 10^{-18}$ \\     
    \hline
  \multicolumn{1}{|C{2.0cm}|}{Peak intensity ratio} & Analytical $r^{\rm ana}_{I}$  & ${0.002 996 599 988 84}$ &  $0.124 844 831 637 85$ \\ \cline{2-4}
  \multicolumn{1}{|C{2.0cm}|}{} & Empirical $r^{\rm emp}_{I}$ &  {$0.002 996 599 988 84$} &   $0.124 844 831 637 85 $  \\   \cline{2-4}
  \multicolumn{1}{|C{2.0cm}|}{{}}  & Fractional difference$(r^{\rm emp}_{I}-r^{\rm ana}_{I})/r^{\rm ana}_{I}$ & $  4.34172933267 \times 10^{-16}$  & $ 1.06713879307 \times 10^{-14}$ \\     
    \hline   
  \multicolumn{1}{|C{2.0cm}|}{Power-law index of } & Analytical $m^{\rm ana}$ &  \multicolumn{2}{c|}{${3.0000}$}    \\ \cline{2-4}
  \multicolumn{1}{|C{2.0cm}|}{$(I_{\rm peak,0}^{z_{\rm ori}=5.94}/$} & Empirical {$m^{\rm emp}$}  &   \multicolumn{2}{c|}{$3.000 000 000 000 99$}   \\   \cline{2-4}
  \multicolumn{1}{|C{2.0cm}|}{{$I_{\rm peak,0}^{z_{\rm ori}=1.00})$}}  & Fractional difference $(m^{\rm emp}-m^{\rm ana})/m^{\rm ana}$ &  \multicolumn{2}{c|}{$-3.29218134236 \times 10^{-13}$}  \\     
    \hline    
 \end{tabular}
 \caption{Results of the multiray code-verification test where two Gaussian profiles, originating at $z_{\rm ori} = 5.94$ and at $z_{\rm ori} =1.00$ respectively, are cosmologically transported in a vacuum but an expanding flat space--time. Four parameters are compared against their theoretical values; the empirical results are found to be consistent with the expected values up to machine floating-point precision. 
 }
 \label{tab:Veri_table}
\end{table}

\vspace{-0.5cm}
\section{Applications}\label{sec:cprtdemo}

Here, we present a set of CPRT calculations to demonstrate the ability of the algorithm in tracking the change of polarization on astrophysical and cosmological scales. 
Changes in polarization features caused by the frequency shift of the radiation, 
or those caused by the evolution of intervening cosmic plasmas can be separately investigated; direct studies of their combined effects can also be directly carried out. 

We start with a set of single-ray calculations, showing in our case studies how polarization changes over cosmological distances with and without a bright line-of-sight point source. 
Then we demonstrate how to incorporate cosmological MHD simulation results into CPRT calculations to make polarization maps. We compute the polarization of a simulated galaxy cluster. We also compute the entire polarized sky using a model magnetized universe. Polarization maps generated in such a way, i.e. by CPRT calculations with an interface of simulation results, encapsulate theoretical predictions. They are crucial to aid our interpretation of observational data. Model templates of the entire sky are particularly important for comparison with future observational data, such as those from all-sky surveys of polarized emission with the SKA. 
%

\subsection{Cosmological evolution of polarization}\label{sec:cprt_SR}

We perform a set of ray-tracing calculations for 
radiation with observed frequency $\nu_{\rm obs} = \nu_{0}= 1.42$~GHz propagating from $z = 6.0$ through some distributions of $n_{\rm e}(z)$ and $|\mathbfit{B}(z)|$ as described below. 
The $z$-sampling scheme follows the recipe described in Section \ref{subsec:zsampling}. 

\subsubsection{Point-source emissions}\label{sec:cprt_ptsource}
\begin{table*}
\begin{minipage}[t]{\hsize}
\begin{center}
\begin{tabular}{|c|c|c|c|}
 \hline
{} & {Without point source} &{Point source at $z=6.0$} &{Point source at $z=0.206$} \\
\hline
  initial Stokes parameters 
$
\left[ \begin{array}{c}
I \\
Q\\
U\\
V \end{array} \right]_{z=6.0}$&
$\left[ \begin{array}{c}
0.0000 \\
0.0000\\
0.0000\\
0.0000 \end{array} \right]$& 
$\left[ \begin{array}{c}
8.7096 \times 10^{-14} \\ 
-4.5372 \times 10^{-16}\\ 
2.5731 \times 10^{-15}\\ 
4.3548 \times 10^{-17} \end{array} \right]$ & 
$\left[ \begin{array}{c}
0.0000 \\
0.0000\\
0.0000\\
0.0000 \end{array} \right]$\\
\hline
    final Stokes parameters 
$
\left[ \begin{array}{c}
I \\
Q\\
U\\
V \end{array} \right]_{z=0.0}$&
$\left[ \begin{array}{c}
1.0438 \times 10^{-23} \\ 
6.7617 \times 10^{-25} \\ 
5.1681 \times 10^{-26} \\ 
-1.7613 \times 10^{-32}  \end{array} \right]$&  
$\left[ \begin{array}{c}
2.5392 \times 10^{-16} \\ 
-2.8095  \times 10^{-18}\\ 
7.0807 \times 10^{-18}\\ 
1.2696 \times 10^{-19} \end{array} \right]$&  
$\left[ \begin{array}{c}
2.5392 \times 10^{-16} \\ 
-1.3221 \times 10^{-18}\\ 
7.5021  \times 10^{-18}\\
1.2696 \times 10^{-19}  \end{array} \right]$\\ 
  \hline
initial $\varphi({z=6.0})$&
0.0000 & 
0.8727 & 
0.0000 \\  
  \hline
final $\varphi({z=0.0})$&
$3.8142 \times 10^{-2}$ &   
0.9731 & 
0.8726 \\  
  \hline
initial $\Pi_{{\rm l}}(z=6.0)$&
0.0000 & 
$3.0000$ &  
0.0000 \\  
  \hline
final $\Pi_{{\rm l}}(z=0.0)$&
$6.4969$&  
$3.0000$ &  
$3.0000$ \\  
  \hline
initial $\Pi_{{\rm c}}(z=6.0)$&
0.0000 & 
$5.0000 \times 10^{-2}$ & 
0.0000 \\  
  \hline
final $\Pi_{{\rm c}}(z=0.0)$&
$1.6874 \times 10^{-7}$ &  
$5.0000 \times 10^{-2}$&  
$5.0000 \times 10^{-2}$ \\  
  \hline
initial $\Pi_{{\rm tot}}(z=6.0)$&
0.0000 & 
$3.0004 $ & 
0.0000 \\  
  \hline
final $\Pi_{{\rm tot}}(z=0.0)$&
$6.4969$  &  
$3.0004$  & 
$3.0004$ \\   
  \hline
  \end{tabular}
\caption{\small{Numerical results of the CPRT calculations for the demonstrative cases where bright point source is (i) absent, (ii) located at $z=6.0$ or (iii) located at $z=0.206$; magnetic fields orientate along the line-of-sight at random angles (see Section \ref{sec:cprt_ptsource}). 
The Stokes parameters are in units of ${\rm erg}\,{\rm s}^{-1}\,{\rm cm}^{-2}\,{\rm Hz}^{-1}\,{\rm str}^{-1}$, $\varphi$ is measured in radian, and $\Pi_{\rm l}$, $\Pi_{\rm c}$, and $\Pi_{\rm tot}$ are expressed in percentages. Note that for case (i) the resulting $I$ has an order of magnitude $10^{-23}$, which is much smaller than the specific intensity of the cosmic microwave background of $10^{-18}$\,${\rm erg}\,{\rm s}^{-1}\,{\rm cm}^{-2}\,{\rm Hz}^{-1}\,{\rm str}^{-1}$ at the same observed frequency. This suggests that emission and polarization signals would be overwhelmed by the CMB background in real observations. 
}\label{table:quasarnumerical_randbori}} 
\end{center}
\end{minipage}
\end{table*}

Bright polarized emitters such as quasars and radio galaxies may lie along the line-of-sight acting as back-light illuminating the foreground. 
Here, we calculate how the polarization and intensity of a fiducial quasar-like point source changes over a cosmological distance. 
Emissions of such a point source at $z$ observed at $1.42$~GHz is given by 
$[{I}, {Q}, {U}, {V}]|_{z}= [{I}, {Q}, {U}, {V}]|_{z=0} (1+z)^{3}$, where 
$[{I}, {Q}, {U}, {V}]|_{z=0} = 
[\,2.54 \times 10^{-16}, -1.32  \times 10^{-18}, 7.50 \times 10^{-18}, 1.27 \times 10^{-19}]$
\,${\rm erg}\,{\rm s}^{-1}\,{\rm cm}^{-2}\,{\rm Hz}^{-1}\,{\rm str}^{-1}$, where we have assumed the degree of linear polarization to be $3.00\%$ 
\citep[][]{Jagers82}, the degree of circular polarization to be $0.05\%$ \citep{Conway71},  
and the polarization angle $\varphi = 0.87$~rad. For demonstrative purposes, we adopt such a simple interpolation of 
$[{I}, {Q}, {U}, {V}]|_{z}$ from $[{I}, {Q}, {U}, {V}]|_{z=0}$, focusing on polarization effects caused by our input plasma of known properties. 
 
Three cases are investigated, including 
(i) the control experiment where there is no bright point source lying along the line-of-sight, no radiation background, but the intervening medium is a self-emitting, absorbing, Faraday-rotating and Faraday-converting medium, 
(ii) the fiducial point source is placed at $z = z_{\rm init} = 6.0$, serving as a bright distant radio back-light, and (iii) the fiducial point source is located much nearer, at $z = 0.206$ (cf. \citealt{Jagers82}). 
The prescription of the intervening plasma at $z=0$ follows model A-II described in Table~\ref{tab:plasmaprop}; simple cosmological evolutions of $n_{\rm e}(z)$,  $T_{\rm e}(z)$ and $|\mathbfit{B}(z)|$ described in Section \ref{subsubsec:CosmoEvo} are now accounted for while the fraction of non-thermal relativistic electrons ${\mathcal F}_{\rm nt}$, their energy spectral index $p$ and the Lorentz factor of low-energy electron cut-off $\gamma_{i}$ are assumed to be constant over all redshifts. The results of the three different scenarios are displayed in parallel in Figs.~\ref{fig:CPRT_SR_CaseQuasarInv_RandBori} -- \ref{fig:CPRT_SR_CaseQuasarPol_RandBori} for comparison purposes. Numerical results are summarized in Table~\ref{table:quasarnumerical_randbori}. 

Differences in the results of the three cases indicate that on cosmological scale, polarized radiative transfer of light traveling through a foreground cosmologically-evolving IGM-like plasma, with or without a bright point source, can impart unique polarization features. 
Also, it can be readily seen from Figs.~\ref{fig:CPRT_SR_CaseQuasarInv_RandBori} and \ref{fig:CPRT_SR_CaseQuasarCo_RandBori} that 
both the total emission and the polarized emission from the fiducial point source
dominate over the contributions from the foreground plasma, as expected. 
The invariant intensity ${\mathcal I}$ of the radiation 
stays by and large constant from where the bright point source is positioned 
with a very small increase over increasing $z$ due to the emission of the line-of-sight plasma, which is  calculated in case (i). 
Fluctuations in Stokes parameters are induced by random field orientations along the line-of-sight. 

The observed change of polarization angle $\Delta \varphi$, which is a measure of the amount of Faraday rotation and is sensitive to the magnetic field directions along the line-of-sight, depends on the $z$-position of the point-source, as is seen in Fig.~\ref{fig:CPRT_SR_CaseQuasarPol_RandBori} and Table~\ref{table:quasarnumerical_randbori}. In all three cases, we obtained  $\Delta \varphi < \pi$. This indicates that the effect of Faraday rotation is weak, as is expected for a line-of-sight plasma that is threaded with a weak magnetic field of nG and has a low electron number density. 
Insignificant Faraday conversion is also observed in case (i), for which there is only the plasma but no bright sources lying along the line-of-sight. Note that ${\Pi}_{\rm c}$ is much weaker than ${\Pi}_{\rm l}$ by an order of magnitude of $10^{5}$. For case (ii), ${\Pi}_{\rm l}$ and ${\Pi}_{\rm c}$ are dominated by the contributions of the bright point source over the foreground plasmas. 
For case (iii), the sudden drops in ${\Pi}_{\rm l}$ and in ${\Pi}_{\rm tot}$, and the large rise in $\Delta \varphi$, and ${\Pi}_{\rm c}$ show the effects of having a foreground (nearby) source. Understanding the foregrounds, particularly any bright line-of-sight sources and their locations, is crucial for scientific inference of magnetic fields and their evolution. 

In addition, depolarization effect is observed: there is a net drop in $\Pi_{\rm tot}$ as $z$ decreases (i.e. as path-length increases). By the experimental set up, this is mainly due to differential Faraday rotation (i.e. emission at different $z$ is rotated by different amount due to their magneto-ionized foreground, thus reducing the net polarization). 
Random magnetic fields has also been identified in the literature as another cause of depolarization \citep[see e.g.][]{Burn66}. Investigation of the effects of random fields is beyond the scope of this demonstration, but our results here illustrate how the effects on polarization can be quantified by performing a full cosmological polarized radiative transfer. 

\begin{landscape}
\begin{figure}
\begin{minipage}{.42\linewidth}
\centering
\rotatebox[origin=c]{90}{$\mathbf{\mathcal{I}(z)}$}\quad
\subfloat{\includegraphics[trim={0.0cm 0.0cm 0.0cm 0.0cm},clip,width=0.75\textwidth,valign=c]{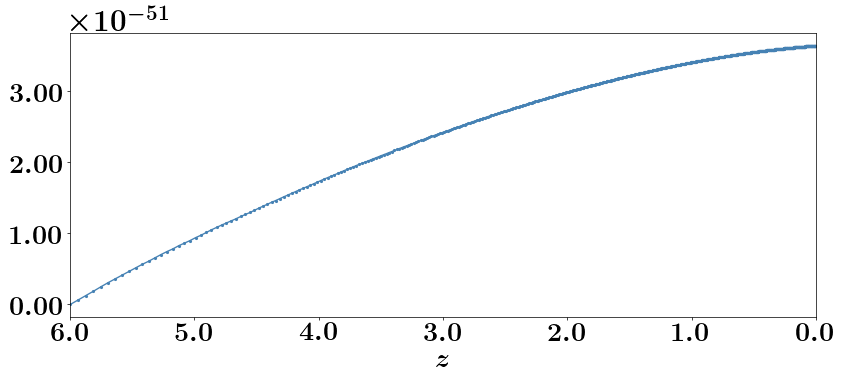}}\vspace{0.0em}\\
\rotatebox[origin=c]{90}{$\mathbf{\mathcal{Q}(z)}$}\quad
\subfloat{\includegraphics[trim={0.0cm 0.0cm 0.0cm 0.0cm},clip,width=0.75\textwidth,valign=c]{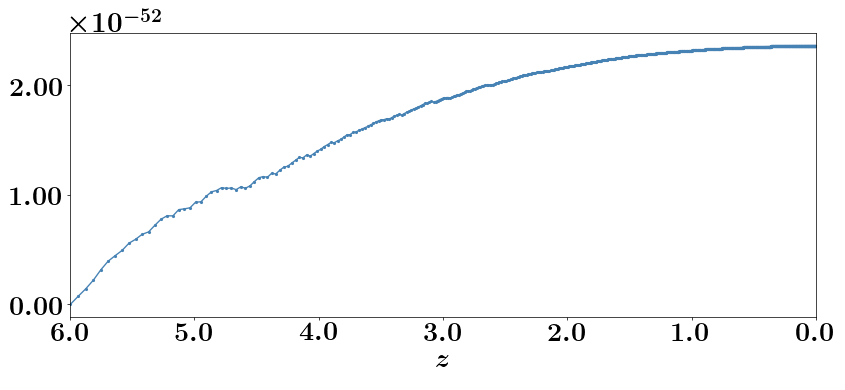}}\vspace{0.0em}\\
\rotatebox[origin=c]{90}{$\mathbf{\mathcal{U}(z)}$}\quad
\subfloat{\includegraphics[trim={0.0cm 0.0cm 0.0cm 0.0cm},clip,width=0.75\textwidth,valign=c]{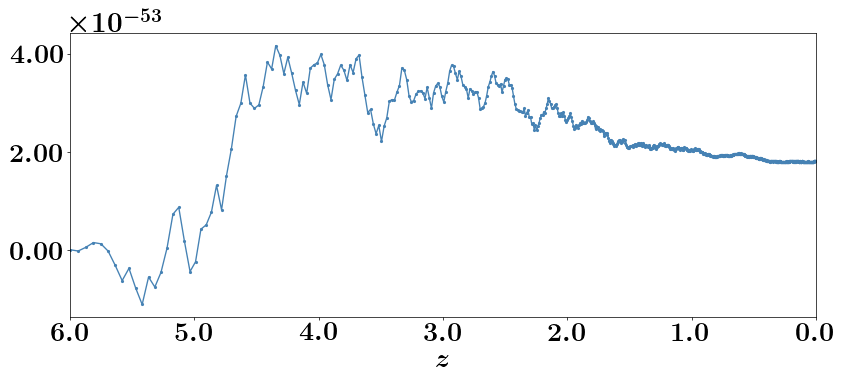}}\vspace{0.0em}\\
\stepcounter{figure}\addtocounter{figure}{-1}  
\rotatebox[origin=c]{90}{$\mathbf{\mathcal{V}(z)}$}\quad
\subfloat[]{\includegraphics[trim={0.0cm 0.0cm 0.0cm 0.0cm},clip,width=0.75\textwidth,valign=c]{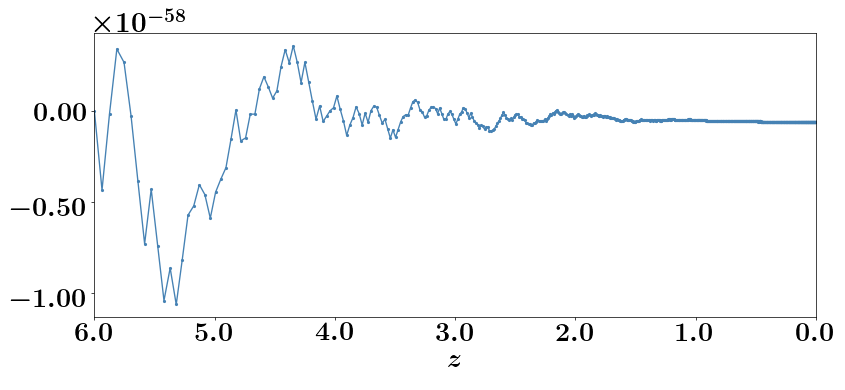}}
\end{minipage}%
\hspace{-8.2em}
\begin{minipage}{.42\linewidth}
\centering
\rotatebox[origin=c]{90}{}\quad
\subfloat{\includegraphics[trim={0.0cm 0.0cm 0.0cm 0.0cm},clip,width=0.75\textwidth,valign=c]{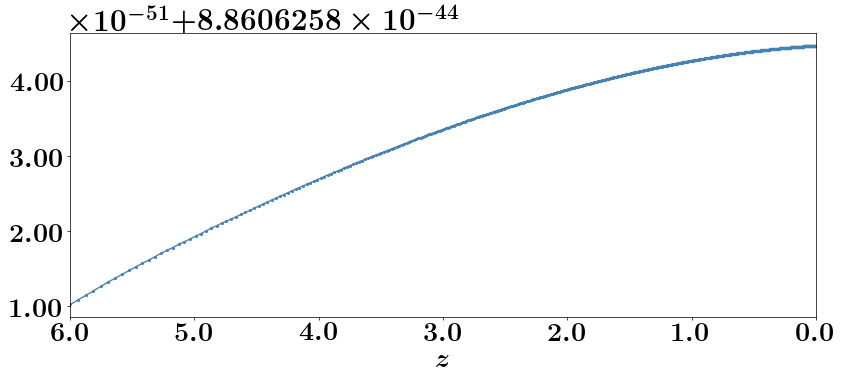}}\vspace{0.0em}\\
\rotatebox[origin=c]{90}{}\quad
\subfloat{\includegraphics[trim={0.0cm 0.0cm 0.0cm 0.0cm},clip,width=0.75\textwidth,valign=c]{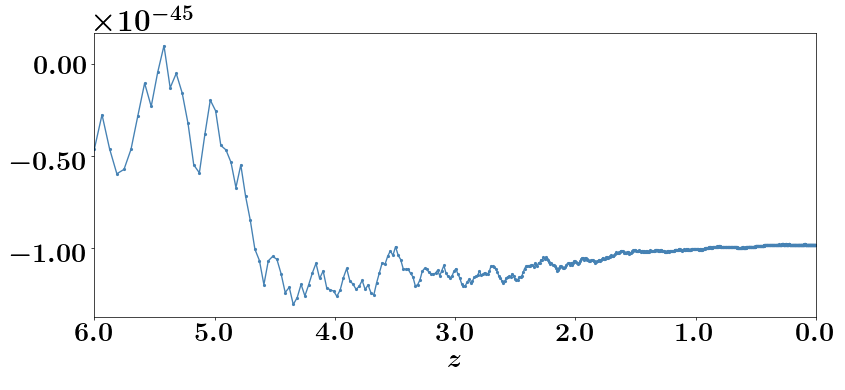}}\vspace{0.0em}\\
\stepcounter{figure}\addtocounter{figure}{-1}  
\rotatebox[origin=c]{90}{}\quad
\subfloat{\includegraphics[trim={0.0cm 0.0cm 0.0cm 0.0cm},clip,width=0.75\textwidth,valign=c]{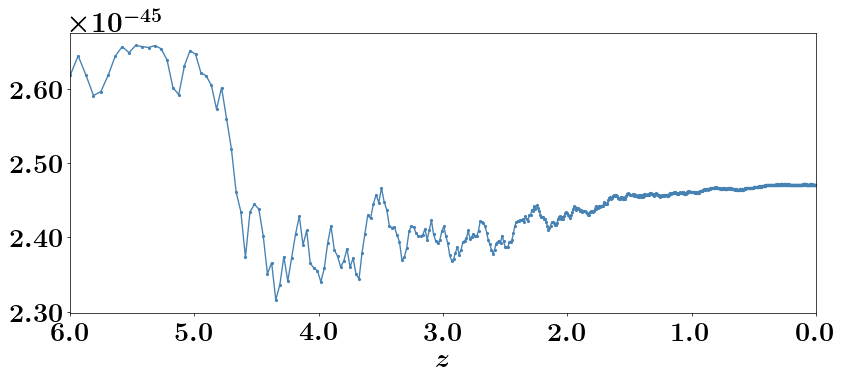}}\vspace{0.0em}\\
\rotatebox[origin=c]{90}{}\quad
\subfloat[]{\includegraphics[trim={0.0cm 0.0cm 0.0cm 0.0cm},clip,width=0.75\textwidth,valign=c]{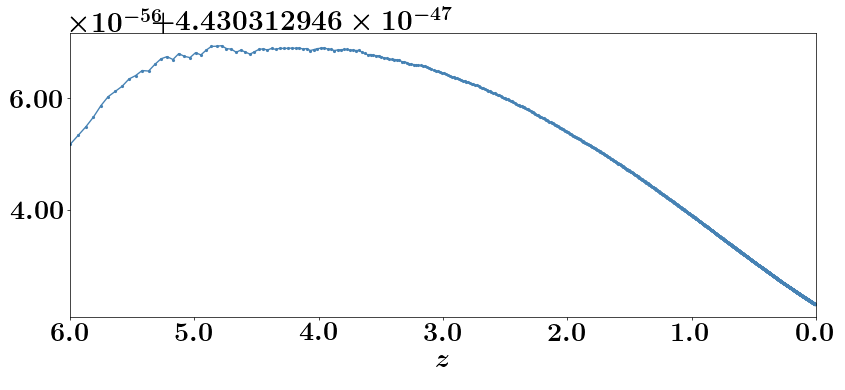}}
\end{minipage}%
\hspace{-9.2em}
\begin{minipage}{.42\linewidth}
\centering
\rotatebox[origin=c]{90}{}\quad
\subfloat{\includegraphics[trim={0.0cm 0.0cm 0.0cm 0.0cm},clip,width=0.75\textwidth,valign=c]{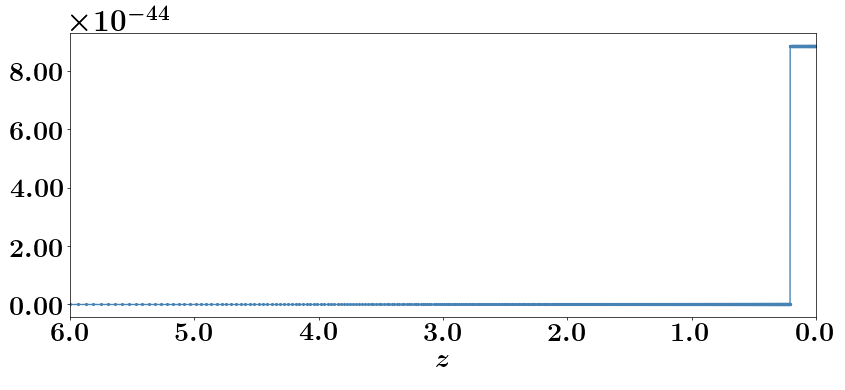}}\vspace{0.0em}\\
\rotatebox[origin=c]{90}{}\quad
\subfloat{\includegraphics[trim={0.0cm 0.0cm 0.0cm 0.0cm},clip,width=0.75\textwidth,valign=c]{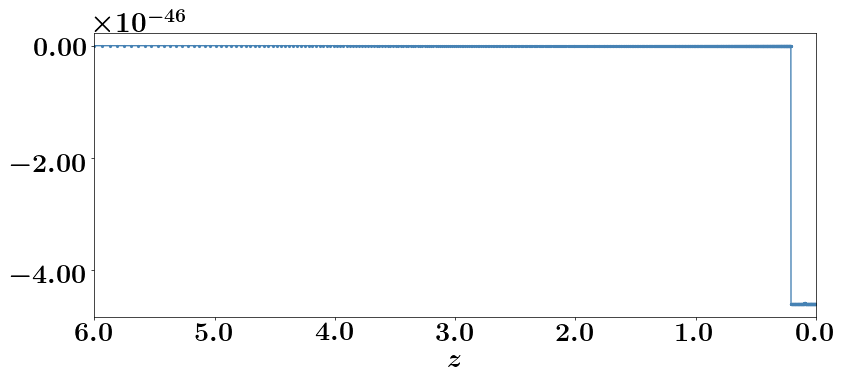}}\vspace{0.0em}\\
\stepcounter{figure}\addtocounter{figure}{-1}  
\rotatebox[origin=c]{90}{}\quad
\subfloat{\includegraphics[trim={0.0cm 0.0cm 0.0cm 0.0cm},clip,width=0.75\textwidth,valign=c]{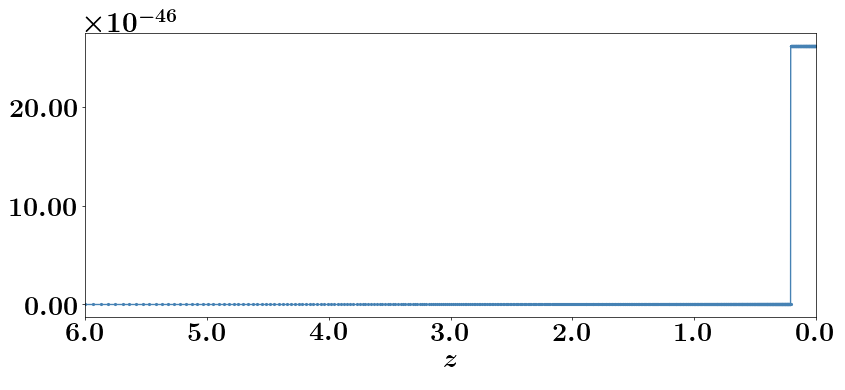}}\vspace{0.0em}\\
\rotatebox[origin=c]{90}{}\quad
\subfloat[]{\includegraphics[trim={0.0cm 0.0cm 0.0cm 0.0cm},clip,width=0.75\textwidth,valign=c]{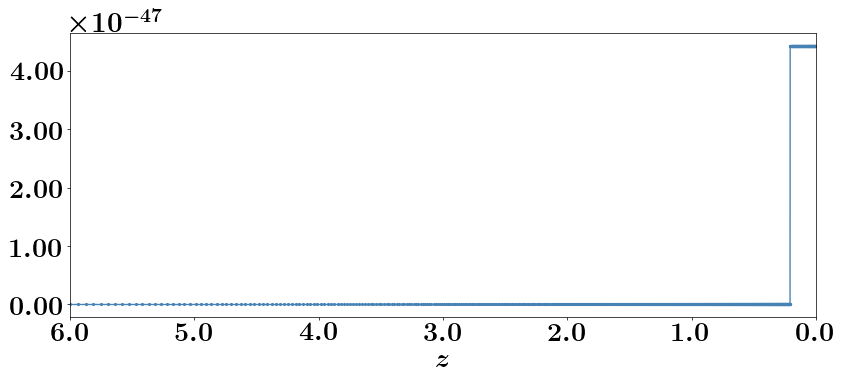}}
\end{minipage}%
 \caption{Cosmological evolution of the {\it invariant} Stokes parameters (in units of\,${\rm erg}\,{\rm s}^{-1}\,{\rm cm}^{-2}\,{\rm Hz}^{-4}\,{\rm str}^{-1}$) for $\nu_{\rm obs} = 1.42$~GHz for the cases where the radio bright point source is (i) absent, (ii) located at $z= 6.0$, and (iii) located at $z=0.206$; line-of-sight magnetic field orientations, simulated from a single realization, are random (see Section \ref{sec:cprt_ptsource}). Emission, absorption, Faraday rotation, and Faraday conversion for thermal bremsstrahlung and non-thermal synchrotron radiation process are taken into account. 
Note that fluctuations caused by random field directions in the results of case (iii) can be seen in zoom-in figures, where at $z$ prior to the point-source location the Stokes parameters evolve as in those of case (i). Here, we display the results over the full-redshift range for comparison purposes.
}
\label{fig:CPRT_SR_CaseQuasarInv_RandBori}
 \end{figure}
\end{landscape}

\begin{landscape}
\begin{figure}
\begin{minipage}{.42\linewidth}
\centering
\rotatebox[origin=c]{90}{$\mathbf{{I}(z)}$}\quad
\subfloat{\includegraphics[trim={0.0cm 0.0cm 0.0cm 0.0cm},clip,width=0.75\textwidth,valign=c]{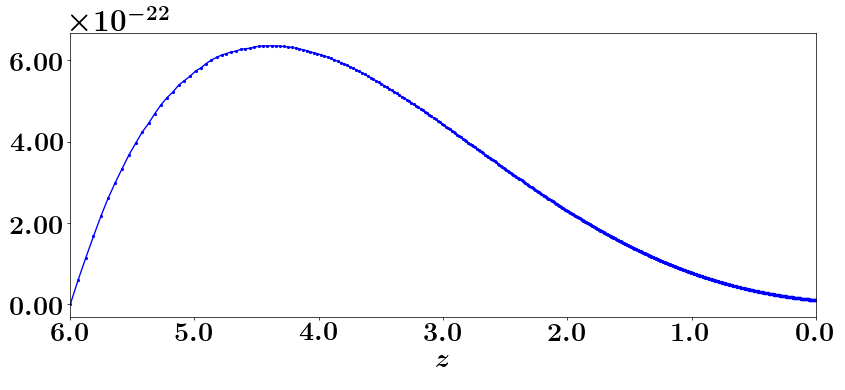}}\vspace{0.0em}\\
\rotatebox[origin=c]{90}{${Q}(z)$}\quad
\subfloat{\includegraphics[trim={0.0cm 0.0cm 0.0cm 0.0cm},clip,width=0.75\textwidth,valign=c]{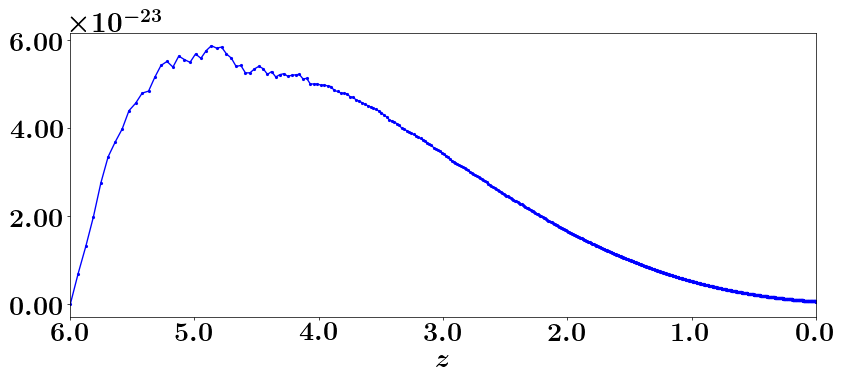}}\vspace{0.0em}\\
\rotatebox[origin=c]{90}{$\mathbf{{U}(z)}$}\quad
\subfloat{\includegraphics[trim={0.0cm 0.0cm 0.0cm 0.0cm},clip,width=0.75\textwidth,valign=c]{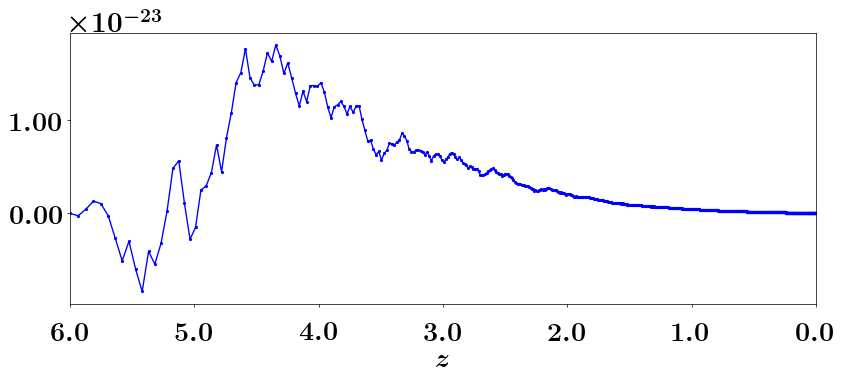}}\vspace{0.0em}\\
\stepcounter{figure}\addtocounter{figure}{-1}  
\rotatebox[origin=c]{90}{$\mathbf{{V}(z)}$}\quad
\subfloat[]{\includegraphics[trim={0.0cm 0.0cm 0.0cm 0.0cm},clip,width=0.75\textwidth,valign=c]{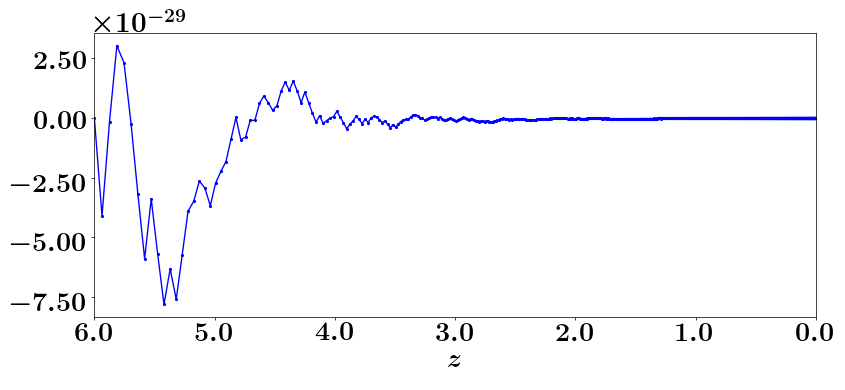}}
\end{minipage}%
\hspace{-8.2em}
\begin{minipage}{.42\linewidth}
\centering
\rotatebox[origin=c]{90}{}\quad
\subfloat{\includegraphics[trim={0.0cm 0.0cm 0.0cm 0.0cm},clip,width=0.75\textwidth,valign=c]{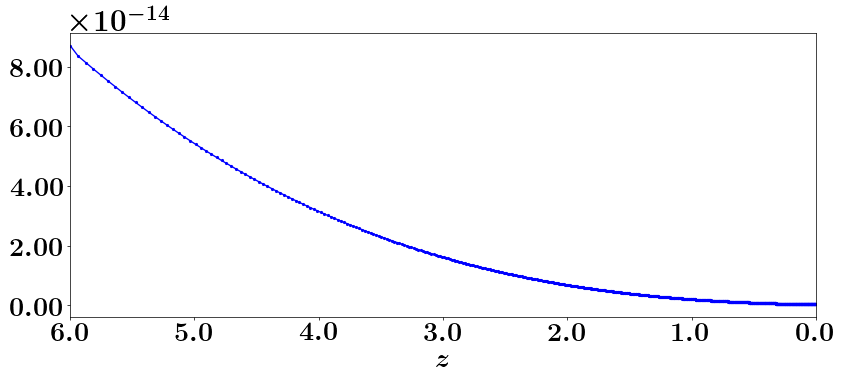}}\vspace{0.0em}\\
\rotatebox[origin=c]{90}{}\quad
\subfloat{\includegraphics[trim={0.0cm 0.0cm 0.0cm 0.0cm},clip,width=0.75\textwidth,valign=c]{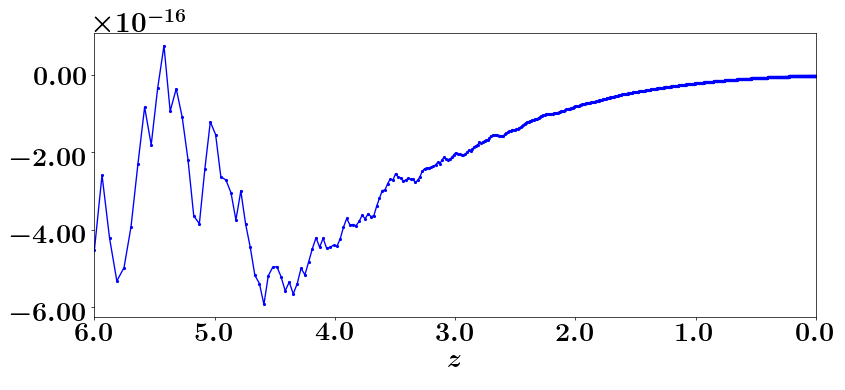}}\vspace{0.0em}\\
\stepcounter{figure}\addtocounter{figure}{-1}  
\rotatebox[origin=c]{90}{}\quad
\subfloat{\includegraphics[trim={0.0cm 0.0cm 0.0cm 0.0cm},clip,width=0.75\textwidth,valign=c]{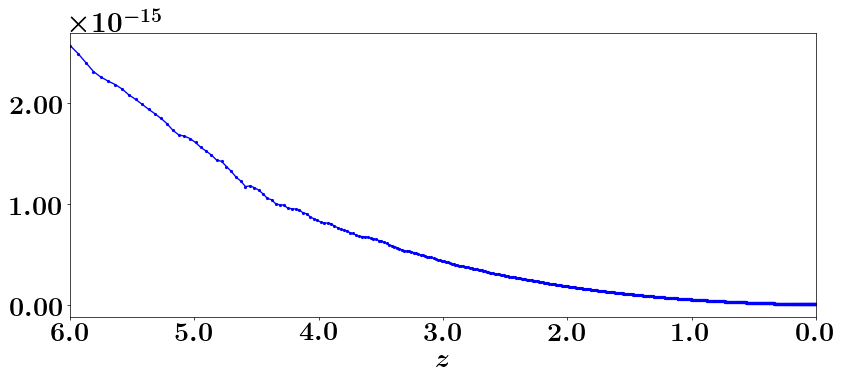}}\vspace{0.0em}\\
\rotatebox[origin=c]{90}{}\quad
\subfloat[]{\includegraphics[trim={0.0cm 0.0cm 0.0cm 0.0cm},clip,width=0.75\textwidth,valign=c]{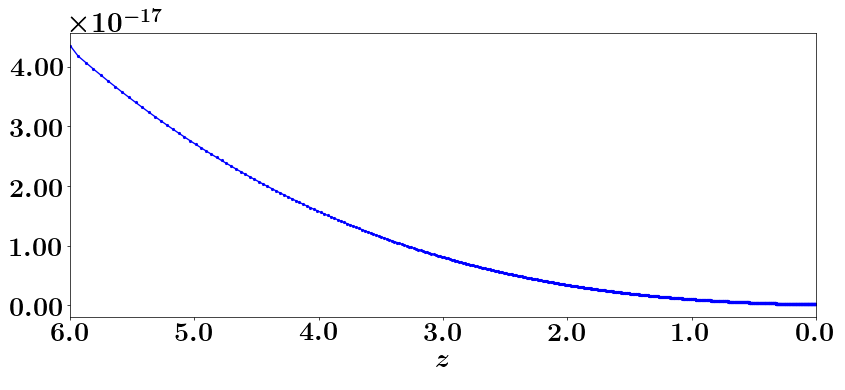}}
\end{minipage}%
\hspace{-9.2em}
\begin{minipage}{.42\linewidth}
\centering
\rotatebox[origin=c]{90}{}\quad
\subfloat{\includegraphics[trim={0.0cm 0.0cm 0.0cm 0.0cm},clip,width=0.75\textwidth,valign=c]{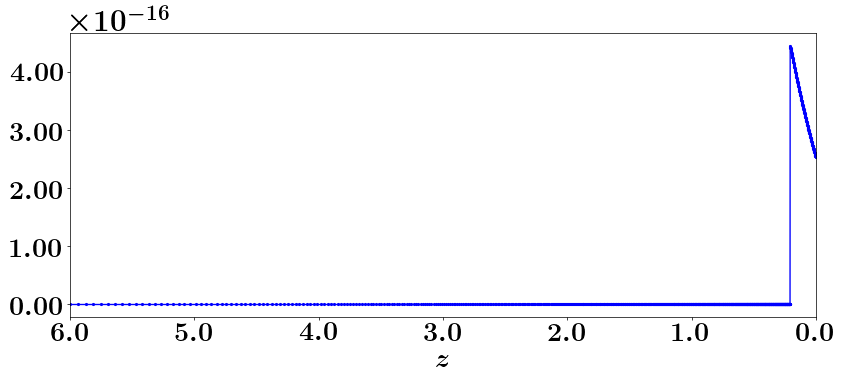}}\vspace{0.0em}\\
\rotatebox[origin=c]{90}{}\quad
\subfloat{\includegraphics[trim={0.0cm 0.0cm 0.0cm 0.0cm},clip,width=0.75\textwidth,valign=c]{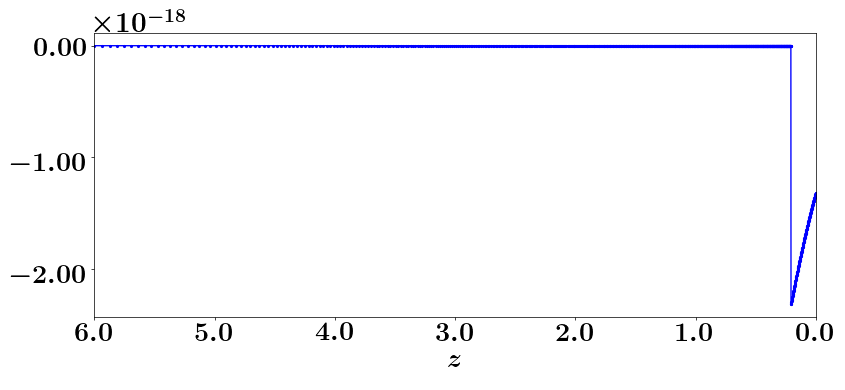}}\vspace{0.0em}\\
\stepcounter{figure}\addtocounter{figure}{-1}  
\rotatebox[origin=c]{90}{}\quad
\subfloat{\includegraphics[trim={0.0cm 0.0cm 0.0cm 0.0cm},clip,width=0.75\textwidth,valign=c]{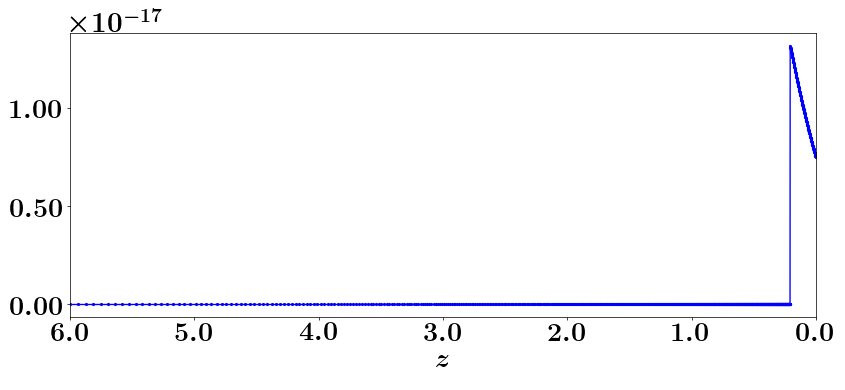}}\vspace{0.0em}\\
\rotatebox[origin=c]{90}{}\quad
\subfloat[]{\includegraphics[trim={0.0cm 0.0cm 0.0cm 0.0cm},clip,width=0.75\textwidth,valign=c]{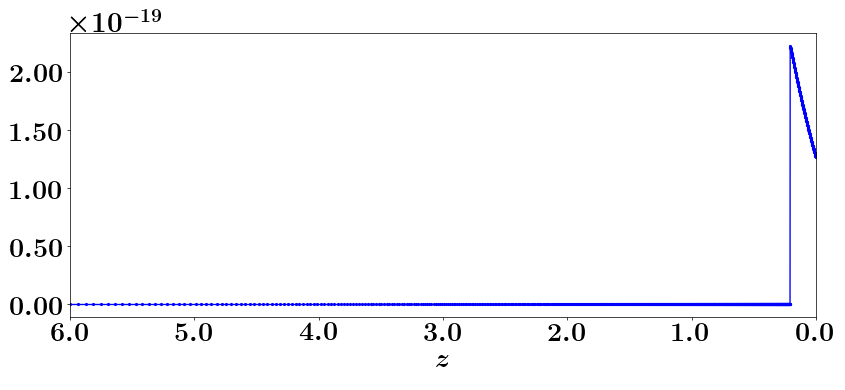}}
\end{minipage}%
 \caption{Cosmological evolution of the {\it comoving} Stokes parameters (in units of\,${\rm erg}\,{\rm s}^{-1}\,{\rm cm}^{-2}\,{\rm Hz}^{-1}\,{\rm str}^{-1}$) for $\nu_{\rm obs} = 1.42$~GHz for the cases where the radio bright point source is (i) absent, (ii) located at $z= 6.0$, and (iii) located at $z=0.206$; line-of-sight magnetic field orientations, simulated from a single realization, are random (see Section \ref{sec:cprt_ptsource}). Emission, absorption, Faraday rotation, and Faraday conversion for thermal bremsstrahlung and non-thermal synchrotron radiation process are taken into account. Note that fluctuations caused by random field directions in the results of case (iii) can be seen in zoom-in figures, where at $z$ prior to the point-source location the Stokes parameters evolve as in those of case (i). Here, we display the results over the full-redshift range for comparison purposes. 
}
 \label{fig:CPRT_SR_CaseQuasarCo_RandBori}
 \end{figure}
\end{landscape}

\begin{landscape}
\begin{figure}

\begin{minipage}{.42\linewidth}
\centering
\rotatebox[origin=c]{90}{$\mathbf{\Delta{\varphi}\,(z)}$}\quad
\subfloat{\includegraphics[trim={0.0cm 0.0cm 0.0cm 0.0cm},clip,width=0.75\textwidth,valign=c]{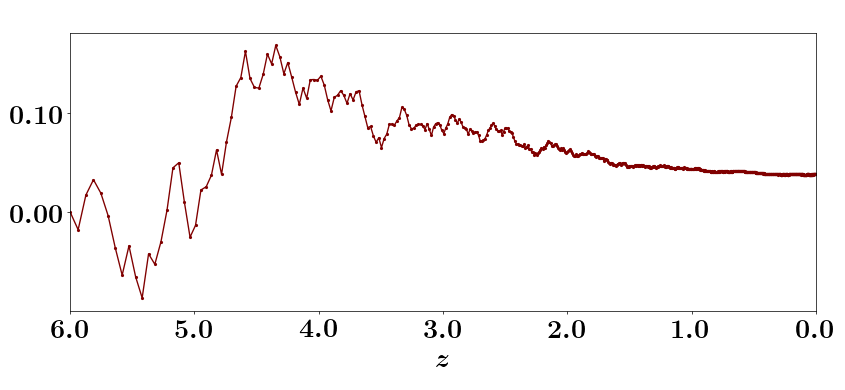}}\vspace{0.0em}\\
\rotatebox[origin=c]{90}{$\mathbf{{\Pi_{\rm l}}(z)}$}\quad
\subfloat{\includegraphics[trim={0.0cm 0.0cm 0.0cm 0.0cm},clip,width=0.75\textwidth,valign=c]{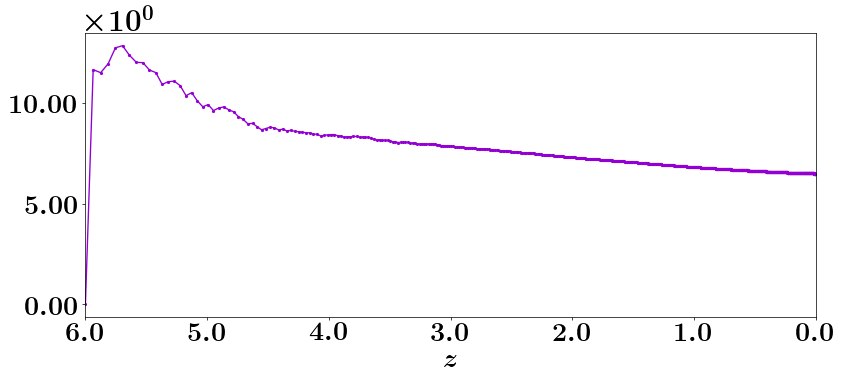}}\vspace{0.0em}\\
\rotatebox[origin=c]{90}{$\mathbf{{\Pi_{\rm c}}(z)}$}\quad
\subfloat{\includegraphics[trim={0.0cm 0.0cm 0.0cm 0.0cm},clip,width=0.75\textwidth,valign=c]{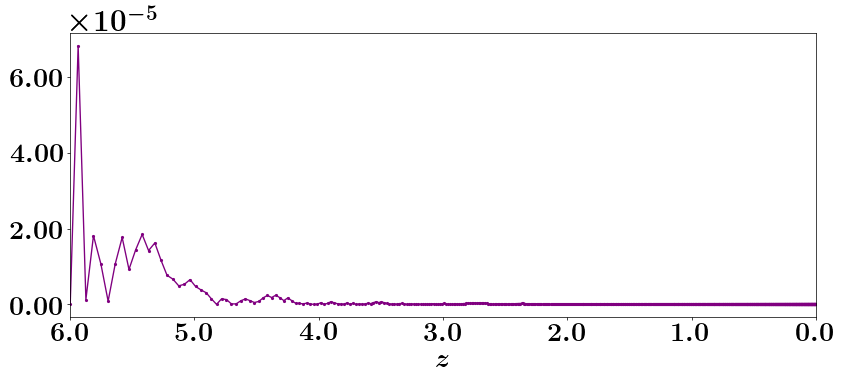}}\vspace{0.0em}\\
\stepcounter{figure}\addtocounter{figure}{-1}  
\rotatebox[origin=c]{90}{$\mathbf{{\Pi_{\rm tot}}(z)}$}\quad
\subfloat[]{\includegraphics[trim={0.0cm 0.0cm 0.0cm 0.0cm},clip,width=0.75\textwidth,valign=c]{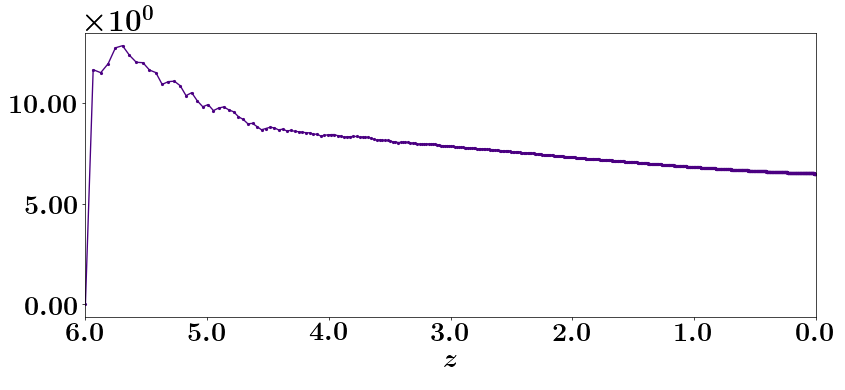}}
\end{minipage}%
\hspace{-8.2em}
\begin{minipage}{.42\linewidth}
\centering
\rotatebox[origin=c]{90}{}\quad
\subfloat{\includegraphics[trim={0.0cm 0.0cm 0.0cm 0.0cm},clip,width=0.75\textwidth,valign=c]{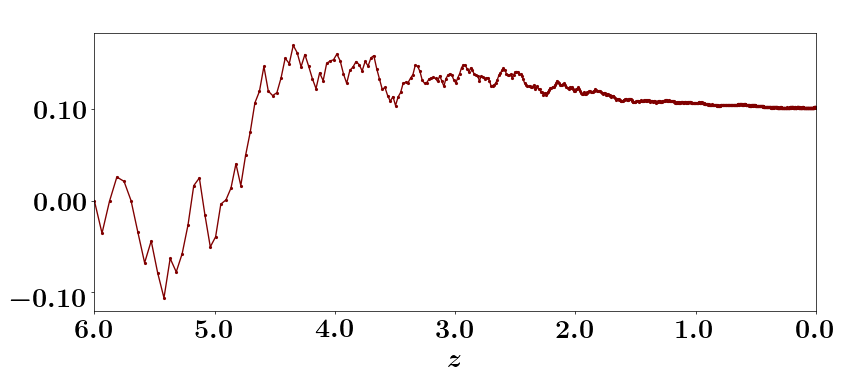}}\vspace{0.0em}\\
\rotatebox[origin=c]{90}{}\quad
\subfloat{\includegraphics[trim={0.0cm 0.0cm 0.0cm 0.0cm},clip,width=0.75\textwidth,valign=c]{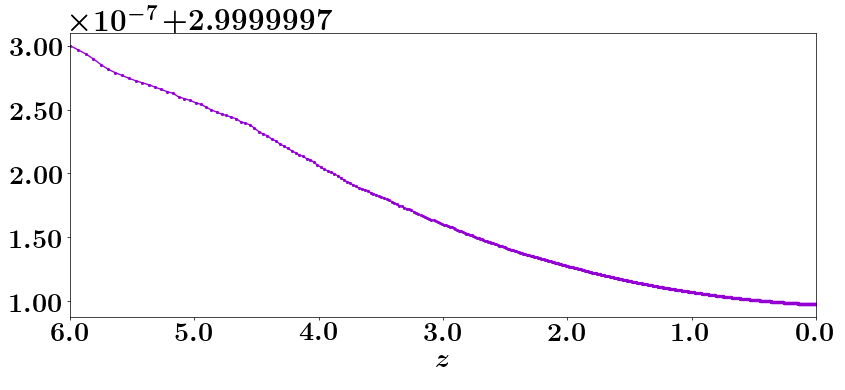}}\vspace{0.0em}\\
\stepcounter{figure}\addtocounter{figure}{-1}  
\rotatebox[origin=c]{90}{}\quad
\subfloat{\includegraphics[trim={0.0cm 0.0cm 0.0cm 0.0cm},clip,width=0.75\textwidth,valign=c]{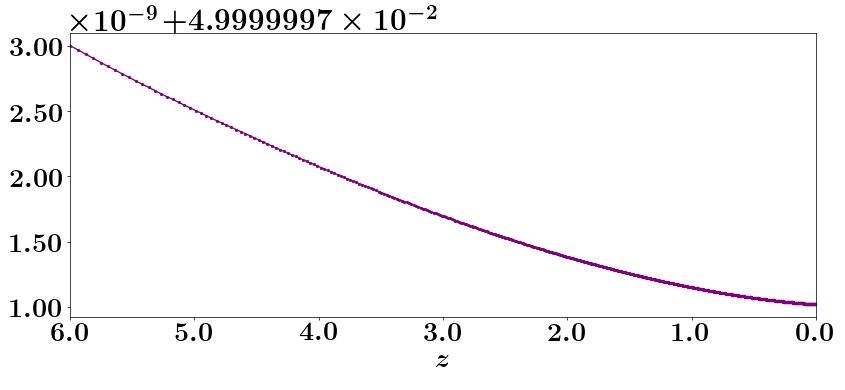}}\vspace{0.0em}\\
\rotatebox[origin=c]{90}{}\quad
\subfloat[]{\includegraphics[trim={0.0cm 0.0cm 0.0cm 0.0cm},clip,width=0.75\textwidth,valign=c]{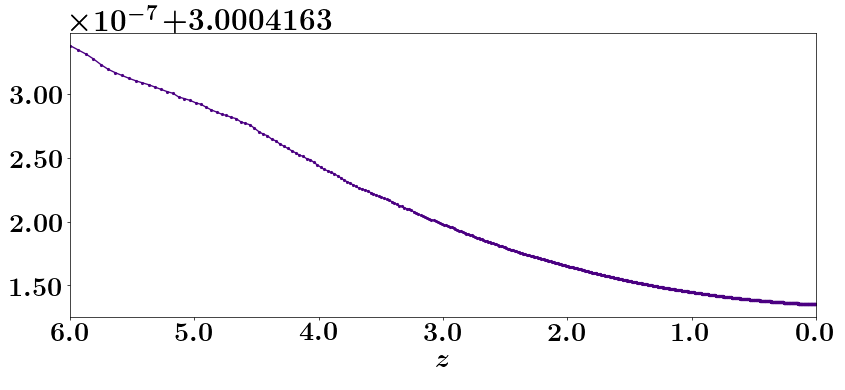}}
\end{minipage}%
\hspace{-9.2em}
\begin{minipage}{.42\linewidth}
\centering
\rotatebox[origin=c]{90}{}\quad
\subfloat{\includegraphics[trim={0.0cm 0.0cm 0.0cm 0.0cm},clip,width=0.75\textwidth,valign=c]{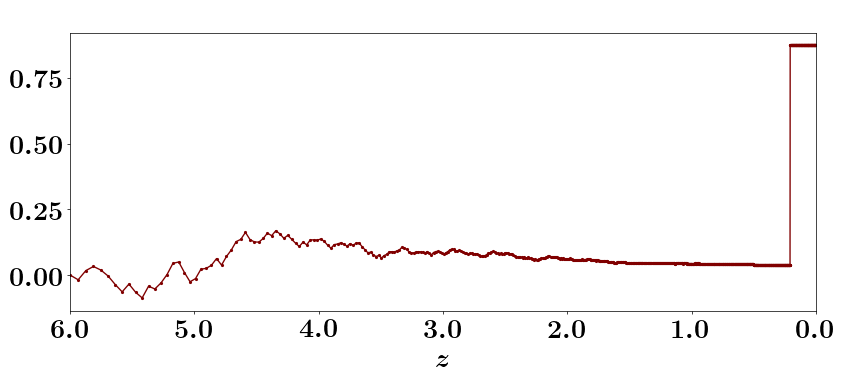}}\vspace{0.0em}\\
\stepcounter{figure}\addtocounter{figure}{-1}  
\rotatebox[origin=c]{90}{}\quad
\subfloat{\includegraphics[trim={0.0cm 0.0cm 0.0cm 0.0cm},clip,width=0.75\textwidth,valign=c]{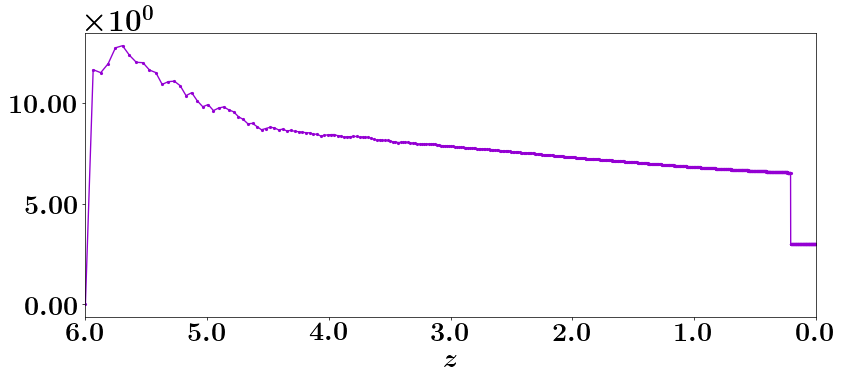}}\vspace{0.0em}\\
\rotatebox[origin=c]{90}{}\quad
\subfloat{\includegraphics[trim={0.0cm 0.0cm 0.0cm 0.0cm},clip,width=0.75\textwidth,valign=c]{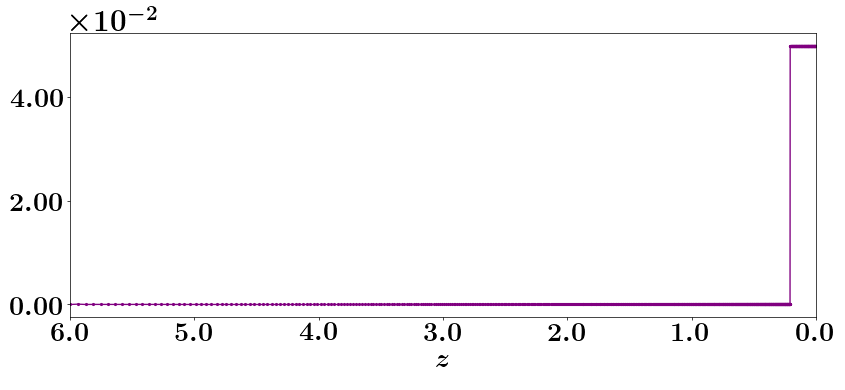}}\vspace{0.0em}\\
\rotatebox[origin=c]{90}{}\quad
\subfloat[]{\includegraphics[trim={0.0cm 0.0cm 0.0cm 0.0cm},clip,width=0.75\textwidth,valign=c]{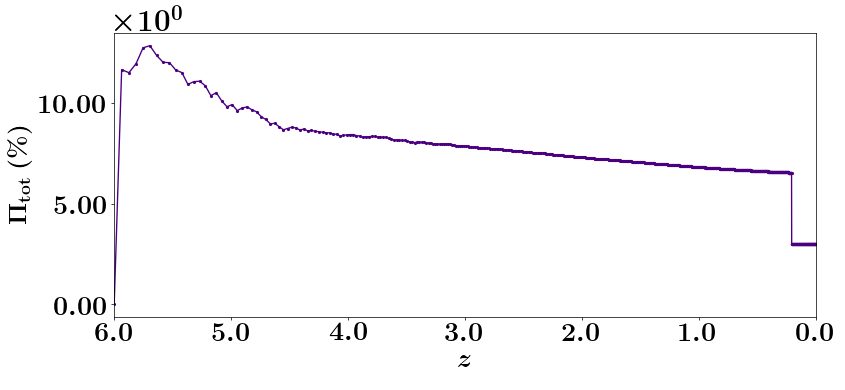}}
\end{minipage}%
 \caption{Cosmological evolution of $\Delta \varphi$ (in radian), $\Pi_{\rm l}$, $\Pi_{\rm c}$ and $\Pi_{\rm tot}$ (in per cent) for the cases where the radio bright point source is (i) absent, (ii) located at $z= 6.0$, and (iii) located at $z=0.206$; line-of-sight magnetic field orientations, simulated from a single realization, are random (see Section \ref{sec:cprt_ptsource}). Emission, absorption, Faraday rotation, and Faraday conversion for thermal bremsstrahlung and non-thermal synchrotron radiation process are taken into account. Note that the change of polarization angle is sensitive to the randomness of the magnetic field angle along the line-of-sight.}
 \label{fig:CPRT_SR_CaseQuasarPol_RandBori}
\end{figure}
\end{landscape}

\subsection{Single galaxy cluster}\label{sec:cprtcluster}

Here, we illustrate the making of intensity and polarization maps of an astrophysical object by carrying out pencil-beam (post-processing) CPRT calculations, where results obtained from a cosmological MHD simulation are incorporated. Each pixel of the maps corresponds to a solution obtained by the radiative transfer calculation. 
 
\begin{figure}    
\centering  
        \mbox{\includegraphics[width=.33\linewidth]{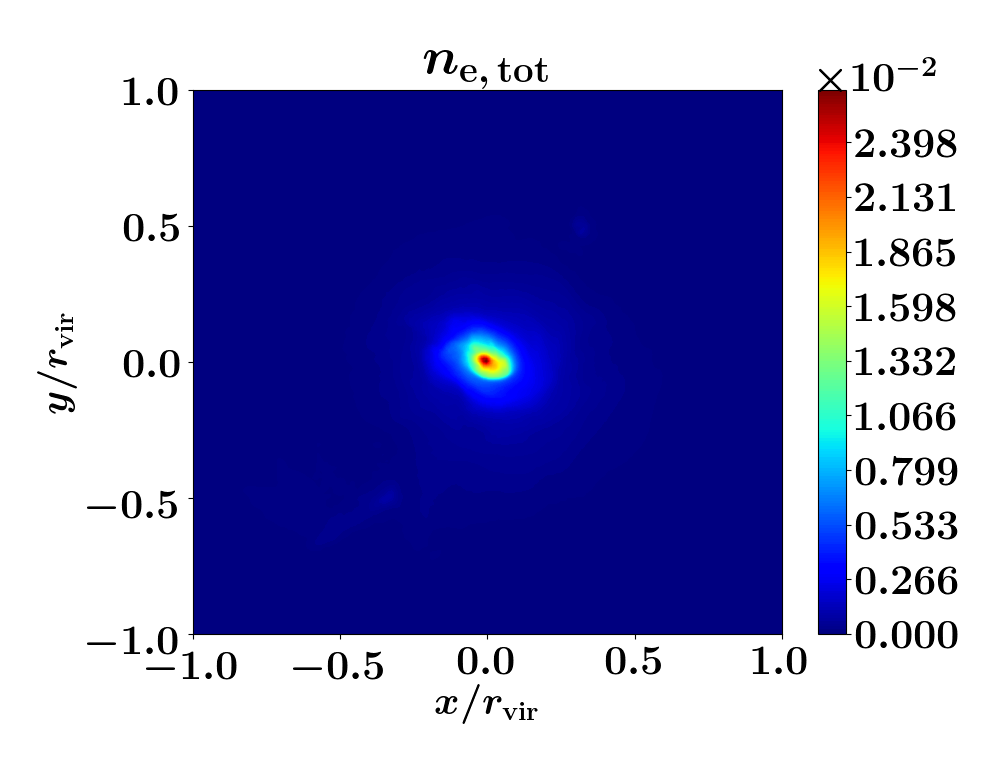}}
        \mbox{\includegraphics[width=.33\linewidth]{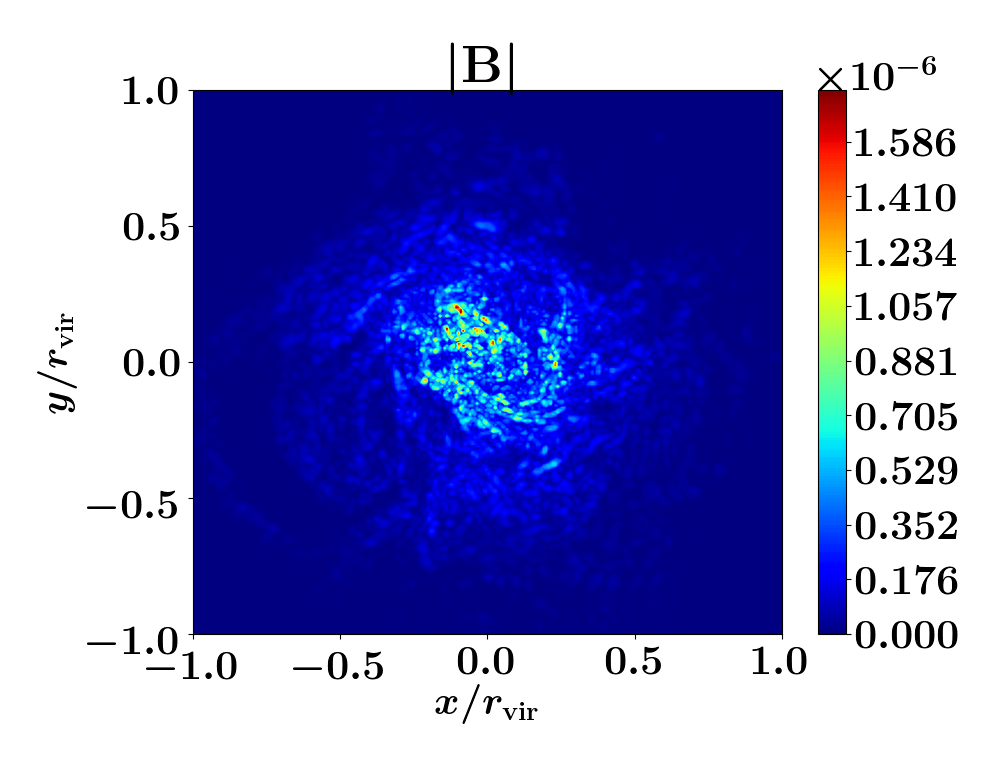}}
        \mbox{\includegraphics[width=.33\linewidth]{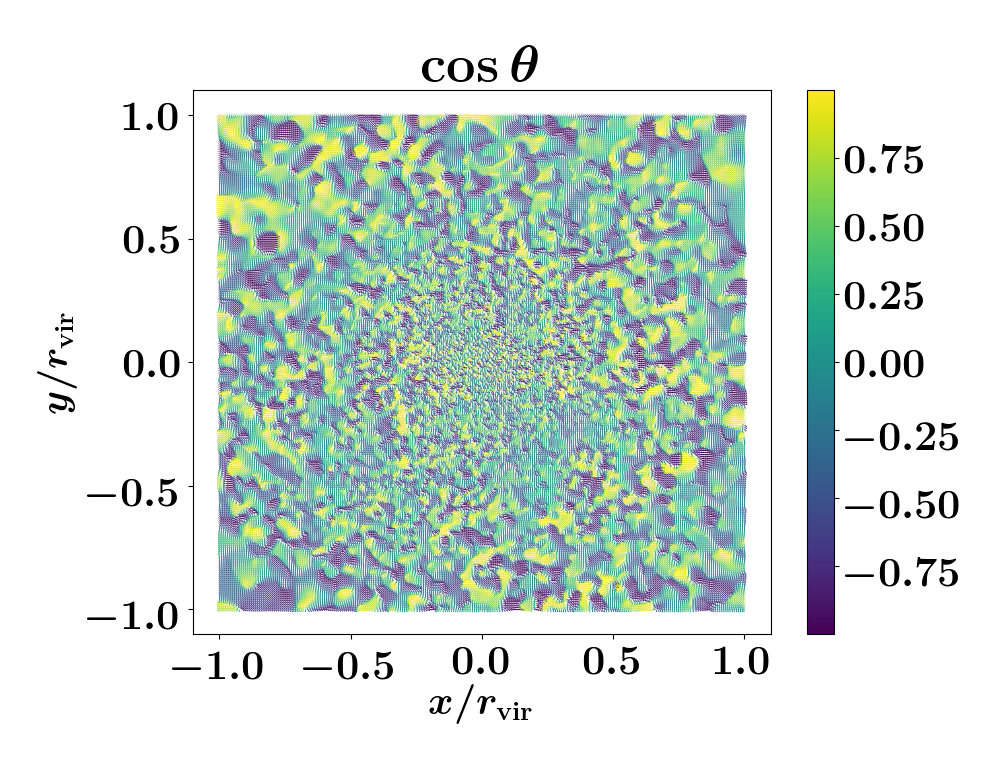}}        
    \caption
     {\small{The line-of-sight view of the central slices of a simulated galaxy cluster obtained from a GCMHD+ simulation, showing the structure of electron number density (left), magnetic field strength (middle) and magnetic field orientations along the line-of-sight as defined by $\cos{\theta}$ (right). The whole galaxy cluster data of dimension $256 \times 256 \times 256$ are used for the demonstrative pencil-beam calculation (see Section \ref{sec:cprtcluster}). Note that the GCMHD+ simulation is adiabatic, so to focus on the evolution of the magnetic field due to structure formation without the additional complications, e.g. the impacts of star formation.}} 
    \label{fig:clusterinputB}
\end{figure}

We use the data of a simulated galaxy cluster obtained from the ``cleaned" implementation of a higher resolution GCMHD+ simulation \cite[see Section 4 in][]{Barnes18}. The GCMHD+ simulations, designed to focus on the evolution of the magnetic field due to structure formation without the additional complications, are adiabatic, i.e. no radiative cooling, reionization, star formation and feedback from supernovae and Active Galactic Nuclei. The cluster obtained at $z=0$ from the simulation has a virial radius of $R_{\rm vir}=1.4439$~Mpc, 
and a gas mass of $m_{\rm gas} \sim 10^{13}$~${\rm M}_{\odot}$. 
Non-thermal electrons has energy density that amounts to 1\% of the thermal energy density \cite[see][]{Barnes18}. Simple statistics of the properties of the cluster are summarized in Table~\ref{tab:Stat_tableGC}. In Fig.~\ref{fig:clusterinputB} we plot the central slices of the data cube viewing along the $z$-direction, illustrating the 
input structures of electron number density, magnetic field strength and orientation for the CPRT calculation. 
   
Radiative transfer of a total number of $256^2= 65536$ rays is computed from $z=6.0$ to $z=0.0$ through the galaxy cluster centered at $z_{\rm cluster}$. Without loss of generality, we choose $z_{\rm cluster}=0.5$ (i.e. placed between $z=0.500645$ and $z=0.499355$, corresponding to a length scale of $2.89$~Mpc $\approx 2 R_{\rm vir}$). In order to study the intrinsic polarization emission of the cluster, no materials fill the line-of-sight outside the cluster and zero initial radiation background are assumed. Emission, absorption, Faraday rotation, and Faraday conversion by thermal bremsstrahlung and non-thermal synchrotron radiation process are taken into account. 

\begin{figure}    
\centering 
    \mbox{\includegraphics[trim={0.0cm 0.0cm 0.0cm 0.0cm},clip, width=.33\linewidth]{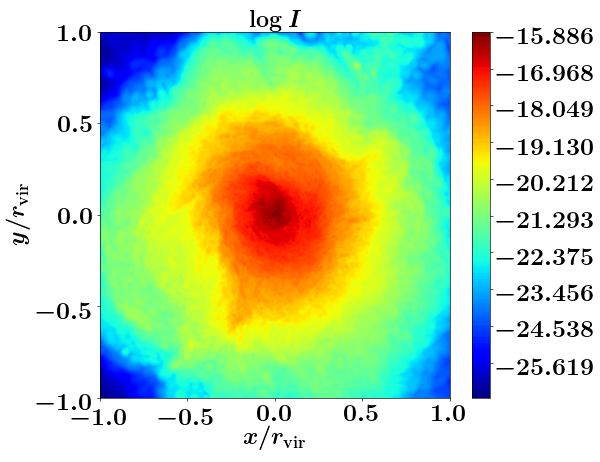}}
     \mbox{\includegraphics[trim={0.0cm 0.0cm 0.0cm 0.0cm},clip, width=.33\linewidth]{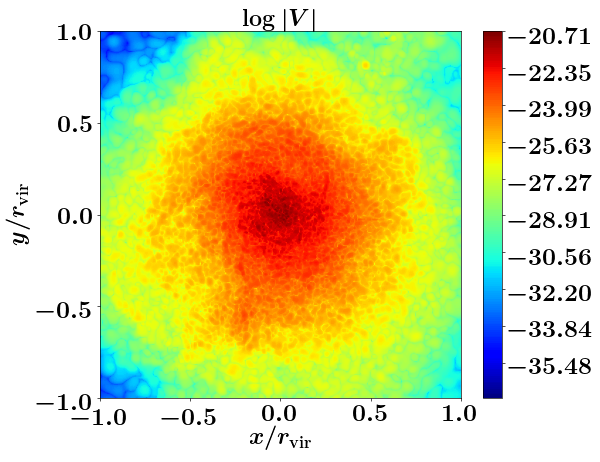}} \\
       \mbox{\includegraphics[trim={0.0cm 0.0cm 0.0cm 0.0cm},clip, width=.33\linewidth]{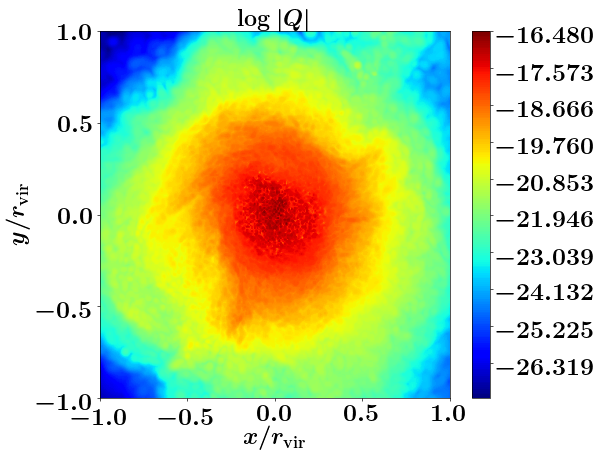}}
           \mbox{\includegraphics[trim={0.0cm 0.0cm 0.0cm 0.0cm},clip, width=.33\linewidth]{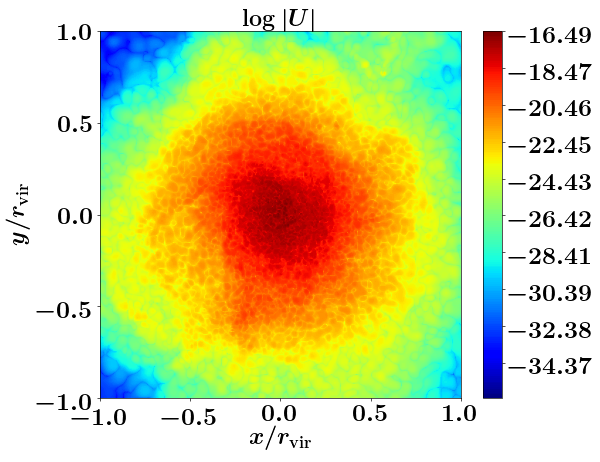}}
      \mbox{\includegraphics[trim={0.0cm 0.0cm 0.0cm 0.0cm},clip, width=.33\linewidth]{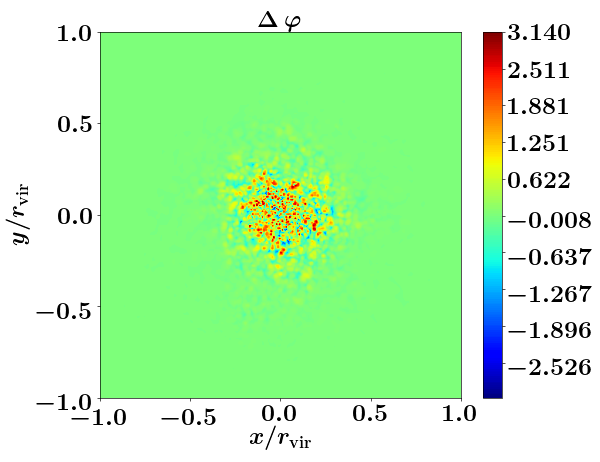}}
         \hspace{0px}
    \mbox{\includegraphics[trim={0.0cm 0.0cm 0.0cm 0.0cm},clip, width=.33\linewidth]{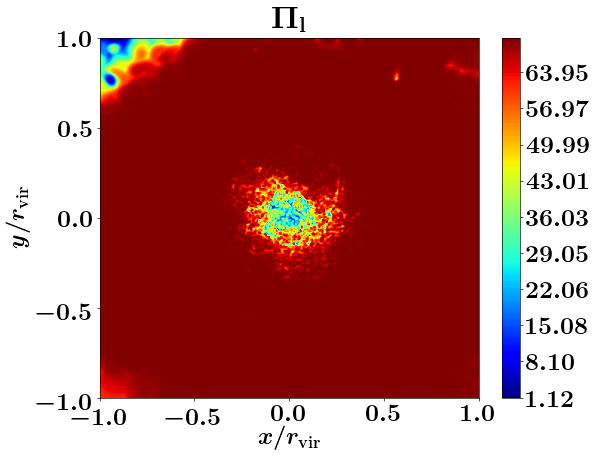}}   
    \mbox{\includegraphics[trim={0.0cm 0.0cm 0.0cm 0.0cm},clip, width=.33\linewidth]{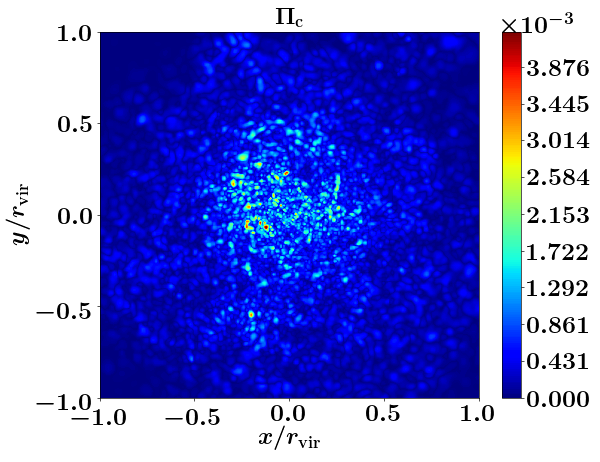}}
    \mbox{\includegraphics[trim={0.0cm 0.0cm 0.0cm 0.0cm},clip, width=.33\linewidth]{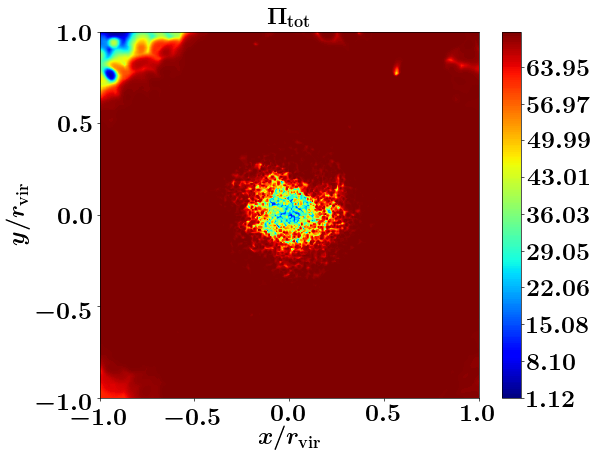}}
    \vspace{-0.0cm}
    \caption
     {\small{Resulting maps of $\log{I}$, $\log{|V|}$, 
     $\log{|Q|}$, and $\log{|U|}$ in units of\,${\rm erg}\,{\rm s}^{-1}\,{\rm cm}^{-2}\,{\rm Hz}^{-1}\,{\rm str}^{-1}$, $\Delta \varphi$ in radian, and the maps of $\Pi_{\rm l}$, $\Pi_{\rm c}$, $\Pi_{\rm tot}$ in per cent,  obtained from the demonstrative CPRT calculation at $z=0$ for a simulated galaxy cluster (see Section \ref{sec:cprtcluster}).}} 
    \label{fig:clustermaps}
\end{figure}

Fig.~\ref{fig:clustermaps} shows the resulting intensity and polarization maps obtained at $z=0$; simple statistics of those maps are summarized in Table~\ref{tab:Stat_tableGC}. The simulated cluster is intrinsically polarized at the $\nu_{\rm obs}=1.42$~GHz with the mean value of degree of total polarization $\sim 68.57$~\%, dominated by linear polarization. Emission is the highest in the cluster's central region, where both magnetic field and electron number density are the highest (see Fig.~\ref{fig:clusterinputB}). Faraday rotation is also strong in the central region, leading to a bigger change of polarization angle, as is seen in the map of $\Delta \varphi$ shown in Fig.~\ref{fig:clustermaps}. At the same time, depolarization in that region is also the most significant, where the degree of polarization is $\lesssim  30 \%$ and the minimum reaches $\sim 1 \%$.  Strong differential Faraday rotation and the effect of random field orientations along the line-of-sight are the causes of depolarization in this demonstration. These results agrees with the observational trends of smaller degree of polarization for sources close to the cluster center \citep[see e.g.][]{Bonafede11, Feretti2012}.

CPRT calculation provides a rich set of data products, enables quantitative measures of polarization and intensity, and its algorithm allows interfacing with simulation results. While here we demonstrate the calculation of a simulated cluster at a fixed redshift and show only the intensity and polarization maps at $z=0$, the CPRT algorithm can generate maps at any sampled redshifts. Comparisons of the statistics of maps generated at different redshifts may provide a useful means to study the cosmological evolution of magnetic fields, as well as giving insights to tomographic studies of large-scale magnetic fields in real data. Mock data set obtained from CPRT calculations can also be used to test analysis tools used for magnetic field structure inference. 

\begin{table*}
\begin{minipage}[t]{\hsize}
\centering
\begin{tabular}{|l|r|r|r|r|l|l|}
\hline
  \multicolumn{1}{|c|}{} &
  \multicolumn{1}{c|}{Mean} &
  \multicolumn{1}{c|}{Standard Deviation} &
  \multicolumn{1}{c|}{Minimum} &
  \multicolumn{1}{c|}{Maximum} \\
\hline
 Input &&&& \\ \hline
  ${n_{\rm e,\,tot}}$ & $6.3561\times 10^{-5}$ &  $2.7001\times 10^{-4}$ & $6.3577\times 10^{-7}$ & $2.6508\times 10^{-2}$\\
  $|\mathbfit{B}|$ & $1.5585\times 10^{-8} $&$5.0621\times 10^{-8}$ &  $2.5227\times 10^{-14}$ & $2.5175\times 10^{-6}$\\
  $\cos{\theta}$ & $-5.9235\times 10^{-3}$ & $5.7434\times 10^{-8}$ & -1.0000 & 1.0000\\
 \hline
 Output &&&& \\   \hline
  $I$ & $7.0217\times 10^{-18}$ & $4.4204\times 10^{-18}$ &  $2.2564\times 10^{-27}$ & $1.3000\times 10^{-16}$\\
  $Q$ & $1.6270\times 10^{-19}$ & $1.0358\times 10^{-18}$ &  $-2.96286\times 10^{-19}$ & $3.3099\times 10^{-17}$\\
  $U$ & $-2.3702\times 10^{-21}$ & $9.3950\times 10^{-19}$ & $-2.9899\times 10^{-17}$ & $3.2602\times 10^{-17}$\\
  $V$ & $-2.6325\times 10^{-25}$ & $4.2232\times 10^{-23}$ & $-1.5752\times 10^{-21}$ & $1.9561\times 10^{-21}$\\
    $\Delta {\varphi}$& $1.1957\times 10^{-3}$ & $0.2543$ & $-3.1240$ & $3.1401$\\
  $\Pi_{\rm l}$ & $68.5725$ & $8.0302$ & $1.1191$ & $70.5876$ \\
  $\Pi_{\rm c}$& $2.3278\times 10^{-4}$& $2.8580\times 10^{-4}$  & $6.4526\times 10^{-10}$ & $4.2846\times 10^{-3}$\\
  $\Pi_{\rm tot}$& $68.5725$ &  $8.0302$ & $1.1191$ & $70.5876$\\
\hline\end{tabular}
\caption{\small{Statistics of the input and output parameters at $z = 0$ of the demonstrative pencil-beam CPRT calculation using the simulated galaxy cluster obtained from a GCMHD+ cosmological MHD simulation; see Section \ref{sec:cprtcluster}. $n_{\rm e,\,tot}$ is in units of ${\rm cm}^{-3}$, while $|{\mathbfit{B}}|$ is in G. 
The Stokes parameters are in units of\,${\rm erg}\,{\rm s}^{-1}\,{\rm cm}^{-2}\,{\rm Hz}^{-1}\,{\rm str}^{-1}$, $\Delta \varphi$ is in radian, and $\Pi_{\rm l}$, $\Pi_{\rm c}$, $\Pi_{\rm tot}$ are in per cent. All values are corrected to four decimal places for compactness.}\label{tab:Stat_tableGC}} 
\end{minipage}
\end{table*}

\subsection{All-sky calculation}\label{sec:cprtallsky}

With the advent of the SKA surveys over a very large fraction of the celestial sky will be enabled. Here, we describe how applying the all-sky CPRT algorithm allows us to compute theoretical polarization maps of the radio sky, with a model magnetized universe obtained from a cosmological MHD simulation with GCMHD+ code \citep{Barnes12, Barnes18} as an input structure. Mock data of such a kind can be statistically characterized for comparison with observations, as well as serving as testbeds for validating analysis methods used for scientific inference. 

Ray-tracing CPRT calculations are carried out for a total number of $N_{\rm ray}= 12 \times 64^2 = 49152$ rays distributed on $z$-spheres according to the HEALPix sampling scheme \citep[][]{gorski:2005}. We consider radiation frequency of $\nu_{\rm obs} = 1.42$~GHz. 
Contribution from the redshifted CMB photons to the radiation background is neglected, and radio polarization is arisen from sources consisting both thermal and non-thermal electrons distributed across the entire universe in the post-reionization epoch\footnote{We expect that non-linear growth in magnitudes and structures of electron number density during the reionization epoch would have imparted observational signatures to the traveling radiation, varying the statistics such as the polarization power spectrum. However, for demonstrative purpose we do not consider such an effect in this paper.} (i.e.\,$\,z \leq 6.0$). Both thermal bremsstrahlung and non-thermal synchrotron radiation process are taken into account. To isolate the polarization signatures imparted by magnetic structures, electron number density $n_{\rm e, tot}(z, \theta, \phi)$ is assumed to be uniform across the entire sky at each $z$; its cosmological evolution over $z$ underwent a dilution in an expanding universe, i.e. $n_{\rm e, tot}(z)=n_{\rm e, tot, 0}(1+z)^{3}$, where $n_{\rm e, tot, 0} = 2.1918 \times 10^{-7}$~cm$^{-3}$ (see Appendix \ref{app:neIGM} for details). We assume that non-thermal relativistic electrons amounts to 1\% of the total electron number density. 
We also assume that their energy spectrum follows a power law with a spectral index of $p=4.0$ (i.e. the non-thermal electrons have aged, steepening the spectrum), corresponding to a radiation power-law spectrum with index $\alpha = (p-1)/2= 1.5$. 
The low cutoff of the electron energy is set to $\gamma_{i}=10.0$, and the high cut-off is set to infinity. 

\begin{figure}
\centering       
\mbox{ \includegraphics[trim={0.0cm 0.0cm 0.0cm 0.0cm},clip,width=0.457\textwidth]{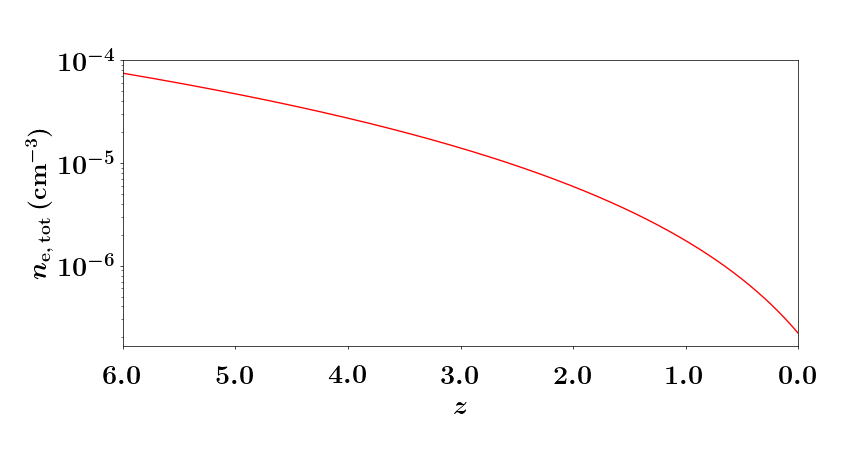}} 
\mbox{\includegraphics[trim={0.0cm 0.0cm 0.0cm 0.0cm},clip,width=0.457\textwidth]{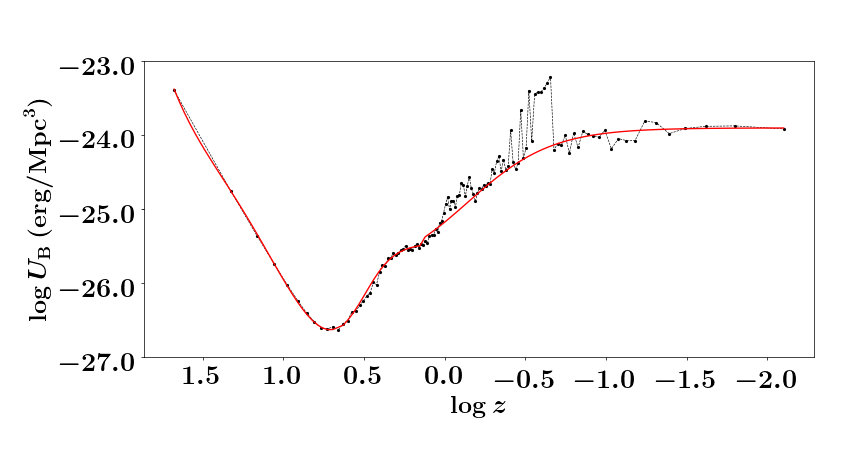}}  
\caption{\small{Plots of the cosmological evolution of the $n_{\rm e,\,tot}$ (left) and  that of the logarithmic of magnetic energy density $U_{\rm B} = |\mathbfit{B}|^{2}/8\pi$ (right) outputted from a GCMHD+ cosmological simulation. The solid red line in the right diagram shows the piecewise function that fits to the data, ignoring the anomalous bump caused by instantaneous infall and outflow of the simulation box. Note that smoothing via the 21-point averaging method is applied to obtain $B_{\parallel}(z)$ for the CPRT calculation. Note also that we consider only the post-reionization epoch, i.e. $6.0 \geq z \geq 0.0$ , in our calculation.
}\label{fig:inputneB}}
\end{figure}

We use the GCMHD$+$ cosmological MHD simulation \citep{Barnes12} to determine the evolution of the large-scale magnetic field as structures in the universe assemble. A cubic region of comoving volume ($40$~Mpc)$^3$ was taken from a comoving ($100$~Mpc)$^3$ volume in the simulation, which started at $z = 47.4$ as determined by the initial condition generator grafic$++$. The magnetic field was assumed to be generated at some early epoch via a method that filled the volume of the simulation. It has a configuration of $B = (10^{-11},0,0)$\,${\rm G}$. 
We fit analytically the output of $B_{\parallel}(z)$ obtained from the GCMHD$+$ simulation by the piecewise function: 
 \begin{eqnarray}
 \log_{10} \left( \frac{B^{2}_{\parallel}(z)}{8\pi} \right) =
     \label{eq:Anafit}  
      \left\{  
     \begin{array}{ll}
 8.1737\,x^{4}-40.352\,x^{3}  +73.647\,x^2-55.264\,x -12.16     & :  0.64 <  x  < 1.70   \\    
 0.67\tanh(-x/0.18+2.72)-26.14                                                   & :  0.15 \leq x \leq 0.64     \\
 -\tanh(x/0.52+0.28)-24.91                                                          & :    -2.00 \leq x < 0.15  \  ,
      \end{array}
   \right.
 \end{eqnarray} 
with $x = \log_{10}{z}$. 
This fit\footnote{Note that the anomalous bump in Fig.~\ref{fig:inputneB} at $\log{z} \approx -0.5$ is caused by the instantaneous infall and outflow of the simulation box. 
This structure does not appear in the other four simulations that ran with different initial conditions, and is therefore neglected.}, 
plotted in Fig.~\ref{fig:inputneB}, is smoothed by interpolation using twenty-one-points averages to model the input of $B_{\parallel}(z)$ for the CPRT calculation. 
We assumed a log-normal spatial distribution of $B_{\parallel}(z, \theta, \phi)$ over each $z$-sphere, where the mean value 
is deduced from equation~(\ref{eq:Anafit}) multiplied by a factor of $10^3$ to 
match the expected observed field strength of $1.0$ nG typical to filaments \citep[see e.g.][]{Araya-Melo12}. 
The log-normal distribution ensures the magnetic field strength to be all positive. 
Directions of the magnetic fields, which are defined by the $\cos{\theta}$, are assumed to have random orientations along the line-of-sight.

\begin{figure}   
\centering
    \mbox{\includegraphics[trim={0.0cm 0.0cm 0.0cm 0.0cm},clip, width=.45\linewidth]{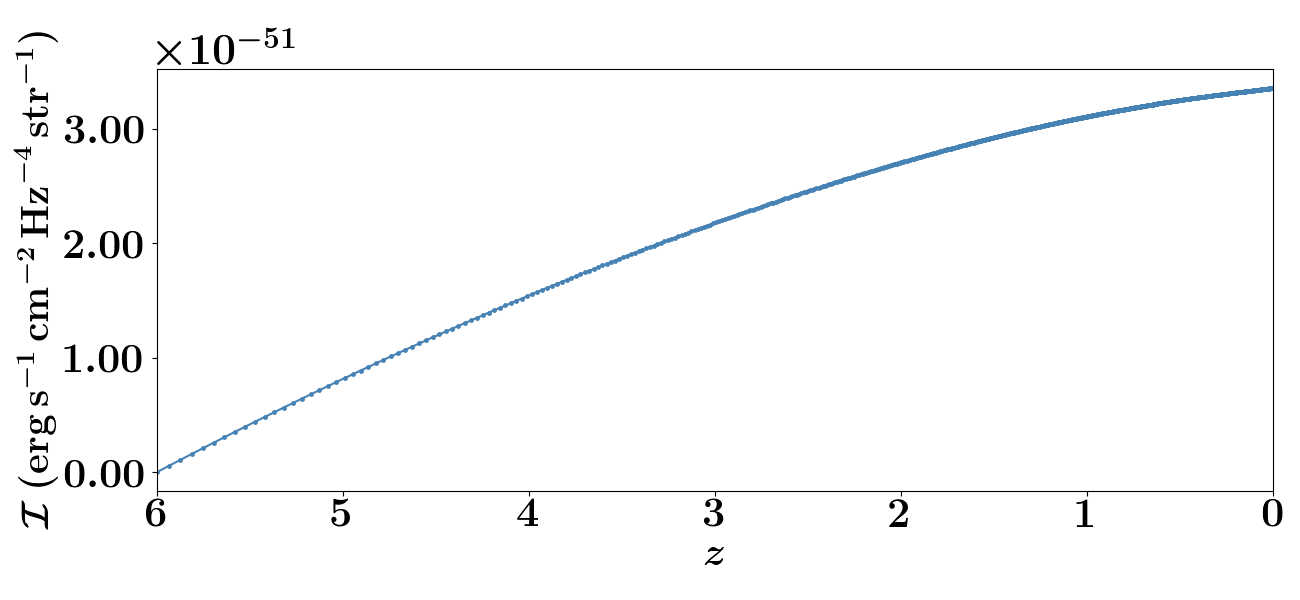}}   
        \mbox{\includegraphics[trim={0.0cm 0.0cm 0.0cm 0.0cm},clip, width=.45\linewidth]{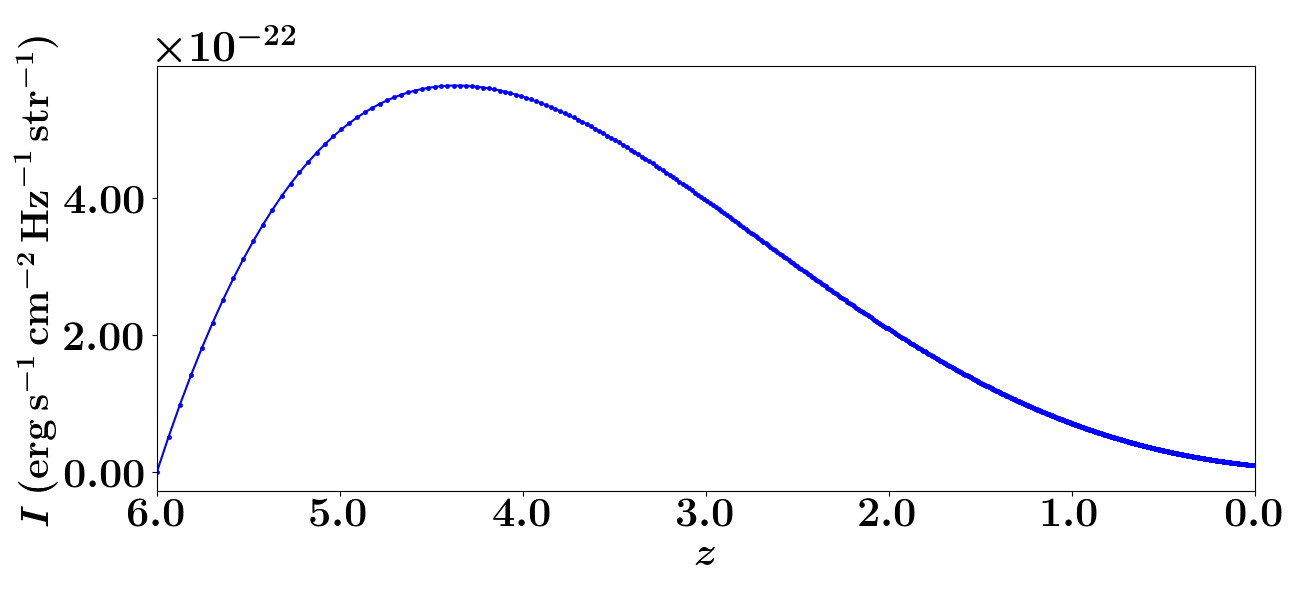}}   
    \mbox{\includegraphics[trim={0.0cm 0.0cm 0.0cm 0.0cm},clip, width=.45\linewidth]{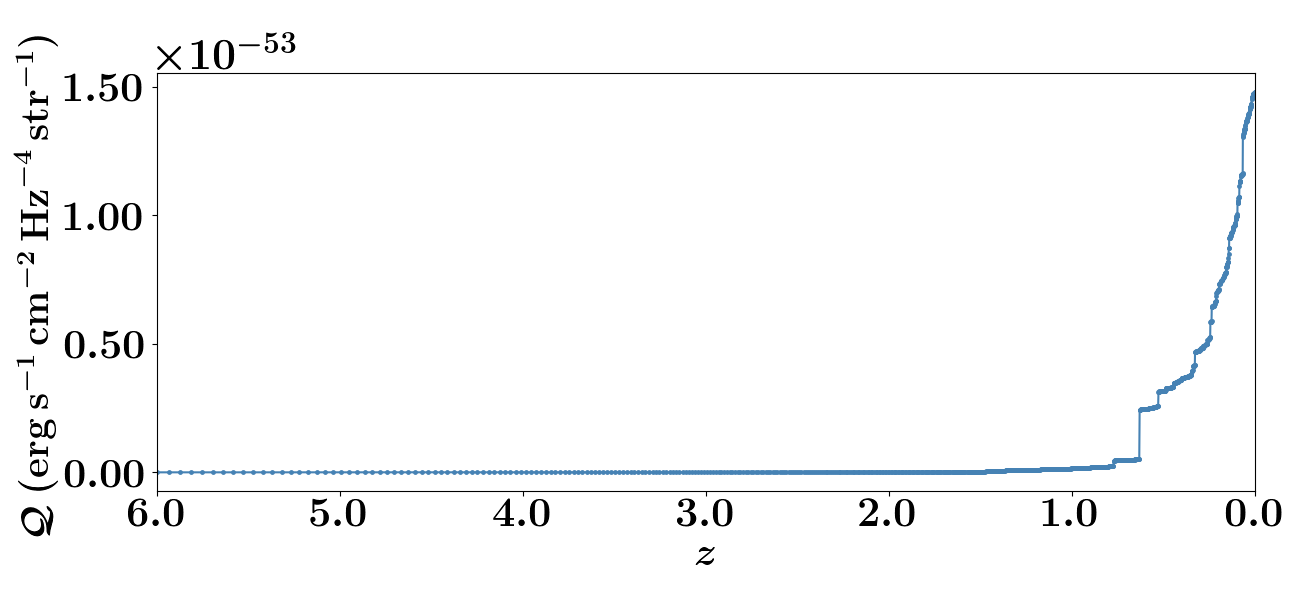}}
        \mbox{\includegraphics[trim={0.0cm 0.0cm 0.0cm 0.0cm},clip, width=.45\linewidth]{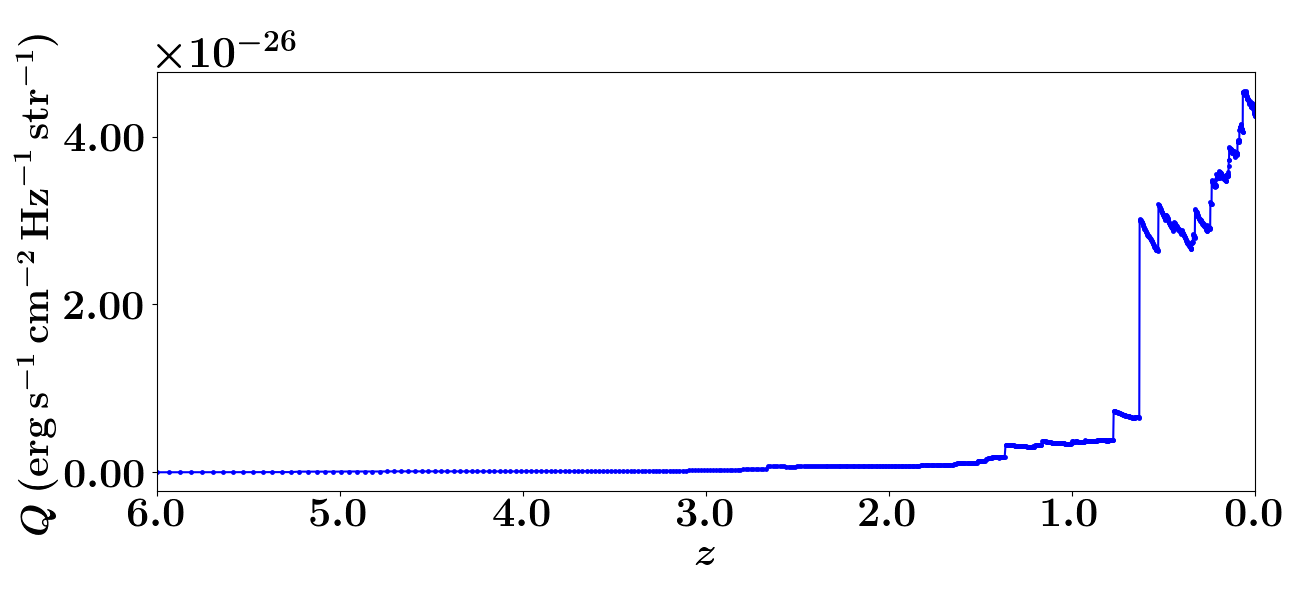}}
    \mbox{\includegraphics[trim={0.0cm 0.0cm 0.0cm 0.0cm},clip, width=.45\linewidth]{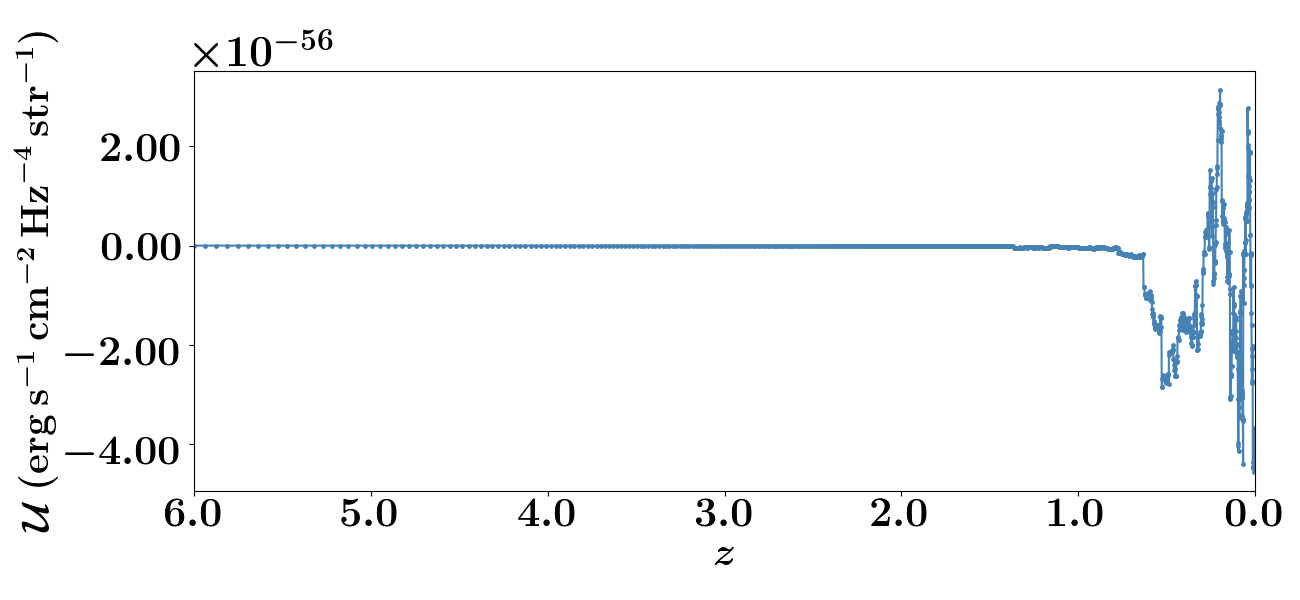}}
        \mbox{\includegraphics[trim={0.0cm 0.0cm 0.0cm 0.0cm},clip, width=.45\linewidth]{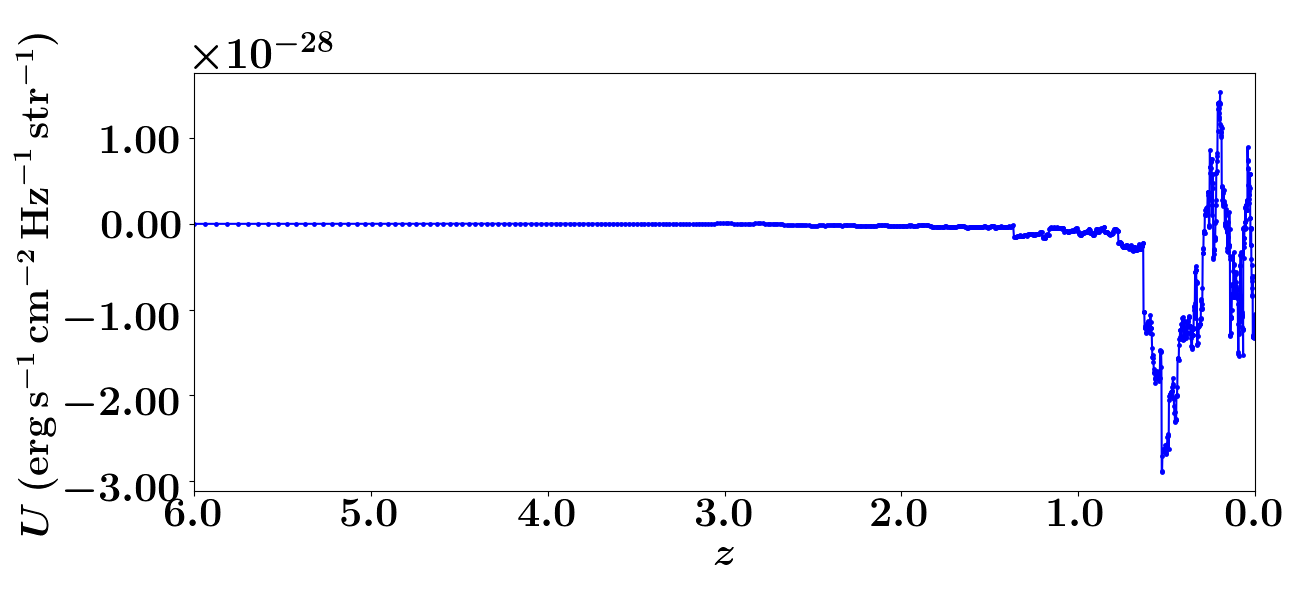}}
    \mbox{\includegraphics[trim={0.0cm 0.0cm 0.0cm 0.0cm},clip, width=.45\linewidth]{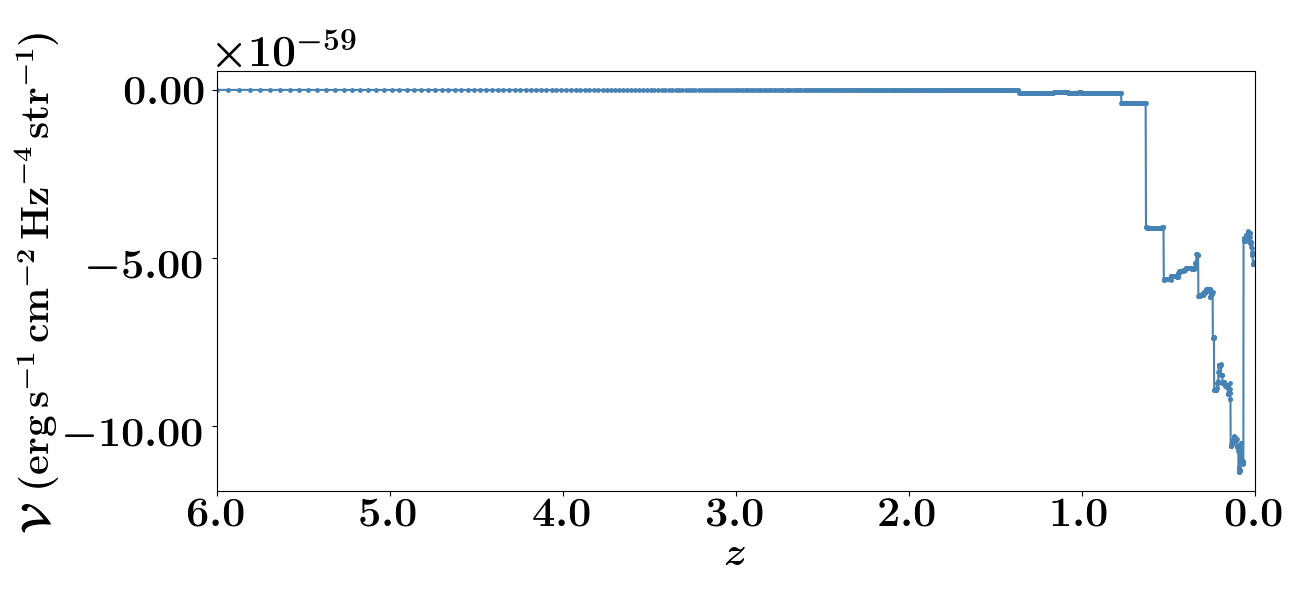}}
    \mbox{\includegraphics[trim={0.0cm 0.0cm 0.0cm 0.0cm},clip, width=.45\linewidth]{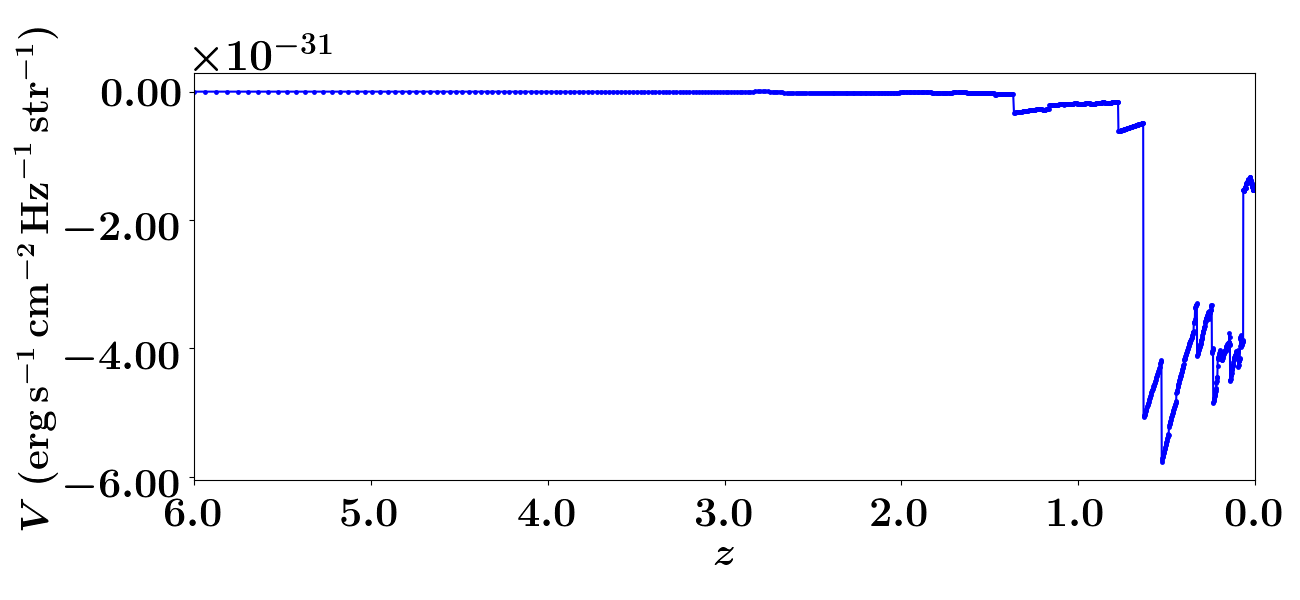}}
        \mbox{\includegraphics[trim={0.0cm 0.0cm 0.0cm 0.0cm},clip, width=.45\linewidth]{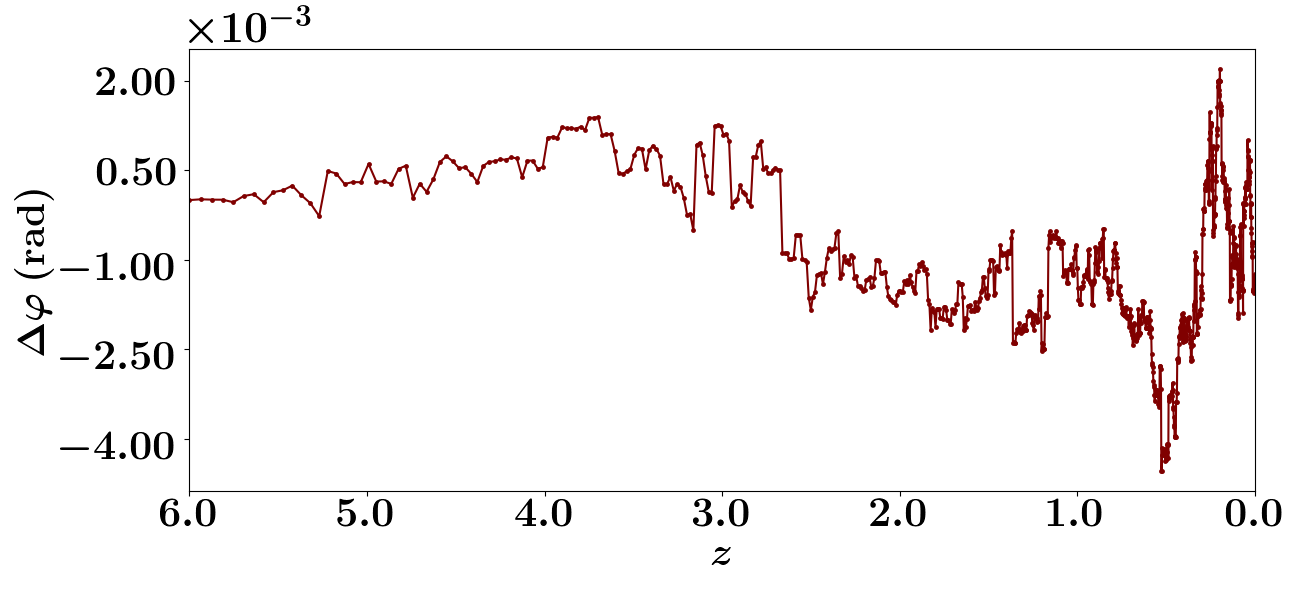}}
    \mbox{\includegraphics[trim={0.0cm 0.0cm 0.0cm 0.0cm},,clip, width=.45\linewidth]{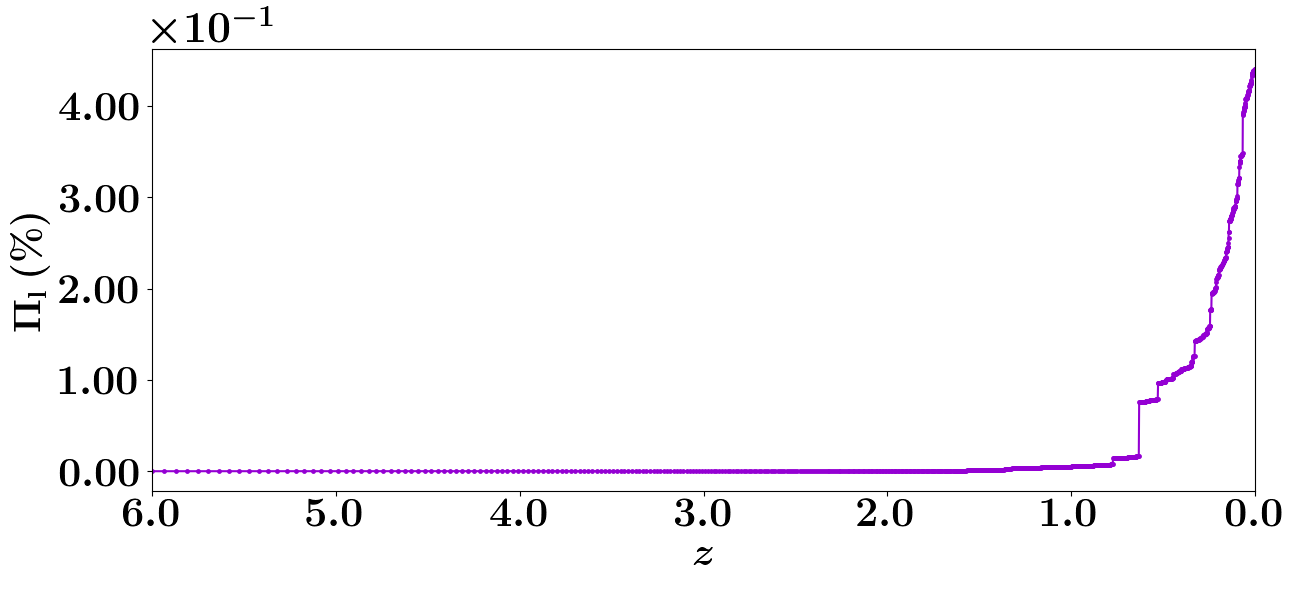}}
    \mbox{\includegraphics[trim={0.0cm 0.0cm 0.0cm 0.0cm},,clip, width=.45\linewidth]{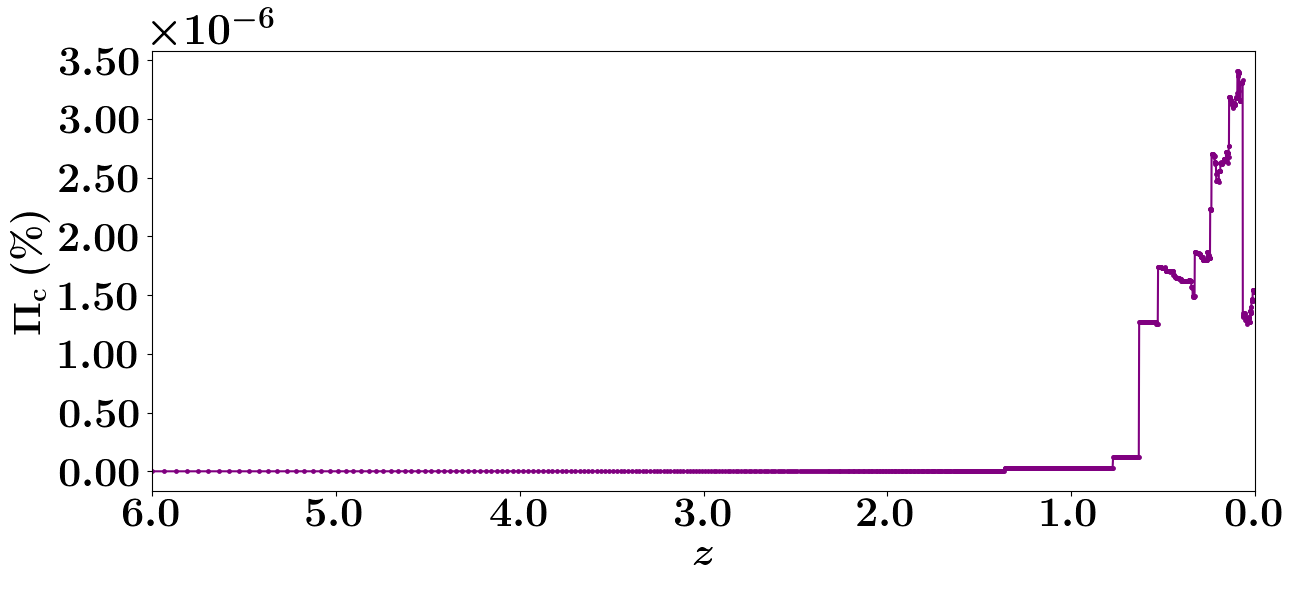}}
    \mbox{\includegraphics[trim={0.0cm 0.0cm 0.0cm 0.0cm},,clip, width=.45\linewidth]{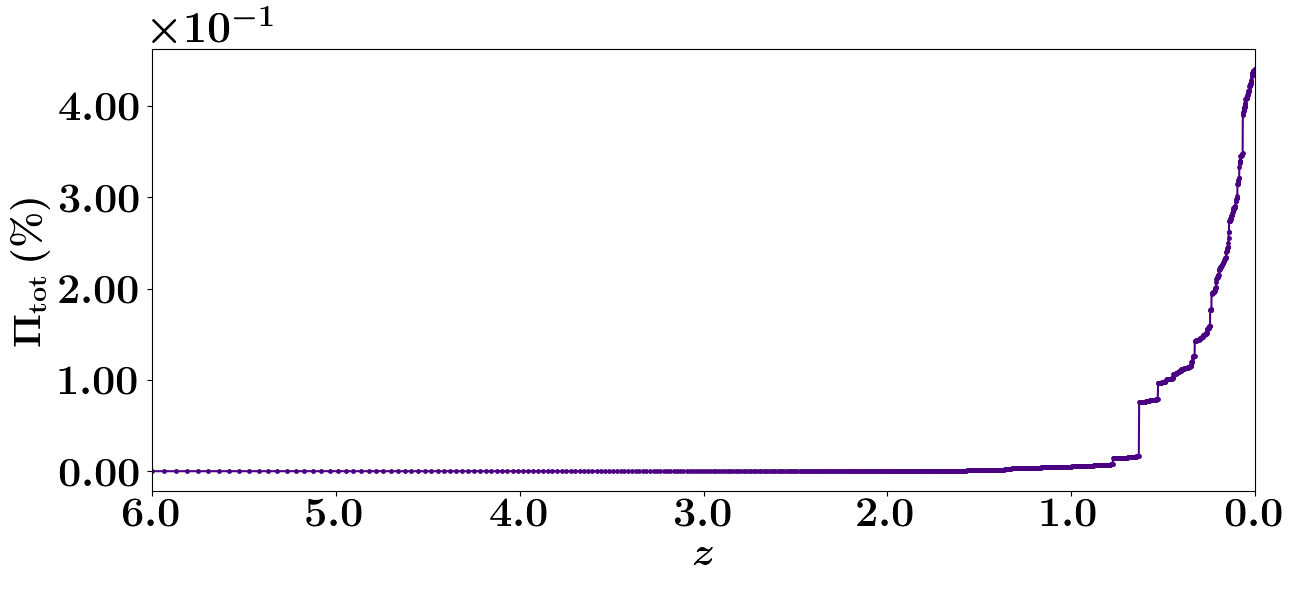}}
    \caption
    {\small{Cosmological evolution of the invariant Stokes parameters, the comoving Stokes parameters, $\Delta \varphi$, $\Pi_{\rm l}$, $\Pi_{\rm c}$ and $\Pi_{\rm tot}$ over the redshifts $6.0 \geq z \geq 0.0$ obtained from the CPRT calculation using a a model magnetized universe obtained from the GCMHD+ simulation as the input structure, see Section \ref{sec:cprtallsky}.}\label{fig:StokesEvo_LogNormB}}
\end{figure}

\subsubsection{Results and discussion} \label{subsubsec:results}
{\bf (I) Along a randomly selected ray}\\
Fig.~\ref{fig:StokesEvo_LogNormB} shows the resulting cosmological evolution of both the invariant and co-moving Stokes parameters, as well as the cosmological evolution of $\Delta \varphi$, $\Pi_{\rm l}$, $\Pi_{\rm c}$ and $\Pi_{\rm tot}$ of a randomly selected ray. 
Notably, one can see that the fluctuations in $Q, U$ and $V$ increase significantly during the late time, 
i.e. when the structure formation and evolution processes (such as the assembly of galaxy clusters) in the cosmological simulation become prominent and that magnetic fields become significantly amplified along with these processes, hence imposing a Faraday screen (i.e. strong Faraday-rotating component). 
In addition, highly volatile behavior is observed in the change of polarization angle over $z$, i.e. throughout the entire radiation path. Volatility in the evolution of polarization angle increases the difficulty to distinguish between different Faraday depth components, limiting the usage of the standard approach to infer magnetic field properties using RM synthesis \citep[see e.g.][]{Brentjens05} in some cases. Results showing similar trends in polarization evolution are observed commonly in all the other randomly selected rays.\\ 

\begin{figure}
  \centering
    \subfloat{\includegraphics[trim={0.2cm 0.0cm 0.0cm 0.0cm},clip, width=.43\linewidth]{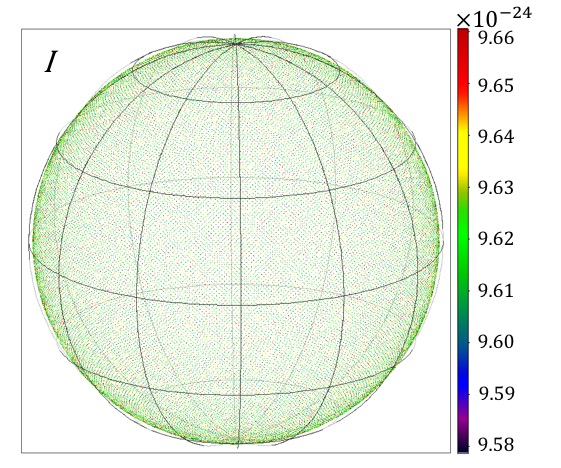}}   
    \subfloat{\includegraphics[trim={0.2cm 0.0cm 0.0cm 0.0cm},clip, width=.43\linewidth]{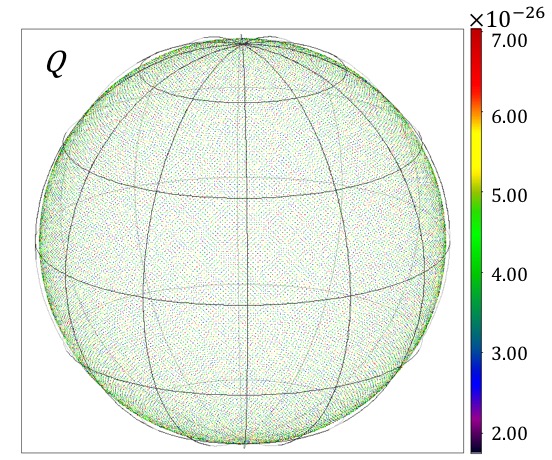}}   
    \\
    \subfloat{\includegraphics[trim={0.2cm 0.0cm 0.0cm 0.0cm},clip, width=.43\linewidth]{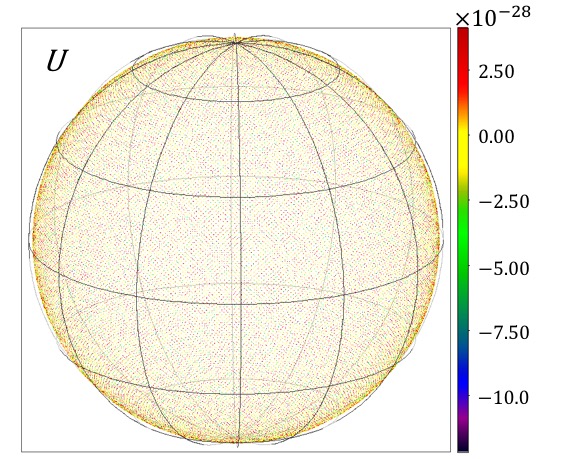}}  
    \subfloat{\includegraphics[trim={0.2cm 0.0cm 0.0cm 0.0cm},clip, width=.43\linewidth]{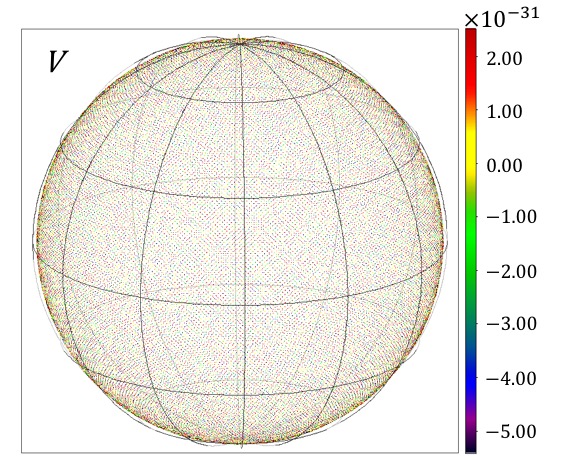}}   
\caption[] {\label{fig:Mapz0} All-sky maps of the Stokes parameters $I, Q, U, V$ at $z=0$ obtained from the demonstrative CPRT calculation in which cosmological GCMHD+ simulation results of the cosmological evolution of magnetic field strength is incorporated, log-normal distribution of the field strength over the redshift spheres are assumed, and the electron number density is diluted by $1/(1+z)^3$ due to the expansion of the universe; see Section \ref{sec:cprtallsky}. The scale of the colorbar is adopted to make the fluctuations in the Stokes maps apparent. The full dynamical range of the data is given in Table~\ref{tab:Stat_table}.}
\end{figure}

{\bf (II) All-sky maps} \\
Theoretical all-sky polarization maps of $I$, $Q$, $U$, and $V$ can be generated at any chosen redshifts. 
In Fig.~\ref{fig:Mapz0} we show the Stokes maps obtained at $z=0$. 
Their statistics are summarized in Table~\ref{tab:Stat_table}. As pointed out from the previous discussion on the single-ray results, 
the evolution of the change of polarization angle, which serves as a probe to Faraday rotation effect, is highly volatile and complex, 
demanding advanced statistical analyses at different redshifts to be performed for science extraction from map data. \\

\begin{table}
\begin{minipage}[H]{\hsize}
\centering
\begin{tabular}{|l|r|r|r|r|l|l|}
\hline
  \multicolumn{1}{|c|}{} &
  \multicolumn{1}{c|}{Mean} &
  \multicolumn{1}{c|}{Standard Deviation} &
  \multicolumn{1}{c|}{Minimum} &
  \multicolumn{1}{c|}{Maximum} \\
\hline
 Input &&&& \\ \hline
  ${n_{\rm e,\,tot}}$ & $2.1918\times 10^{-7}$ &  0.0000  & $2.1918\times 10^{-7}$ & $2.1918\times 10^{-7}$\\
  $|\mathbfit{B}|$ & $5.2855\times 10^{-9}$ &$ 6.8763\times 10^{-9}$ &  $4.5734\times 10^{-11}$ & $1.9587\times 10^{-7}$\\
  $\cos{\theta}$ & $-1.8850\times 10^{-3} $ & $0.5769$ & $-1.0000$ & $0.9999$\\
 %
 \hline
 Output &&&& \\   \hline
  $I$ & $9.6274 \times 10^{-24}$ & $3.3006 \times 10^{-26}$ &  $9.5894\times 10^{-24}$ & $1.1695\times 10^{-23}$\\
  $Q$ & $4.7385\times 10^{-26}$ & $2.6055\times 10^{-26}$ &  $1.7357\times 10^{-26}$ & $1.6790 \times 10^{-24}$\\
  $U$ & $2.9300\times 10^{-31}$ & $3.5445\times 10^{-28}$ & $-2.7505\times 10^{-26}$ & $1.0346\times 10^{-26}$\\
  $V$ & $-1.0209\times 10^{-33}$ & $2.0177\times 10^{-30}$ & $-2.1578\times 10^{-28}$ & $8.4035\times 10^{-29}$\\
  $\Delta {\varphi}$& $1.1597\times 10^{-5}$ & $4.7925\times 10^{-1}$  & $-3.0689\times 10^{-1}$ &$3.4868\times 10^{-1}$\\
  $\Pi_{\rm l}$ & $4.9131 \times 10^{-3}$ & $2.5886\times 10^{-3}$ & $1.8100\times 10^{-3}$ & $1.4359\times 10^{-1}$\\
  $\Pi_{\rm c}$& $5.4946\times 10^{-8}$& $1.8629\times 10^{-7}$  & $1.5424\times 10^{-12}$ & $1.8452\times 10^{-5}$\\
  $\Pi_{\rm tot}$& $4.9131\times 10^{-3}$& $2.5886\times 10^{-3}$ & $1.8100\times 10^{-3}$ & $1.4359\times 10^{-1}$\\
\hline\end{tabular}
\captionof{table}{\small{Statistics of the input and output parameters at $z = 0$ of the demonstrative all-sky CPRT calculation using a model magnetized universe obtained from a cosmological GCMHD+ simulation; see Section \ref{sec:cprtallsky}. $n_{\rm e,\,tot}$ is in units of ${\rm cm}^{-3}$, while $|{\mathbfit{B}}|$ is in G. 
The Stokes parameters are in units of\,${\rm erg}\,{\rm s}^{-1}\,{\rm cm}^{-2}\,{\rm Hz}^{-1}\,{\rm str}^{-1}$, $\Delta \varphi$ is in radian, and $\Pi_{\rm l}$, $\Pi_{\rm c}$, $\Pi_{\rm tot}$ are in per cent. All values are corrected to four decimal places for compactness.}\label{tab:Stat_table}} 
\end{minipage}
\end{table}

{\bf (III) Remarks on the CPRT method and the existing RM techniques} \\

Our demonstrative calculation results have two major implications in the study of large-scale magnetic fields, 
firstly on the future power spectrum analysis, and secondly on the validity of the current methodologies to investigate large-scale magnetic fields. 

Our results show that a Faraday screen can be introduced when structure formation and evolution processes in the universe becomes prominent, as is seen in Fig.~\ref{fig:StokesEvo_LogNormB} where significant polarization fluctuations happened during the late time when galaxy clusters started to assemble in the simulation, boosting the mean magnetic field strength. 
This finding means that cosmological contributions from line-of-sight IGM-like media will likely be screened (or shielded) by fluctuations sourced from astrophysical structures like a galaxy cluster (i.e. ionized systems with  relatively high magnetic field strengths and electron number density). This further implies that the polarization power spectrum of an all-sky map will be dominated by high frequency (small scale) signals. 
At the same time, it is worth noting that the morphology of ionized bubbles during the Epoch of Reionization, which has not been investigated in this paper, may imprint observable signatures onto the polarization maps, contributing to the power in low frequency (large scale) in polarization power spectrum as those ionized regions overlapped. 

The highly volatile cosmological evolution of Stokes parameters suggests that analysis methods using RM is likely to be deemed inappropriate to study inter-galactic magnetic fields, particularly those that permeate emitting cosmic filaments. This is because Faraday rotation will no longer be the single important process that imprints the polarization signals, but also the emissions from filaments themselves, as well as the absorption processes along the line-of-sight. However, the quantity of RM is derived from a restrictive case of polarized radiative transfer, as is described in Section \ref{sec:Intro} and reviewed in details in our related paper \citep{On18}. Interpretation of polarization signals in those cases, therefore, requires full CPRT consideration, so to correctly and accurately determine how large-scale magnetic fields have evolved and where they came from.

\section{Discussion and Summary}

In this paper, a covariant formulation of cosmological polarized radiative transfer, 
which provides a solid theoretical foundation to use polarized light as a probe of large-scale magnetic fields, is presented. Such a formulation naturally accounts for the space--time metric of an arbitrary cosmological model with a flat geometry. It is derived based on a covariant general relativistic radiative transfer formulation, 
which is derived from the first principles of conservation of phase--space volume and photon number. 
Without loss of generality, the corresponding polarized radiative transfer equations derived using a flat FRW space--time metric are constructed. In addition, we developed the (all-sky) CPRT algorithm that allows incorporation of the results from cosmological MHD simulation to the CPRT calculations, henceforth, a straightforward generation of theoretical polarization maps. Those maps serve as model templates, crucial for interpreting all-sky polarized data which will be measured by the next generation radio telescopes such as the SKA.  

Sets of CPRT calculations are performed to validate the code implementation of the ray-tracing algorithm and to 
demonstrate its applications for practical astrophysical studies. We summarize below the richness of the polarization data product the CPRT algorithm offers, as well as the findings from our demonstrative sets of calculations. 

Solving the CPRT equation yields the evolution of the Stokes parameters of radiation as a function of $z$, 
allowing tracking of how the intensity and polarization of radiation are modified on its way by 
local radiation processes (thermal bremsstrahlung and non-thermal synchrotron radiation process in this paper) in a cosmologically evolving universe. 
From the set of single-ray calculations presented in Section \ref{sec:cprt_SR}, 
we showed the resulting evolution of the polarization for cases where a bright radio point source is present or absent, and magnetic field orientations are random along the line-of-sight. It is seen that line-of-sight bright radio point sources dominate the intensity and polarization, and their locations lead to different signatures in the polarization evolution of the radiation. 
CPRT calculations provide quantitative studies of the intensity and polarization of radiation in their transport. They allow direct tracking of the change of polarization angle, which will help to resolve $n\pi$-ambiguity problem, aiding our interpretation of observational data. 
Evolution of the degree of linear, circular and total polarization can also be computed, allowing further investigation of Faraday rotation and depolarization. 

Carrying out multiple-ray CPRT calculation yields data maps of intensity and polarization, 
where spatial fluctuations across the sky plane can be statistically studied and characterized 
for comparison with observational data for magnetic field structure inference. 
We performed such a calculation using simulated cluster data obtained from a GCMHD+ simulation as the input. 
Contributions from the galaxy cluster dominated over those from the inter-galactic space. This is as expected due to their much higher electron number density and magnetic field strength. Faraday screening effect may be dominated when performing common analysis methods that use RM as a quantitative measure to study intra-cluster magnetic fields. We highlight that carrying out a full CPRT calculation, which does not assume the relative strength of radiative transfer effect in emission, absorption, Faraday rotation, and Faraday conversion, 
allows a reliable assessment of the validity of the standard RM methods in different astrophysical scenarios in an expanding Universe. 

In full cosmological settings, we performed an all-sky CPRT calculation using a model magnetized universe obtained from a cosmological GCMHD+ simulation as the input. Our results show that the cosmological evolution of the polarization components of propagating radiation is highly volatile, suggesting that full CPRT consideration is needed for accurate large-scale inter-galactic magnetic field studies, particularly for the fields that permeate emitting cosmic filaments. 
Another implication is that polarization power spectra obtained from all-sky measurements are likely to be dominated by the high frequency (small scale) signals caused by strong Faraday-rotating components, 
such as galaxy clusters. 
Impacts on the polarization signals due to the morphology of the cosmic reionization, which are not addressed in this paper but are important research problems, will be considered in our future work. 

All in all, the CPRT formulation provides a reliable platform to compute polarized sky. 
Furthermore, with known input distributions of $n_{\rm e}(z)$ and $\mathbfit{B}(z)$, 
and full radiative transfer processes taken into account, 
results obtained from the forward computation of the CPRT algorithm 
will provide valuable data sets that may also serve as a testbed for assessing analysis tools used for large-scale magnetic field studies. 
Also, since the cosmological terms in the CPRT equation can be easily switched off in our algorithm and its code implementation, calculations in astrophysical contexts can be easily carried out; calculations for foreground contributions in cosmology studies can also be performed. 

With the current version of the implementation, our next step is to characterize the polarization fluctuations for different input magnetic field and electron number density distributions, developing statistical methods for reliable scientific inference from data. Alongside, implementation of more accurate transfer coefficient expressions for a broader class of synchrotron distributions, as well as the inclusion of cyclotron process, is to be carried out and tested. Solving a stiff set of CPRT equations is foreseen to be one of the biggest numerical challenges. Nonetheless, the CPRT formulation, and its algorithm provide a solid theoretical foundation and a reliable platform to study large-scale magnetic fields. The CPRT formulation derived and the (all-sky) algorithm that has been developed enable more straightforward comparisons between theories and observations, 
ultimately guiding us to answers about the origins and the co-evolution of magnetic fields with structures in the Universe. 
%

\section*{Acknowledgments}

We thank Ziri Younsi for discussions of covariant radiative transfer. 
Jennifer Y. H. Chan is supported by UCL Graduate Research Scholarship and UCL Overseas Research Scholarship. 
Alvina Y. L. O is supported by the Government of Brunei under the Ministry of Education Scholarship. 
Thomas D. Kitching acknowledges support from a Royal Society University Research Fellowship.


\bibliography{References/references} 

\bibliographystyle{mn2e_Michael}
\renewcommand{\bibname}{References} 

\appendix
\section[]{Adopted coordinate system} \label{app:adoptedCoordsys}
%
\tikzset{
  > = {stealth},
  inertial frame/.style = {x={(-20:2cm)}, y={(-160:2cm)}, z={(90:2cm)}},
  local frame/.style = {shift={(local origin)}, x={(40:.7cm)}, y={(150:.7cm)}, z={(105:.7cm)}}
}

\begin{figure}
\centering
\begin{tikzpicture}[scale=2, inertial frame]
  \draw[->] (0,0,0) -- (1,0,0) node[anchor=east,above] (y){$\tilde{y}$};
  \draw[->] (0,0,0) -- (0,0.5,0) node[above] (x){$\tilde{x}$};
  \draw[->] (0,0,0) -- (0,0,0.9) node[left](z){$\tilde{z}$};
 \node[text width=1cm] at (0.03,0,0.8)(B){$\mathbfit{B}$};
 \draw[thick, ->] (0,0,0) -- (0,0,0.8);
 
  \draw[dotted] (0,0,0) -- ++(0:0.6) coordinate (projection) -- ++(0,0,0.9) coordinate (local origin);
   \draw[thin] (0,0.18,0) -- (0,0.08,.05) -- (0,-0.06,0.05);

  \draw[thick, ->] (0,0,0) -- (local origin);
  \node[text width=1cm] at (0.6,0,0.7)(k){$\mathbfit{k}$};
  \draw[blue, y={(0,0,1)}, z={(projection)}, ->] (90:.29) arc (90:65:.4) node[below, pos=.3] {$\theta$};

  \begin{scope}[local frame]
    \draw[->] (0,0,0) -- (0.88,0.19,0) node[above](3){$z$};
    \draw[->] (0,0,0) -- (0,-0.9,-0.18) node[above](2){$y$};
    \draw[densely dashed, ->] (0,0,0) -- (0,-0.7,-0.7) node[above](b){$b$};
    \draw[->] (0,0,0) -- (0,1.25,-0.94) node[left](1){$x$};
   \draw[densely dashed, ->] (0,0,0) -- (0,1.1,-0.6) node[above](a){$a$};
    \draw[red, y={(a)}, z={(projection)}, ->] (133:.465) arc (155:105:.3) node[right, pos=-0.34] {\small $\chi$};
     \draw[red, y={(2)}, z={(projection)}, <-] (111:.66) arc (79:59:.5) node[right, pos=-0.20] {\small $\chi$};

  \end{scope} 
\end{tikzpicture}
\caption{Coordinate systems adopted and the geometry of the magnetic field considered in this work.} 
\label{fig:Coord}
\end{figure}
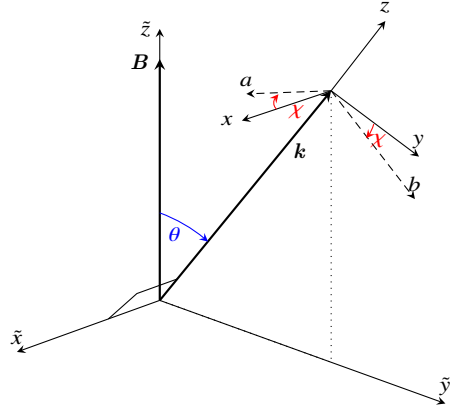

In this work, we adopt right-handed coordinate systems as depicted in Fig.~\ref{fig:Coord} following 
\citet{HuangShcherbakov11}. In our notations, magnetic field $\mathbfit{B}$ is directed along the $\tilde{z}$-axis, 
making an angle $\theta$ clockwise to the propagation direction of the radiation $\mathbfit{k}$.   
An orthonormal $(x, y, z)$ basis is defined such that 
$\mathbfit{z} \parallel \mathbfit{k}$, 
$\mathbfit{x} = C (\mathbfit{B} \times \mathbfit{k})$, where $C$ is a scalar that can be positive or negative, 
and $\mathbfit{x} \parallel \tilde{\mathbfit{x}}$, and  
$\mathbfit{y} = (\mathbfit{k} \times \mathbfit{x})$. 
Here, $\mathbfit{x}$ is perpendicular to the plane of $(\mathbfit{B}, \mathbfit{k})$, and $(\mathbfit{B}, \mathbfit{k}, \mathbfit{y})$ are coplanar. 
Electric field of an electromagnetic wave traveling along $\mathbfit{k} \parallel \mathbfit{z}$ oscillates in the $(x,y)$-plane. 
By such a choice of configuration (or by the choice of $\mathbfit{y} \parallel \tilde{\mathbfit{y}}$ in the systems defined in \citet{Sazonov69a, Pacholczyk77}), 
absorption coefficient $u_{\nu}$, conversion coefficient $g_{\nu}$ and emission coefficient $\epsilon_{U, \nu}$ are zeros. 

Note that the transfer coefficient matrices are commonly derived in the   
``magnetic-field" system, i.e. first in the $(\tilde{x}, \tilde{y}, \tilde{z})$ basis, and then projecting them onto $(x, y)$ for $\mathbfit{k} \parallel \mathbfit{z}$ and $\cos{\theta} = (\mathbfit{k} \cdot \mathbfit{B})/(|\mathbfit{k}| |\mathbfit{B}|)$
\citep[see e.g.][]{Sazonov69a, Pacholczyk70, Pacholczyk77, JonesOdell77_transfer, HuangShcherbakov11}. 
Transformation between the coordinate systems $\tilde{e}_{i}= (\tilde{x}, \tilde{y}, \tilde{z})$ and ${e}_{j} = (x, y, z)$ is given by 
${e}_{j} =  \tilde{e}_{i}{M}_{ij}$, where 
\begin{eqnarray} 
{M}_{ij} = 
\left( \begin{array}{ccc}
1 & 0 & 0  \\
0 & \cos{\theta} & \sin{\theta}  \\
0 & -\sin{\theta} & \cos{\theta}  \end{array} \right)  \   , 
\end{eqnarray} 
i.e. 
\begin{eqnarray} 
\left( \begin{array}{c}
{x}\\
{y}\\
{z} \end{array} \right)  
= 
\left( \begin{array}{ccc}
1 & 0 & 0  \\
0 & \cos{\theta} & -\sin{\theta}  \\
0 & \sin{\theta} & \cos{\theta}  \end{array} \right) 
\left( \begin{array}{c}
{\tilde{x}}\\
{\tilde{y}}\\
{\tilde{z}} \end{array} \right)   \  .   
\label{eq:coordsystrans}
\end{eqnarray} 
It follows that the rotation of vectors is given by $A_{i}=({M}^{\rm T})_{ij} \tilde{A}_{j}$, 
and the rotation of tensors is given by $\sigma_{ij} = ({M}^{\rm T})_{ik} \tilde{\sigma}_{km} ({M})_{mj} $ \citep{HuangShcherbakov11}. 
In future studies where observational data are confronted with theoretical predictions obtained by CPRT calculations, 
it is also useful to introduce the ``observer's" (or polarimeter's) system $(a,b)$, which is defined by rotating the $(x, y)$-plane about 
the $k$-direction. 
Such a transformation, i.e. between the local system (given by the local projection of the magnetic field) in the comoving frame 
and the frame in which polarimetric data are measured, invokes the use of rotational matrix ${\bf {\mathcal R}}(\chi)$, which follows 
the definition given in equations~(50) and (51) in \citet{HuangShcherbakov11}, 
where the angle 
$\chi$ relates $\mathbfit{a}$ and $\mathbfit{b}$ to the magnetic field components perpendicular to $\mathbfit{k}$, i.e. 
$\mathbfit{B}_{\perp} = \mathbfit{B} - \mathbfit{k}(\mathbfit{k} \cdot \mathbfit{B})/k^2$, by 
$\sin{\chi} = (\mathbfit{a} \cdot \mathbfit{B}_{\perp})/|\mathbfit{B}_{\perp}|$ and 
$\cos{\chi} = -(\mathbfit{b} \cdot \mathbfit{B}_{\perp})/|\mathbfit{B}_{\perp}|$ respectively. 

\section[]{Convention of polarization} \label{app:Convention} 

Stokes parameters $I_\nu$, $Q_\nu$, $U_\nu$ are defined unambiguously once the $(x, y)$ coordinate system is specified. 
The different definitions of polarization angle adopted in the cosmic microwave background community and 
the International Astronomical Union (IAU) can be reconciled by a sign flip of $U_\nu$. 
However, interpretation of the sign of $V_\nu$ (and consequently the signs for the corresponding transfer coefficients 
$\epsilon_{V, \nu}$, $v_\nu$ and $h_\nu$) in the literature is often ambiguous. 
This is because the sign of $V_\nu$ depends not only 
on the definition of the senses of circular polarization (which also 
depends on the handedness of the coordinate systems used) and the definition of $V_\nu$, 
but also on the choice of sign in the time-dependent description of the electromagnetic wave, 
as well as the definition of the relative phase between the $x$ and $y$-components of the electric vector of the wave. 
Much variation in these dependences exist in the literature, or sometimes this information is inexplicitly assumed or left unstated. 
Another source of variation comes from the choice of the attachment of the sense of circular polarization to the helicity of the photon. 
Any confusion and ambiguity can easily cause a slip in the interpretation of $V_\nu$. 

Here, we first describe the circular polarization sense 
defined by the Institute of Electrical and Electronics Engineers (IEEE) \citep{IEEE97}, which is 
commonly adopted by radio astronomers (but opposite to classical physicists and optical astronomers' common practice\footnote{The right-handed circular polarization convention by the IEEE corresponds to the left-handed circular polarization convention in the classical sense, i.e. IEEE-RCP $=$ classical-LCP.}), and the International Astronomical Union (IAU) convention of Stokes $V_{\nu}$ \citep{IAU}. 
Then we discuss the intricacies to test the conformity to the IEEE/IAU polarization convention. 
Finally we remark on the magnetic field direction of the system and state explicitly the Stokes V convention used in this paper. \\

{\bf IEEE/IAU polarization convention:} \\
\begin{figure}
\centering       
\begin{tikzpicture}[scale=0.6, inertial frame]
  \draw[->] (0,0,0) -- (-1.5,0,0) node[above] (y){$y$ (East)};
  \draw[->] (0,0,0) -- (0,1.6,0) node[left] (z){$z$ (Observer)};
  \draw[->] (0,0,0) -- (0,0,2) node[left](x){$x$ (North)};

\begin{turn}{100}
\foreach \t in {-226,-224,...,-58}{%
\draw[line width=1.5pt,color=red!60] ({1.0*cos(\t)},{1.0*sin(\t)},{-1.60+\t/360})--({1.0*cos(\t+7)},{1.0*sin(\t +7)},{-1.58+\t/360});
}

\foreach \t in {-50,-48,...,128}{%
\draw[line width=2.0pt,color=red] ({1.0*cos(\t)},{1.0*sin(\t)},{-1.60+\t/360})--({1.0*cos(\t+7)},{1.0*sin(\t +7)},{-1.58+\t/360});
}
\foreach \t in {130,135,...,300}{%
\draw[line width=1.5pt,color=red!60] ({1.0*cos(\t)},{1.0*sin(\t)},{-1.60+\t/360})--({1.0*cos(\t+7)},{1.0*sin(\t +7)},{-1.58+\t/360});
}
\foreach \t in {310,315, ..., 495}{%
\draw [line width=2.0pt,color= red]({1.0*cos(\t)},{1.0*sin(\t)},{-1.60+\t/360})--({1.0*cos(\t+7)},{1.0*sin(\t +7)},{-1.58+\t/360});
}
\foreach \t in {500,505,...,650}{%
\draw [line width=1.5pt,color= red!60]({1.0*cos(\t)},{1.0*sin(\t)},{-1.60+\t/360})--({1.0*cos(\t+7)},{1.0*sin(\t +7)},{-1.58+\t/360});
}
\end{turn}
\draw[very thick, ->] (0,0,0) -- (-0.8, 0.0, 1.22) node[above] (E){$\mathbfit{E}$};  
\draw[y={(-1,0,0)}, x={(0,0,1)}, ->] (0:.4) arc (0:34:.4) node[above, pos=.5] {$\varphi$};
\end{tikzpicture}
\vspace{-35pt}
\caption{A right-handed circularly polarized wave, as defined by the IEEE, 
in the adopted right-handed coordinate system (cf. Fig.~1 in \citet{Straten10_RCPfig} but angle and electric field notations are made consistent to the notations used in this paper). The electric vector rotates counter-clockwise as seen by the observer, i.e. at a fixed position as time advances (note that at fixed time the electric vector along the line-of-sight rotates clockwise i.e. forms a left-handed screw in space). 
\label{fig: RCPdef}}
\end{figure}
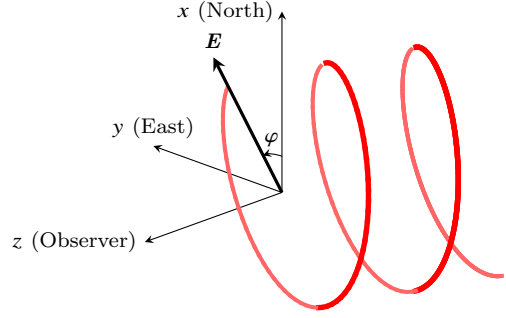

The exact quote of the \citet{IEEE97}'s definition\footnote{The same definition was first introduced in 1942 when the IEEE was still known as the Institute of Radio Engineers (IRE).} of a right-handed polarized wave reads 
``a circularly or an elliptically polarized electromagnetic wave for which the electric field vector, when viewed with the wave approaching the observer, rotates counter-clockwise in space". 
As pointed out by \citet{HamakerBregman96_Stokes}, 
such a definition stipulates that the position angle $\varphi$ of the electric vector of the wave at any point {\it increases} with time, implying that the $y$-component of the filed, $\mathbfit{E}_{y}$, to {\it lag} the $x$- component, $\mathbfit{E}_{x}$. 
In other words, the electric field traces out a counter-clockwise helix (right-hand screw) in time at fixed position, 
whereas in space at any instant in time it forms a clock-wise helix (left-hand screw) \citep[see e.g.][]{Rochford2001521}. 
The IAU endorses the sense of circular polarization defined by IEEE and defines $V_{\nu} = {\rm (RCP-LCP)}$, i.e. 
$V_{\nu}$ is positive for RCP \citep{IAU}. 
The $x$- and $y$- axes of a right-hand triad align with North and astronomical East, and the $z$- axis points towards the observer for standard IAU convention. \\
%

{\bf Conformity to IEEE/IAU convention:} \\

It is important to note that even when the Stokes parameters are uniquely defined 
when combined the IEEE/IAU definition with the standard formulae for 
$I_{\nu}= \langle |E_x|^{2} + |E_y|^{2}  \rangle$,  
$Q_{\nu}= \langle |E_x|^{2} - |E_y|^{2} \rangle$,  and 
$U_{\nu}= 2 \langle |E_x|\,|E_y|\,\cos{\delta} \rangle$,  
two similar but distinct mathematical representations are allowed for the same physics of the problem, 
as is shown by \citet{HamakerBregman96_Stokes}. 
One has the choice\footnote{
Another choice is related to the attachment of the RCP and LCP to positive and negative helicity \citep[see also Appendix III in][for a complete table of different conventions of RCP, including those that do not comply to IEEE/IAU convention]{simmons1970states}.
} 
of the sign of the time dependence of the electromagnetic wave, i.e. $e^{+i\omega t}$ or $e^{-i\omega t}$, for $\omega > 0$. 
Both choices are equally valid, but once the sign is chosen for 
\begin{eqnarray} 
\mathbfit{E}(z, t) = \mathbfit{E}_{0}\,e^{\pm i (\omega t - kz)}
=
\left( \begin{array}{c}
E_{x}(z, t) \\ 
E_{y}(z, t)
\end{array} \right) \  
=
\left( \begin{array}{c}
E_{x0} \,e^{\pm i (\omega t -kz + \phi_x)} \\ 
E_{y0} \,e^{\pm i (\omega t -kz + \phi_y)}    \end{array} \right) \  ,
\label{eq:ElectricFieldVector}
\end{eqnarray} 
the following quantities must have the sign adjustments such that 
$\mathbfit{E}_{y}$ lags $\mathbfit{E}_{x}$ for a (unit amplitude) RCP wave:  
\begin{eqnarray} 
\mathbfit{E}^{\rm RCP} = \frac{1}{\sqrt{2}} \left( \begin{array}{c}
1 \\ 
\mp i  \end{array} \right)  \    ,
\label{eq:ylagsx}
\end{eqnarray} 
\citet{HamakerBregman96_Stokes}, and that 
\begin{eqnarray} 
V_\nu =& 2 \langle |E_x|\,|E_y|\sin{\delta} \rangle     \label{eq:Vdef}    \\
    =& \mp i \langle E_x E^{*}_y - E_y E^{*}_x \rangle  \   
\label{eq:Vdef2}
\end{eqnarray} 
\citet{HamakerBregman96_Stokes}, so that $V$ is positive for RCP, i.e. IEEE/IAU compliant. 
Note that the sign adjustment in equation~(\ref{eq:Vdef2}) is equivalent to defining the sign of $\delta =\pm(\phi_y- \phi_x)$ in equation~(\ref{eq:Vdef}) for $\delta \in (0, \pi)$, where time delays correspond to negative (positive) values of the phases $\phi_x$ and $\phi_y$ for 
$e^{\pm i (\omega t - kz)}$ according to Equation (\ref{eq:ElectricFieldVector}).
It is apparent that one differing convention of the above would lead to a sign reversal. In brief, an unambiguous interpretation of the circular polarization from $V_{\nu}$ requires a clear specification of the adopted handedness of the coordinate systems, the convention of circular polarization, the definition of Stokes $V_{\nu}$, as well as the chosen mathematical representation of the traveling plane wave. \\

{\bf Remark on the B-field convention:} \\

Given the coordinate systems and the geometry of the problem presented in Fig.~\ref{fig:Coord}, 
let's consider the simple case where a uniform magnetic field $\mathbfit{B}$ aligns with $\mathbfit{k}$, so ${\theta} = 0$. 
An electron would then precess about $\mathbfit{B}$ in the $(\tilde{x}-\tilde{y})$-plane, moving counter-clockwise as viewed along $\mathbfit{k} \parallel \mathbfit{B}$. The electric vector of the electromagnetic wave follows the electron motion, thus also rotating couter-clockwise as viewed by the observer. This results in IEEE-RCP, and according to the IAU convention, $V_\nu > 0$. 

In this paper, we adopt the conventions conforming to the IEEE/IAU standard and stick to the magnetic field convention where the magnetic field is positive when pointing towards the observer\footnote{This is opposite to the astronomical convention that traditionally defines magnetic field direction as positive when pointing away from the observer (i.e. $\theta = 0$ corresponds to a negative field while $\theta = \pi$ corresponds to a positive field).}. We follow the same coordinate systems as \citet{HuangShcherbakov11} and use it as the main reference paper to check against the signs of the Stokes parameters and their corresponding transfer coefficients. The transfer coefficients therefore all have positive signs in their expressions. 

\section[]{Transfer coefficients}\label{App:transferCoef}

In this Appendix we present the transfer coefficients for both thermal bremsstrahlung and non-thermal synchrotron radiation process. The non-thermal relativistic electrons gyrating around magnetics field lines has a power-law energy spectrum.  
We adopt the expressions given in \citet{Pacholczyk77} and \citet{JonesOdell77_transfer} respectively, 
but the sign of the circular polarization described by Stokes $V$ are made to be consistent and complied to the IEEE/IAU convention, given the coordinate system explicitly shown in Appendix~\ref{app:adoptedCoordsys}. 
The emission coefficients have units of ${\rm erg}\,{\rm s}^{-1}\,{\rm cm}^{-3}\,{\rm Hz}^{-1}\,{\rm str}^{-1}$ and the absorption and Faraday coefficients have units of ${\rm cm}^{-1}$. 

\subsection{Thermal bremsstrahlung}\label{sec:therm} 
Transfer coefficients of thermal bremsstrahlung have been presented in \citet{Pacholczyk77, Meggitt82, Wickramasinghe85, Rybicki86}. 
In this paper, we adopt the expressions given in \citet{Pacholczyk77} and make certain changes such that the set of coefficients would follow the same conventions of polarization we have specified. 

For a magnetized thermal plasma, 
the coefficients of Faraday rotation and Faraday conversion are respectively,  
\begin{eqnarray}
f_{\rm th} &=& \frac{\left({\omega_{\rm p}^{2}}/{c\,\omega_{\rm B}}\right)\cos\theta}{\left({\omega^{2}}/{\omega_{\rm B}^{2}}\right)-1}   \   , {\rm and}
\label{eq:f}    \\
h_{\rm th} &=& \frac{\left(\omega_{\rm p}^{2}/c\,\omega_{\rm B}\right)\sin^{2}\theta}{2\,\left(\omega^{3}/\omega_{\rm B}^{3}-\omega/\omega_{\rm B}\right)}  
\label{eq:h} 
\end{eqnarray} 
\citep[][]{Pacholczyk77}, 
where  $\omega = 2\pi \nu$ is the radiation angular frequency, 
$\omega_{\rm p}= (4\pi n_{\rm e}e^{2}/m_{\rm e})^{{1}/{2}}$ is the plasma frequency,  
$\omega_{\rm B} = (eB/m_{\rm e}c)$ is the electron gyrofrequency, 
and $\theta$ is the angle between the radiation propagation and the magnetic field. 
The thermal bremsstrahlung components of the absorption coefficients are given by
\begin{eqnarray}
\kappa_{\rm th} &=& \frac{\omega_{\rm p}^{2}\left(2\omega^{4}+2\omega^{2}\omega_{\rm B}^2-3\omega^2\omega_{\rm B}^{2}\sin^{2}\theta+\omega_{\rm B}^4\sin^{2}{\theta} \right)}{2\,c\,\omega^2 \left(\omega^2-\omega_{\rm B}^2\right)^{2}}\,\nu_{\rm c}   
\label{eq:kappa_t}\   , \\
q_{\rm th} &=& \frac{\omega_{\rm p}^{2}\,\omega_{\rm B}^{2}\sin^{2}{\theta}\,\left(3\omega^2-\omega_{\rm B}^2 \right)}{2\,c\,\omega^2 \left(\omega^2-\omega_{\rm B}^2 \right)^2}\,\nu_{\rm c}
\label{eq:q_t}\   , {\rm and} \\
v_{\rm th} &=& \frac{2\,\omega_{\rm p}^{2}\,\omega\,\omega_{\rm B}\cos\theta}{c\,\left(\omega^2-\omega_{\rm B}^2 \right)^2}\,\nu_{\rm c}    
\label{eq:v_t}
\end{eqnarray} 
\citep[][]{Pacholczyk77}, where the collisional frequency is
\begin{eqnarray}
\nu_{\rm c} 
&= \frac{4\sqrt{2\pi}e^{4}n_{\rm e}}{3\sqrt{m_{\rm e}}\,(k_{\rm B}T_{\rm e})^{3/2}}\ln{\Lambda} 
&\approx 3.64\,n_{\rm e}\,T_{\rm e}^{-3/2}\ln{\Lambda} \   , 
\label{eq:collisionalfreq}
\end{eqnarray} 
with the Coulomb logarithm factor, for $\omega \gg \omega_{\rm p}$, 
\begin{eqnarray}
\Lambda = 
 \Biggl\{ \begin{array}{ll} 
  \ 
 \Big(\frac{2}{1.781}\Big)^{5/2}\Big( \frac{k_{\rm B}T_{\rm e}}{m_{\rm e}}\Big)^{1/2} \Big( \frac{k_{\rm B}T_{\rm e}}{e^2 \omega}\Big)  \  & ,\:{\rm for}\   T_{\rm e} \le 3.16 \times10^{5}~{\rm K} \\ 
  \   
  \\ \
 \frac{8\pi k_{\rm B}T_{\rm e}}{1.781 h\omega} \  & ,\:{\rm for}\  T_{\rm e} > 3.16 \times10^{5}~{\rm K} \end{array}  
\label{eq:coulomblog}
\end{eqnarray} 
 \citep[][]{KRLang74}.  
$k_{\rm B}$ is the Boltzmann constant and $T_{\rm e}$ is the temperature of the electrons in thermal equilibrium. The emission coefficients in $I$, $Q$ and $V$ can be computed via the Kirchoff's law: 
\begin{eqnarray}
\epsilon_{I, {\rm th}} = \kappa_{\rm th}\,B_{\omega}   \ , \,\,\,\, 
 \epsilon_{Q, {\rm th}} =q_{\rm th}\,B_{\omega}  \    ,{\rm and} \,\,\,\, 
 \epsilon_{V, {\rm th}} = v_{\rm th}\,B_{\omega}    \   ,
 \end{eqnarray} 
where the Planck function $B_{\omega} =k_{\rm B}T_{\rm e}\omega^{2}/(2{\pi}^{2}c^{2})$ by the Rayleigh-Jeans law. 

It is interesting to note that both the frequency dependence and the dependence on the magnetic field are different for 
Faraday rotation and Faraday conversion. 
The strength of the Faraday rotation effect is proportional to 
$\nu^{-2} n_{\rm e, th} |\mathbfit{B}_{\parallel}| \delta s$, and the strength of Faraday conversion is proportional to $\nu^{-3} n_{\rm e, th} |\mathbfit{B}_{\perp}|^{2} \delta s$, where $|\mathbfit{B}_{\parallel}| = |\mathbfit{B}| \cos{\theta}$, $|\mathbfit{B}_{\perp}| = |\mathbfit{B}| \sin{\theta}$, and 
$\delta s$ is the photon propagation length. 

Another useful remark concerns the use of rotation measure (RM) in the literature for quantifying the strength of Faraday rotation. 
RM is defined as ${\mathcal R} \equiv \Delta \varphi \, c^2/ \nu^2$, where $\varphi = 0.5 \arctan({U/Q})$. 
A widely-used formula in RM analysis is ${\mathcal R}(s)  = 0.812 \int^{s}_{s_0} \frac{{\rm d}s'}{\rm pc} \, \frac{n_{e}(s')}{\rm cm^{-3}} \, \frac{ |\mathbfit{B}_{\parallel}(s')|}{\mu{\rm G}}$\,${\rm rad}\,{\rm m}^{-2}$, which it can be shown that this is derived from the polarized radiative transfer equation (equation \ref{eq:PRT}), under the assumptions that the effects of emission, absorption, Faraday conversion and contribution from non-thermal electrons are negligible \citep[see][for details]{On18}. 
In a realistic situation, however, these assumptions do not hold. The intensity of $Q$ and $U$ of the observed polarized light is not solely dictated by Faraday rotation process. An accurate inference of magnetic field properties from the polarization signatures of observed light, therefore, demands a full polarized radiative transfer treatment.  

\subsection{Non-thermal synchrotron radiation}\label{sec:nontherm} 

We adopt the expressions of the transfer coefficients for cosmic synchrotron sources from \citet{JonesOdell77_transfer}, and 
make appropriate sign changes for the transfer coefficients at $V_\nu$ to keep a self-consistent polarization convention defined explicitly in this paper. 
For relativistic electrons following a power-law energy distribution with an index $p$ , 
\begin{eqnarray} 
\d n = [n_{\gamma}\gamma^{p}]\gamma^{-p} \Theta (\gamma - \gamma_{i}) g(\Psi) \d\gamma \d\Omega_{\Psi} \     
\label{eq:n_rel_e}
\end{eqnarray}
\citep{JonesOdell77_transfer}, where $\Theta (\gamma - \gamma_{i})$ is the step function, 
$\gamma_{i}$ is the low-energy cutoff of electrons, and 
$g(\Psi)$ is the pitch-angle distribution, normalized to $\int \d\Omega_{\Psi} g(\Psi)  =1$.
The corresponding number density of electron is 
\begin{eqnarray} 
n_{\gamma}= \int_{\gamma_i}^{\infty}\d\gamma[n_{\gamma}\gamma^{p}]\gamma^{-p} =[n_{\gamma}\gamma^{p}]\gamma_{i}^{-(p-1)}/(p-1) \,\,\,\,\,\,\,\,\,\, ,\:{\rm for}\  (p > 1)  \   .
\end{eqnarray} 
The normalization factor $[n_{\gamma}\gamma^{p}]$ and 
the index $p$ are related to the spectral index of the radiation by $\alpha = (p-1)/2$. 
The transfer coefficients for non-thermal synchrotron radiation are 
\begin{eqnarray}
f_{\rm nt} &=&  f_{\alpha}\kappa_{\perp}\left( \frac{\omega_{B_{\perp}}}{\omega}\right)^{2} (\ln \gamma_i)\,\gamma_{i}^{-2(\alpha+1)}\cot\theta   \left[1+\frac{\alpha+2}{2\alpha+3}\frac{\d\,{\left(\ln{g(\theta)}\right)}}{\d\,{\left(\ln{(\sin{\theta})}\right)}} \right]   \    , \\
\label{eq:f_nt}
h_{\rm nt}  &=&  h_{\alpha}\kappa_{\perp}\left(\frac{\omega_{B_{\perp}}}{\omega}\right)^{3} \gamma_{i}^{-(2\alpha-1)} 
 \left[\frac{1-(\omega_{i}/\omega)^{\alpha-1/2}}{\alpha-1/2} \right]   \  ,  \  {\rm for} \   (\alpha > 1/2)  \    , \\
\label{eq:h_nt}
\kappa_{\rm nt}  &=&  \kappa_{\alpha}\kappa_{\perp}\left(\frac{\omega_{B_{\perp}}}{\omega}\right)^{\alpha + 5/2}      \    , \\
\label{eq:k_nt}
q_{\rm nt}  &=& q_{\alpha}\kappa_{\perp}\left(\frac{\omega_{B_{\perp}}}{\omega}\right)^{\alpha + 5/2}   \       , \\
\label{eq:q_nt}
v_{\rm nt}  &=&  v_{\alpha}\kappa_{\perp}\left(\frac{\omega_{B_{\perp}}}{\omega}\right)^{\alpha+3} \cot\theta  
 \left[1+\frac{1}{2\alpha+3}\frac{\d\,{\left(\ln{g(\theta)}\right)}}{\d\,{\left(\ln{(\sin{\theta})}\right)}} \right]\   , \\
\label{eq:v_nt}
\epsilon_{I, {\rm nt}} &=& \epsilon_{\alpha}^{\,\,I}\epsilon_{\perp}\left(\frac{\omega_{B_{\perp}}}{\omega}\right)^{\alpha}       \    , \\
\label{eq:emiI_nt}
\epsilon_{Q, {\rm nt}}&=&  \epsilon_{\alpha}^{\,\,Q}\epsilon_{\perp}\left(\frac{\omega_{B_{\perp}}}{\omega}\right)^{\alpha}    \       ,{\rm and} \\
\label{eq:emiQ_nt}
\epsilon_{V, {\rm nt}} &=&  \epsilon_{\alpha}^{\,\,V}\epsilon_{\perp}\left(\frac{\omega_{B_{\perp}}}{\omega}\right)^{\alpha+1/2} \cot\theta  
\left[1+\frac{1}{2\alpha+3}\frac{\d\,{\left(\ln{g(\theta)}\right)}}{\d\,{\left(\ln{(\sin{\theta})}\right)}} \right]  \   ,
\label{eq:emiV_nt}
\end{eqnarray} 
\citep[and references therein]{JonesOdell77_transfer}, where 
$\omega_{B_{\perp}} = \omega_{\rm B}\,\sin \theta$,
$\kappa_{\perp} = (2\pi r_{\rm e}c)\,{\omega^{\,\,\,-1}_{B_{\perp}}}[4\pi g(\theta)][n_{\gamma}\gamma^{p}] $, 
$\epsilon_{\perp} = (m_{\rm e}c^{2})(r_{\rm e}/2 \pi c)\,{\omega_{B_{\perp}}}[4\pi g(\theta)][n_{\gamma}\gamma^{p}] $ 
with the classical electron radius $r_{\rm e} = e^{2}/{m_{\rm e}c^2}$, and 
the fiducial frequency $\omega_{i} = \gamma_{i}^{2}\omega_{B_{\perp}}$. 
The dimensionless functions in the transfer coefficients are 
\begin{eqnarray}
f_{\alpha} & = & 2\,\frac{\left(\alpha+3/2 \right)}{\alpha +1}     \    , \\ 
h_{\alpha} & = &1   \  ,        \\
\kappa_{\alpha} & = & \frac{3^{\alpha+1}}{4} \Gamma \left( \frac{\alpha}{2} +\frac{25}{12}   \right) \,  \Gamma\left( \frac{\alpha}{2} +\frac{5}{12}   \right)   \    ,  \\
q_{\alpha} & = &  \frac{\left(\alpha+3/2 \right)}{\left(\alpha+13/6 \right)}  \kappa_{\alpha}  \    ,   \\
v_{\alpha} & = & \frac{3^{\alpha+1/2}}{2} \, \frac{\left(\alpha+2 \right)}{\left(\alpha+1 \right)}\,\left( \alpha+ \frac{3}{2}\right)  
 \Gamma\left( \frac{\alpha}{2} +\frac{7}{6}   \right) \,  \Gamma\left( \frac{\alpha}{2} +\frac{5}{6}   \right)  \  , \\
\epsilon_{\alpha}^{\,\,I} & = &   \frac{3^{\alpha+1/2}}{4\left(\alpha+1 \right)} \, \Gamma\left( \frac{\alpha}{2} +\frac{11}{6}   \right) \,  \Gamma\left( \frac{\alpha}{2} +\frac{1}{6}   \right)       \    , \\
\epsilon_{\alpha}^{\,\,Q} & = &   \frac{\left(\alpha+1 \right)}{\left(\alpha+5/3 \right)} \,\epsilon_{\alpha}^{\,\,I}     \    , {\rm and}  \\
\epsilon_{\alpha}^{\,\,V} & = &   \frac{3^{\alpha}}{2} \, \frac{\left(\alpha+3/2 \right)}{\left(\alpha+1/2 \right)}\,\Gamma\left( \frac{\alpha}{2} +\frac{11}{12}   \right) \,  \Gamma\left( \frac{\alpha}{2} +\frac{7}{12}   \right)     \   .    
\end{eqnarray} 
The transfer coefficients are derived from a nearly isotropic dielectric tensor, 
appropriate for cosmic plasmas with low electron densities and weak magnetic fields, such that $\omega > \omega_{i}$ and both $\omega$ and $\omega_{i}$ are above the gyro-frequency $\omega_{\rm B}$. 
The condition $\gamma_{i}^{2} > \cot^{2}\theta$ also has to be satisfied. 
In addition, dielectric suppression is assumed to be negligible, which generally holds valid for cosmic media \citep[see][for details]{Jones74b, Melrose91Book}. In this paper isotropic electron distribution is assumed so $g(\theta) = 1/4\pi$. 
Comparing to the thermal bremsstrahlung expression in the high-frequency limit $(\omega \gg \omega_{\rm B})$, the non-thermal synchrotron Faraday rotation coefficient has an extra function factor 
$\zeta (p, \gamma_i) = \frac{(p-1)(p+2)}{(p+1)}\Big( \frac{\ln{\gamma_{\rm i}}}{\gamma_{\rm i}^{2}} \Big)$, implying that Faraday rotation weakens with increasing electron energy \citep[see also][]{Melrose97a, HuangShcherbakov11}. 

\section[]{Derivation of the covariant radiative transfer formulation}\label{App:Covariant}

Derivation of the covariant radiative transfer formulation has been presented in \citet[][]{Rybicki86, Fuerst04, Younsi12}. 
Here, we repeat the derivation for clarity and completeness. 

Consider a bundle of particles filling a phase--space volume element 
${\rm d}\mathcal{V} \equiv \d{\bf x}^{3}\,\d{\mathbfit p}^{3}$,
with 3-spatial volume element $\d{\bf x}^{3} = \d x\,\d y\,\d z$ and the 
3-momentum volume element $\d{\mathbfit p}^{3} = \d p_{x}\,\d p_{y}\,\d p_{z}$ at given time $t$. 
According to the Liouville's theorem, ${{\rm d}\mathcal{V}}/{{\rm d}\lambda_{\rm a}}=0$. 
Since ${\rm d}\mathcal{V}$ is conserved along the affine parameter $\lambda_{\rm a}$, it is Lorentz invariant. 

The distribution function\,(or phase space density) of the particles in the bundle is represented by  
$f(x^{i}, \,p^{i}) = \d N/{\rm d}\mathcal{V}$, 
where $\d{N}$ is the number of particles in $\d V$. 
Since $\d{N}/\d{\mathcal{V}}$ is Lorentz invariant, $f(x^{i}, \,p^{i})$ is also Lorentz invariant.

For photons, $v=c$ and $cp = E$, where $E$ is the photon energy.  
The spatial and momentum volume elements are $\d{\bf x}^{3}= \d A\,c\,\d t$ and $\d{\mathbfit p}^{3}= E^{2}\d E\,\d\Omega$, 
where $\d A$ is the area element through which the photons travel in the time interval $\d t$ 
and $\d\Omega$ corresponds to the direction of photon propagation.
It follows that 
\begin{eqnarray} 
f(x^{i},p^{i})=\frac{\d N}{\d A\,c\,\d t\,E^{2}\,\d E\,\d \Omega}   
\label{eq:distfun}
\end{eqnarray} 
\citep[see][]{Rybicki86}. 
The specific intensity of the radiation is 
\begin{eqnarray} 
I_{E}=\frac{E\d N}{\d A\,c\,\d t\,\d E\,\d\Omega}  \   .
\end{eqnarray} 
Comparing the two expressions yields
\begin{eqnarray} 
f(x^{i},p^{i}) = \frac{I_{E}}{E^{3}} = \frac{I_{\nu}}{\nu^{3}} \equiv \mathcal{I}_{\nu}   \    ,
\end{eqnarray} 
where $\mathcal{I}_{\nu}$ is the Lorentz-invariant intensity. 

The Lorentz-invariant absorption and emission coefficients 
are $\zeta_{\nu} = \nu\,\kappa_{\nu}$ and 
$\xi_{\nu} = {\epsilon_{\nu}}/\nu^{2}$, respectively \citep{Rybicki86}. 
It follows that the covariant radiative transfer equation takes the form
\begin{eqnarray}
\frac{\d{\mathcal I}_{\nu}}{\d\tau_{\nu}}= -{\mathcal I}_{\nu} + {\mathcal S}_{\nu} \   , 
\label{eq:covarRT_tau} 
\end{eqnarray}
where the source function ${\mathcal S}_{\nu} \equiv   \xi_{\nu}/\zeta_{\nu}= {\epsilon}_{\nu}/({\kappa_{\nu}\,\nu^{3}})$. 
Since $\zeta_{\nu}$ and $\xi_{\nu}$ are invariants under the Lorentz transformation, 
the transfer coefficients measured in the observer's frame relates to 
those in the co-moving frame\,(i.e.\,the local rest frame of the medium) via 
$\nu\,\kappa_{\nu} =  \nu_{{\rm co}}\,\kappa_{\nu, {\rm co}}$ and 
${\epsilon_{\nu}}/\nu^{2} = {\epsilon_{\nu, {\rm co}}}/\nu_{{\rm co}}^{2}$. 
Hence, the radiative transfer equation becomes 
\begin{eqnarray} 
\frac{\rm d{\mathcal I}_{\nu}}{\d s} =-\kappa_{\nu}\,{\mathcal I}_{\nu}+\frac{\epsilon_{\nu}}{\nu^{3}}  \    ,
\end{eqnarray} 
\label{eq:modiRT}
\citep[][]{Fuerst04, Younsi12}.

\section[]{Calculation of the total electron number density at the present epoch}\label{app:neIGM}

The Universe is neutral as a whole and the most common atoms in it are Hydrogen and Helium. 
We can approximate $n_{\rm e, tot}=n_{\rm p, tot}=n_{\rm p, He} +n_{\rm p, H}$, 
where ``p" stands for proton, ``H" for Hydrogen and ``He" for Helium; 
$n_{\rm p, He} \approx \rho_{\rm He}/m_{\rm He}$,  
and $n_{\rm p, H}  \approx \rho_{\rm H}/m_{\rm H}$. 
By approximating the density of Hydrogen taking up 75 \% of the density of baryons\,(i.e.\, 
$\rho_{\rm H}=3\rho_{\rm b}/4$), and the density of Helium taking up the remainder, it gives 
$n_{\rm e} = 7\rho_{\rm b}/8m_{\rm p}$. 
The value of $\rho_{\rm b, 0}$ can be calculated from $\Omega_{\rm b, 0} =\rho_{\rm b, 0}/\rho_{\rm crit}$, 
with $\Omega_{\rm b, 0}h^{2} = 0.02230$ \citep{Planck15}, and $\rho_{\rm crit} = 3H_{0}/(8\pi G) =1.87882 \times 10^{-29} h^2$. 
This gives $n_{\rm e, 0} = 2.1918 \times 10^{-7}$~cm$^{-3}$.

\section[]{Remarks on finding an appropriate scale length}\label{app:extrainfor}

Here, in Table \ref{tab:emi_absOnly}, we present the numerical values of the absorption, emission and Faraday rotation coefficients used in the calculations presented in Section \ref{subsec:SRTest}. In general, the very different properties of cosmic media lead to a wide range of orders of magnitude spanned by transfer coefficients in the CPRT equation, resulting in a stiff set of coupled differential equations to solve. It is therefore essential and important to test the capability of the equation solver and the stability of the numerical solution (see Section \ref{subsec:SRTest}). We emphasize that finding an appropriate scale length is crucial to overcoming the stiffness issue. In this work, the very small order of magnitude of the transfer coefficients computed using parameters typical to an IGM and an ICM at $\nu_{\rm obs} = 1.42$~GHz suggests a scale length of a few Mpc when determining the $z$-sampling scheme. 

In addition, note that all the CPRT calculations for the situations discussed in this paper are optically thin (i.e. $\tau \ll 1$). While the media are optically thin, they can be Faraday thick at the same time, such as in the cases of ICM-like environments. Numerical values of the optical depths and Faraday conversion coefficients obtained using the IGM-like model A-I and the ICM-like model B-I are included in Table \ref{tab:emi_absOnly}. Note also that the effect of Faraday conversion is usually much weaker than that of Faraday rotation. Hence, $V$ is nearly always zero in the cases of our interests. 

\begin{table}
\centering
 \begin{tabular}{|p{2.0cm}|C{2.5cm}|*2{C{5.0cm}|}} 
  \hline
  \multicolumn{2}{|c|}{}& IGM-like model A-I & ICM-like model B-I \\ 
  \hline
  \multicolumn{2}{|c|}{$\epsilon_{I,{\rm tot}}$} & \multicolumn{1}{l |}{$\epsilon_{I, {\rm th}} + \epsilon_{I,{\rm nt}}$}  &   \multicolumn{1}{l |}{$\epsilon_{I, {\rm th}} + \epsilon_{I, {\rm nt}}$}  \\     
  \multicolumn{2}{|c|}{(${\rm erg}\,{\rm s}^{-1}\,{\rm cm}^{-3}\,{\rm Hz}^{-1}\,{\rm str}^{-1}$)} & \multicolumn{1}{l |}{$ =  2.59 \times10^{-53} + 2.62 \times 10^{-55}$} &  \multicolumn{1}{l |}{$=  6.91 \times10^{-47} + 1.25 \times 10^{-38}$} \\   
   \multicolumn{2}{|c|}{} & \multicolumn{1}{l |}{$ =2.62 \times10^{-53}$} &  \multicolumn{1}{l |}{$= 1.25 \times10^{-38}$} \\ 
  \hline
  \multicolumn{2}{|c|}{$\kappa_{\rm tot}$} & \multicolumn{1}{l |}{$\kappa_{{\rm th}} + \kappa_{{\rm nt}}$}  &   \multicolumn{1}{l |}{$\kappa_{{\rm th}} + \kappa_{{\rm nt}}$}  \\
  \multicolumn{2}{|c|}{(${\rm cm}^{-1}$)} & \multicolumn{1}{l |}{$ =  2.23 \times 10^{-38} + 8.64 \times 10^{-52}$} &  \multicolumn{1}{l |}{$=  
  2.23 \times10^{-34} + 7.11 \times10^{-34}$} \\    
   \multicolumn{2}{|c|}{} & \multicolumn{1}{l |}{$ =2.23 \times10^{-38}$} &  \multicolumn{1}{l |}{$= 9.34 \times10^{-34}$} \\ 
  \hline
  \multicolumn{2}{|c|}{$\tau = {\int_{z_{\rm init}}^{0.0} \kappa_{\rm tot}(z)} \cdot \d{s}$} & \multicolumn{1}{l |}{$2.72 \times10^{-13}$}  &   \multicolumn{1}{l |}{$1.14 \times10^{-8}$}  \\
  \hline
  \multicolumn{2}{|c|}{$f_{{\rm tot}}$} & \multicolumn{1}{l |}{$f_{{\rm th}} + f_{{\rm nt}}$}  &   \multicolumn{1}{l |}{$f_{{\rm th}} + f_{{\rm nt}}$}  \\     
  \multicolumn{2}{|c|}{(${\rm cm}^{-1}$)} & \multicolumn{1}{l |}{$ =  2.54 \times 10^{-30} + 1.52 \times 10^{-33}$} &  \multicolumn{1}{l |}{$=  1.16 \times 10^{-23} + 8.54 \times 10^{-28}$} \\   
   \multicolumn{2}{|c|}{} & \multicolumn{1}{l |}{$ =2.54 \times10^{-30}$} &  \multicolumn{1}{l |}{$= 1.16 \times10^{-23}$} \\ 
  \hline
  \multicolumn{2}{|c|}{$h_{\rm tot}$} & \multicolumn{1}{l |}{$h_{{\rm th}} + h_{{\rm nt}}$}  &   \multicolumn{1}{l |}{$h_{{\rm th}} + h_{{\rm nt}}$}  \\
  \multicolumn{2}{|c|}{(${\rm cm}^{-1}$)} & \multicolumn{1}{l |}{$ =  3.76 \times 10^{-42} + 8.12 \times10^{-43}$} &  \multicolumn{1}{l |}{$=  
  1.72 \times10^{-32} + 3.01 \times10^{-32}$} \\    
   \multicolumn{2}{|c|}{} & \multicolumn{1}{l |}{$ =4.57 \times 10^{-42}$} &  \multicolumn{1}{l |}{$= 4.72\times 10^{-32}$} \\ 
  \hline 
 \end{tabular}
 \caption{Values of the transfer coefficients and optical depths computed using parameters of models A-I and B-I at radiation frequency $\nu = 1.4$~GHz. The transfer coefficients obtained have a very small order of magnitude, suggesting a scale length of a few Mpc.}
 \label{tab:emi_absOnly}
\end{table}

\bsp	
\label{lastpage}
\end{document}